\documentclass[aps,pre,letterpaper,superscriptaddress,nofootinbib]{revtex4}
\usepackage{amsfonts,amssymb,amsmath,bm,bbm,euscript,mathrsfs}
\usepackage[utf8]{inputenc}
\usepackage[T1]{fontenc}
\usepackage{textcomp}
\usepackage{graphicx}
\usepackage[tight,sf]{subfigure}
\usepackage{enumerate}
\usepackage{hyperref}
\usepackage{color}

\newcommand{\rmd}{\mathrm{d}}
\newcommand{\rmk}{\mathrm{k}}
\newcommand{\rmt}{\mathrm{t}}
\newcommand{\rmY}{\mathrm{Y}}

\newcommand{\rmF}{\mathrm{F}}
\newcommand{\rmJ}{\mathrm{J}}
\newcommand{\rmS}{\mathrm{S}}
\newcommand{\rmE}{\mathrm{E}}
\newcommand{\rmf}{\mathrm{f}}
\newcommand{\rmj}{\mathrm{j}}
\newcommand{\rms}{\mathrm{s}}
\newcommand{\calE}{\mathcal{E}}
\newcommand{\calH}{\mathcal{H}}
\newcommand{\calF}{\mathcal{F}}
\newcommand{\calT}{\mathcal{T}}

\newcommand{\bfY}{\mathbf{Y}}
\newcommand{\bfE}{\mathbf{E}}
\newcommand{\bfe}{\mathbf{e}}
\newcommand{\bfT}{\mathbf{T}}
\newcommand{\bfN}{\mathbf{N}}
\newcommand{\bfB}{\mathbf{B}}
\newcommand{\bfD}{\mathbf{D}}
\newcommand{\bfF}{\mathbf{F}}
\newcommand{\bfM}{\mathbf{M}}
\newcommand{\bfS}{\mathbf{S}}

\begin{document}

\title{First integrals for elastic curves: twisting instabilities of helices}

\author{Didier A. Solis}
\affiliation{Facultad de Matemáticas, Universidad Autónoma de Yucatán,\\
Periférico Norte, Tablaje 13615, C.P. 97110, Mérida, Yucatán, MÉXICO}
\author{Pablo Vázquez-Montejo\footnote{Author to whom any correspondence should be addressed.}}
\email[]{pablo.vazquez@conacyt.mx}
\email[]{pablo.vazquez@correo.uady.mx}
\affiliation{CONACYT - Facultad de Matemáticas, Universidad Autónoma de Yucatán,\\ 
Periférico Norte, Tablaje 13615, C.P. 97110, Mérida, Yucatán, MÉXICO}

%\date{}

\begin{abstract}
We put forward a variational framework suitable for the study of curves whose energies depend on their bend and twist degrees of freedom. By employing the material curvatures to describe such elastic deformation modes, we derive the equilibrium equations representing the balance of forces and torques on the curve. The conservation laws of the force and torque on the curve, stemming from the Euclidean invariance of the energy, allow us to obtain first integrals of the equilibrium equations. To illustrate this framework, we apply it to determine the first integrals for isotropic and anisotropic Kirchhoff elastic rods, whose energies are quadratic in the material curvatures. We use them to analyze perturbatively the deformations of helices resulting from their twisting. We examine three kinds of twisting instabilities on unstretchable helices, characterized by their wavenumbers, depending on whether their boundaries are fixed, displaced along the radial direction or orthogonally to it. We also analyze perturbatively the effect of the bending anisotropy on the deformed states, which introduces a coupling between deformation modes with different wavenumbers.
\end{abstract}

\maketitle

\section{Introduction}

To a good degree of accuracy, the behavior of unstretchable filaments and elastic rods subject to external forces can be described by their centerlines within a harmonic approximation, provided that their thickness is much smaller than their characteristic radius (the inverse of their characteristic curvature) \cite{AudolyBook}. In the simplest model, corresponding to the Euler elastica, their shape is determined by minimizing their bending energy, quadratic in the curvature \cite{LoveBook, AudolyBook}. In a more detailed treatment they are modeled by Kirchhoff rods, whose total energy takes into account not only their bending, but also the twisting about their centerline \cite{LoveBook, LandauBook, AudolyBook, OReillyBook, Dill1992}. This twisting energy is quadratic in the material curvature known as the twist, given by sum of the rate of change of the twist angle and the torsion \cite{LoveBook, Powers2010}.\footnote{In this work, we adopt the usual assumption that the energy required to stretch the rods is much larger than the bending energy, so their length can be considered fixed and the only relevant modes of deformation are their bending and twisting \cite{AudolyBook}. However, there is a further  generalization of the Kirchhoff rods, corresponding to the model known as Cosserat rods, which considers not only these two degrees of freedom, but also the elongation of the rods and the shear of their cross section \cite{Rubin2000, Gazzola2018}.}
\\
The twist of elastic rods gives rise to interesting phenomena. Analogous to the Euler buckling instability of a compressed rod subject to a critical force, there are twisting instabilities of straight or circular rods beyond a critical twist \cite{Thompson1996, GorielyI1997, Goriely2006, Ciarletta2014}. The conformations of biological polymers, relevant to their fundamental functional processes, can be influenced by their twisting, inducing transitions to supercoiled states such as knots or plectonemes, \cite{Marko1994, Marko1995, Schlick1995, Marko1997, Lavery2002, Audoly2007, Clauvelin2008, Charles2019}. Also, the fracture of long elastic rods due to large bending stresses can be controlled through its twisting, which influences the fragmentation cascade and reduces the number of resulting segments \cite{Heisser2018}.
\\
The Kirchhoff equations, establishing the balance of forces and torques along the rod, 
supplemented by linear elastic constitutive relations between torques and strains, provide the standard framework for the study of the dynamics of elastic rods \cite{LoveBook, AudolyBook, OReillyBook, Dill1992, Goriely2000}. These equations have proven appropriate to describe the behavior of a wide range of slender elastic bodies in the harmonic approximation, from microscopic ones such as semiflexible polymers like the DNA, to macroscopic ones such as the human hair, ribbons, plant tendrils and cables \cite{AudolyBook, OReillyBook, Shi1994, Goriely1998, Goriely2000, Starostin2008, Powers2010, Yu2019}. They have been used extensively to study the conformations and dynamics of open and closed elastic rods, \cite{LoveBook, AudolyBook, GorielyI1997, GorielyII1997, GorielyIII1997, Goriely2000}.
\\
However, often additional effects not captured in the harmonic approximation are relevant in the description of filaments, so it is necessary to consider more general energies, which may depend on higher orders of the material curvatures. Some of these energies arise from the asymmetric structure of the filaments, such as the asymmetry in the left and right handed twist of DNA, which introduces a coupling of bend and twist, as well as cubic and quartic terms in the material curvatures \cite{Fuller1971, Marko1994, Eslami2009}. Furthermore, the harmonic approach is accurate for length scales larger than the persistence length of the DNA, whereas for smaller length scales an anharmonic description is required to describe kinks exhibiting high curvature effects \cite{Wiggins2006A, Wiggins2006B}.
\\
In considering the energies of the elastic rods it is more convenient to adopt a variational approach rather than working with the balance equations for the components of forces and torques. In this approach the energy of the elastic rods is minimized to determine the corresponding Euler-Lagrange (EL) equations, which upon integration for appropriate boundary conditions, yield their equilibrium configurations. Both frameworks are equivalent, for the EL equations also represent the balance of forces and torques on the elastic rods, typically described using the Frenet-Serret (FS) frame \cite{CapoChryssGuv2002, Tu2008, Starostin2009}. The EL equations for energies depending only on the curvature and torsion have been derived already \cite{CapoChryssGuv2002, Thamwattana2008A, Thamwattana2008B}. However, as is well known, the FS frame fails to capture the twist of the cross section, so in order to take into account this degree of freedom, one could introduce the twist angle and derive its corresponding EL equation \cite{Tu2008, Starostin2009, Gerhardt2013}.
\\
By exploiting the fact that one has the freedom to use any adapted frame to describe a curve \cite{Bishop1975}, here we present a variational framework in which the geometric quantities of the elastic rod are defined with respect to the material frame adapted to its centerline, rather than with respect to the FS frame. Therefore, the three material curvatures replace the curvature, torsion and the twist angle as the variables characterizing the curve.
\\
 Although several energies depending on the material curvatures have been introduced, very often instead of working with the corresponding equilibrium equations, different approaches, such as discretized models, molecular dynamics simulations or statistical mechanics techniques (mainly Monte Carlo simulations) have been employed to study the corresponding equilibrium configurations. This might be due to the fact that the usual procedure to minimize the energy of the elastic rods requires the determination of the variation of the curvatures as a response to a change of its embedding functions \cite{CapoChryssGuv2002, Thamwattana2008A}, which in general involves lengthy calculations that require a careful treatment \cite{Zhang2004, Zhao2006, Thamwattana2008B, Gerhardt2013}. Here, to simplify the variational principle we use the method of Lagrange multipliers, in which auxiliary variables are introduced to impose the relations between the different geometric quantities of the rod. This method has been used in the derivation of the EL equations of elastic curves confined to surfaces \cite{GuvVaz2012, GuvValVaz2014}, as well as in the derivation of the EL equations of surfaces whose energies depend on their geometry \cite{Guven2004, Guven2018}. In this manner, we can vary independently the material curvatures, the material basis and the embedding functions. This provides us directly the two EL equations representing the force balance along the normal directions and a third equation representing the balance of torques along the curve.
\\
The Euclidean invariance of the energy entails the conservation of the force and torque vectors of the rod \cite{CapoChryssGuv2002}. In turn, these conservation laws imply that the magnitude of the force and the projection of the torque onto the force vector are constant along the rod. These two conserved quantities allow us to obtain first integrals of the EL equations. Furthermore, the conservation of the force and torque vectors single out a cylindrical coordinate basis adapted to the rod \cite{LandauBook, SingerSantiago2008}, which enables us to reconstruct the embedding functions once the material curvatures have been determined.
\\
As a first application of our framework, we are able to determine the first integrals of isotropic and anisotropic\footnote{Anisotropic rods have also been defined with a curvature weighted by a function depending on arc length \cite{Palmer2020A, Palmer2020B}. In this work we consider the usual definition of rods with different bending modes along orthogonal directions.} inextensible Kirchhoff rods with spontaneous curvatures, whose energies are quadratic in the material curvatures. For the anisotropic rods we also consider in passing  a coupling between the bend and twist degrees of freedom. We show that twisted circular helices are solutions of the first integrals for isotropic rods. Since their curvature and torsion are constant, they provide the simplest non-trivial equilibrium states of isotropic rods,  \cite{AudolyBook, GorielyIII1997}, which includes their limit cases with vanishing torsion: linear and circular rods. Taking circular helices without spontaneous curvature as reference configurations, we examine perturbatively their instabilities caused by twisting. To this end, we proceed analogously to the determination of the Euler buckling instability, so we look at the solutions of the linearized first integrals near circular helices. We show that to first order there are three types of solutions depending on whether the boundaries are fixed, displaced radially or along their tangents (orthogonally to the radial direction). We determine the minimum twist required to drive each deformation in terms of the wavenumber of the corresponding perturbation and of the curvature and torsion of the fiducial helices.
\\
We also address perturbatively how the anisotropy in the bending modifies the twisting instabilities of helices. We consider rods with a small bending anisotropy, characterized by the ratio of the bending rigidities, which allows us to linearize the EL equations about twisted isotropic helices without spontaneous curvature. We show that their solutions are given by the sum of two perturbations, both proportional to the small bending anisotropy, but one with the same wavenumber as the initial twist and the other one with thrice the wavenumber. Like the case of instabilities on isotropic helices, there are also three kinds of perturbations depending on the boundary conditions. We also determine the twist, forces and torques required for the onset of each instability, as well as the corresponding intrinsic tension and the total energy of the deformed rods.
\\
This paper is organized as follows. In Sec. \ref{sec:MatFr} we review the description of a curve using an adapted material frame. The variational principle leading to the EL equations  is developed in Sec. \ref{sec:Energy}. The identification of the conservation laws of the force and torque vectors arising from the Euclidean invariance of the energy is presented in Sec. \ref{sec:FMEucInv}. The first integrals of the EL equations are presented in Sec. \ref{sec:1stint}. The reconstruction of the curve in terms of the cylindrical basis adapted to the force and torque vectors is performed in Sec. \ref{sec:Reconstruction}. The specialization of this framework to isotropic and anisotropic Kirchhoff rods, as well as the analysis of twisting instabilities of isotropic and anisotropic helices are presented in Secs. \ref{Sec:Isorods} and \ref{Sec:Anisorods}. Lastly, we close with the discussion and conclusions of our work in Sec. \ref{Sec:conclusions}. In order to compare our framework with previous works, in Appendix \ref{App:FStoMF} we present the connection between the material and FS approaches, whereas in Appendix \ref{App:altderELeqs} we include the variations of the curve required to perform the variational principle in the usual manner.

\section{Kinematics of the curve in terms of the material frame} \label{sec:MatFr}

\begin{figure}[htb] 
 \centering
 \includegraphics[scale=0.5]{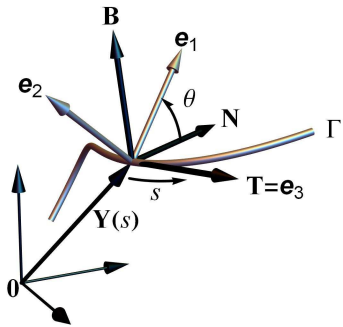}
  \caption{Curve $\Gamma$ parametrized by arc length $s$. The material frame $\{\bfe_1,\bfe_2,\bfe_2\}$, adapted to the curve is obtained by rotating the FS frame $\{\bfT,\bfN,\bfB\}$ about the tangent vector by an angle $\theta$, which captures the twist along the curve.} \label{fig:1}
\end{figure}

A filament or rod whose cross section radius is much smaller than its length admits a description by its centerline, parametrized by arc-length $s$ and embedded in Euclidean space $\mathbb{E}^3$ through the functions $\Gamma : s \rightarrow \bfY(s) = \rmY^i(s) \, \bfE_i$, with $\bfE_i$ the canonical orthonormal basis in $\mathbb{E}^3$, see Fig. (\ref{fig:1}).
Usually the kinematics of the curve are described using the FS frame $\{\bfT, \bfN, \bfB\}$ and the curvature $\kappa$ and torsion $\tau$. However, one might also be interested in the twist of the rod, in which case the appropriate basis for the description of the curve is the material frame adapted to $\Gamma$. This frame is constituted by the right-handed orthonormal basis $\{\bfe_1, \bfe_2,\bfe_3\}$, where  $\bfe_1$ and $\bfe_2$ are any two unit vectors lying on the plane of the cross section, and the third vector is chosen as the unit tangent vector, $\bfe_3 = \bfY'$, where $'=\partial/\partial s$. Thus, they satisfy $\bfe_i \cdot \bfe_j = \delta_{ij}$ and their orientation is given by $\bfe_i = 1/2 \, \epsilon_{ijk} \bfe_j \times \bfe_k$, $i,j,k = 1, 2, 3$, where $\epsilon_{ijk}$ is the Levi-Civita symbol (which can be defined by the scalar triple product $\epsilon_{ijk} = \bfe_i \cdot \left(\bfe_j \times \bfe_k \right)$). The structure equations of the material frame adapted to $\Gamma$, analogous to the FS equations, are
\begin{subequations} \label{def:Streqs}
\begin{eqnarray}
\bfe_1' &=& \kappa_3 \bfe_2 - \kappa_2 \, \bfe_3 \,, \\
\bfe_2' &=& -\kappa_3 \, \bfe_1 + \kappa_1 \, \bfe_3 \,, \\
\bfe_3' &=& \kappa_2 \, \bfe_1 -\kappa_1 \, \bfe_2\,.
\end{eqnarray}
\end{subequations}
These structure equations can be written succinctly as $\bfe_i' = \bfD \times \bfe_i = -\epsilon_{ijk} \kappa_j \bfe_k$, where $\bfD = \kappa_i \, \bfe_i$ is the Darboux vector describing the change of the material frame along the curve \cite{AudolyBook, KreyszigBook}. In particular, by taking the cross product of the tangent vector with its derivative we get the identity 
\begin{equation} \label{def:Dperp}
\bfe_3 \times \bfe_3' = \bfD - (\bfD \cdot \bfe_3 )\bfe_3 = \bfD_\perp := \kappa_I \bfe_I, \quad I=1,2 \,.
\end{equation}
The material curvatures $\kappa_i = 1/2 \, \epsilon_{ijk} \, \bfe'_j \cdot \bfe_k$, quantify the rate of rotation of the material frame about the corresponding material direction $\bfe_i$. The first two curvatures $\kappa_1 = \bfe'_2 \cdot \bfe_3$ and $\kappa_2 = \bfe_3' \cdot \bfe_1$ encode the rotation of the frame about the two normals and the sum of their squares is the squared curvature, i.e. $\kappa^2 = \kappa_1 ^2 + \kappa_2^2$. The third curvature $\kappa_3 = \bfe_1' \cdot \bfe_2$, known as the local twist of the curve, captures the rotation about the tangent. The twist differs with the torsion by the derivative of the twist angle $\theta$ that the normal material frame makes with the normal FS frame measured in the clockwise sense, i. e. $\kappa_3 = \tau + \theta'$. Further details on the connection between the descriptions of the curve in the two frames are presented in Appendix \ref{App:FStoMF}.
\\
The integral of the twist along the curve is known as the twisting number, denoted by $\mathcal{T}w$ and defined by \cite{Fuller1971, Starostin2005}
\begin{equation} \label{def:Tw}
\mathcal{T}w[\bfY] = \frac{1}{2 \pi} \int (\bfe_3 \times \bfe_1) \cdot \rmd \bfe_1= \frac{1}{2 \pi} \int \kappa_3 \, \rmd s \,. 
\end{equation}
Choosing an adapted frame to the curve, i.e. taking the tangential vector as one of the vectors basis, breaks the full rotational invariance in the selection of the orthonormal basis that could be used to describe the curve. However, there is a residual rotational invariance in the selection of the normal vectors. To emphasize this fact, besides the lowercase indices referring to the material frame, $i=1,2,3$, we use uppercase indices referring only to the normal basis, $I=1,2$. The corresponding Levi-Civita symbols for each basis are $\varepsilon_{ijk}$ and $\varepsilon_{IJ}$.
\\
From the structure equations of the normal vectors, we see that the twist $\kappa_3$ plays the role of the normal connection, known also as the spin connection \cite{Kamien2002}: $\omega_{IJ} = \bfe_I'\cdot \bfe_J = \varepsilon_{IJ} \kappa_3$, i.e. $\kappa_3 = \omega_{12} =  -\omega_{21}$. Defining also the antisymmetric normal curvature $K_I =  \varepsilon_{IJ} \kappa_J$ ($K_1 = \kappa_2$, $K_2 = -\kappa_1$), we can recast the structure equations of the material frame in a manner analogous to the Gauss-Weingarten equations for surfaces:
\begin{subequations}
 \begin{eqnarray} 
\bfe_I' &=& \omega_{IJ} \bfe_J -K_I \bfe_3, \label{def:streqsI}\\
\bfe_3' &=& K_I \bfe_I\,. \label{def:streqs3} 
\end{eqnarray}
\end{subequations}
Furthermore, the spin connection permits us to define a covariant derivative of the normal material basis
\begin{equation} \label{def:nrmcovdereI}
\nabla_s \bfe_I := (\delta_{IJ} \partial_s - \omega_{IJ}) \bfe_J = -K_I \bfe_3\,.
\end{equation}
This covariant derivative accounts for the rotational degree of freedom in the normal space to the curve \cite{CapoGuv1995, GuvSanVaz2013}. Moreover, since the normal connection acts only on the normal sector, this covariant derivative acts on the tangent vector in the same manner as for a scalar function, $\nabla_s \bfe_3 := \bfe_3' = K_I \bfe_I$.
\\
If we decompose a vector $\mathbf{A}$ in its tangential and normal parts as $\mathbf{A} = \mathbf{A}_\perp + \mathbf{A}_\parallel$, with  $\mathbf{A}_\perp= A_I \bfe_I$, $I=1,2$ and $\mathbf{A}_\parallel = A_3 \bfe_3$; then, by differentiating the normal part and using the structure equation (\ref{def:streqsI}) we have
\begin{equation} \label{dsVect}
 \mathbf{A}_\perp' = (\partial_s A_I - \omega_{IJ} A_J)  \bfe_I - K_I A_I \bfe_3\,.
\end{equation}
From this equation we define the covariant derivative of the normal components of a vector as
\begin{equation} \label{def:nrmcovder}
\nabla_s A_I \equiv \mathbf{A}_\perp' \cdot \bfe_I =  (\delta_{IJ} \partial_s - \omega_{IJ}) A_J \,.
\end{equation}
Using definitions (\ref{def:nrmcovdereI}) and (\ref{def:nrmcovder}) we can rewrite Eq. (\ref{dsVect}) as $ \mathbf{A}_\perp' = \nabla_s A_I \bfe_I + A_I \nabla_s \bfe_I = \nabla_s \mathbf{A}_\perp$, so the covariant derivative also acts on the normal part of a space vector as it does on a scalar function.
\\
In particular, for the normal part of the Darboux vector $\bfD_\perp$, defined in Eq. (\ref{def:Dperp}), the tangential component of its arc length derivative vanishes due to the contraction of the antisymmetric tensor with a symmetric combination of the normal curvatures, $\bfD_\perp' \cdot \bfe_3 = - \varepsilon_{IJ} \kappa_I \kappa_J =0$. Thus, the covariant derivative of the normal part of the Darboux vector remains normal:
\begin{equation}
\nabla_s \bfD_\perp = \bfD_\perp' = (\partial_s \kappa_I - \omega_{IJ} \kappa_J)  \bfe_I \,.
\end{equation} 
We obtain the covariant derivative of the normal curvatures by projecting this derivative onto the normal vectors
\begin{equation} \label{def:nrmcovderkappa}
\nabla_s \kappa_I := \bfD_\perp' \cdot \bfe_I =  (\delta_{IJ} \partial_s - \omega_{IJ}) \kappa_J \,,
\end{equation}
or explicitly, 
\begin{equation} \label{def:nrmcovderfull}
\nabla_s \kappa_1 = \kappa_1' - \kappa_3 \kappa_2\,, \qquad
\nabla_s \kappa_2 = \kappa_2' + \kappa_3 \kappa_1\,.
\end{equation}
Multiplying these equations by $\kappa_1$ and $\kappa_2$ and adding them up yields $\nabla_s \kappa^2 = (\kappa^2){}'$, verifying that the curvature is a scalar. Multiplying the first one by $\kappa_2$ and the second by $\kappa_1$ and taking their difference provides the identity
\begin{subequations}
\begin{eqnarray} \label{def:nrmcovderiden}
\kappa_2 \nabla_s \kappa_1 - \kappa_1 \nabla_s \kappa_2 &=& \kappa_2 \kappa_1' - \kappa_1 \kappa_2' - \kappa_3 \kappa^2 \,.
\end{eqnarray}
\end{subequations}
Dividing this identity by $\kappa^2$, we can write it as
\begin{equation} \label{def:identarctan}
\nabla_s \arctan \left( \frac{\kappa_1}{\kappa_2} \right) := \arctan \left(\frac{\kappa_1}{\kappa_2}\right)' - \kappa_3\,.
\end{equation}
Taking into account that the twist angle is given by the relation $\theta =\arctan (\kappa_1 / \kappa_2)$ (see Appendix  \ref{App:FStoMF}), we get that the covariant derivative of $\theta$ is given by the negative of the torsion: 
\begin{equation} \label{def:nrmcovderomega}
\nabla_s \theta = \theta' -\kappa_3 = -\tau\,.
\end{equation}
This reflects the fact that the twist angle is a gauge degree of freedom rather than a scalar.
\\
This quantities appear for example in the second derivative of the tangent vector
\begin{equation} \label{eq:e3pp}
 \bfe_3''= -\epsilon_{ij3} \nabla_s \kappa_i \bfe_j - \kappa^2 \bfe_3 =  \nabla_s \kappa_2 \bfe_1-\nabla_s \kappa_1 \bfe_2 -\kappa^2 \bfe_3
\end{equation}
Furthermore, using Eq. (\ref{def:nrmcovderomega}) we can write the cross product of the first and second derivatives of $\bfe_3$ as
\begin{equation} \label{eq:e3pce3pp}
\bfe_3' \times \bfe_3'' = \kappa^2 \bfD_\perp + \left( \kappa_1 \nabla_s \kappa_2 -\kappa_2 \nabla_s \kappa_1\right) \bfe_3 =\kappa^2 (\bfD_\perp + \tau \bfe_3)\,.
\end{equation}

\section{Energy and stresses} \label{sec:Energy}

In general, the energy density of the rod will be a scalar function $\calF$ depending on the material curvatures of $\Gamma$ and their derivatives, so the total energy is given by the line integral \cite{Kamien2002, Starostin2009, Powers2010, GuvValVaz2014}
\begin{equation} \label{def:H}
E[\bfY] = \int_\Gamma \calF(\kappa_i,\kappa_i') \, \rmd s\,.
\end{equation}
For instance, the energy densities penalizing the bending and twisting of an isotropic rod (with circular cross section) are quadratic in the curvature and twist \cite{Goldstein1998, Wolgemuth2000, Powers2010},
\begin{equation}\label{def:TBEnden}
\calF_{IB} = \frac{\rmk}{2} \left(\kappa - c_0 \right)^2\,, \quad \calF_{Tw} = \frac{\rmk_3}{2} \left(\kappa_3 - c_3 \right)^2 \,,
\end{equation}
where $\rmk$ and $\rmk_3$ are the bending and twisting moduli. The bending modulus is given by $\rmk=\rmE I$, where $\rmE$ is the Young modulus and $I=\pi R^4/4$ is the principal moment of inertia, with $R$ the radius of the cross section, whereas the twist modulus is $\rmk_3=\pi \mu R^4/2$, with $\mu$ a Lamé coefficient \cite{LandauBook, AudolyBook}. Thus, both moduli scale with the fourth power of thickness of the rods. For ordinary materials the ratio of twisting to bending moduli is close to one, $\chi_3 = \rmk_3/\rmk \sim 1$, \cite{LandauBook, Neukirch2004}, but in general it can take values in the range $2/3 \leq \chi_3 \leq 1$, \cite{Schlick1995, Goriely2006}. Here $c_0$ and $c_3$ denote the spontaneous curvature and twist, respectively. In the case of an anisotropic rod it bends differently along two orthogonal directions, known as the principal directions of curvature \cite{AudolyBook}. Moreover, the rod might have different spontaneous curvatures along such directions, $c_i$ \cite{Goldstein2000, Miller2014}, or different material moduli, $\rmk_I$, $I=1,2$ \cite{Langer1996, Kamien2009}. For instance, for a rectangular cross section with lenghts $l_1$ and $l_2$, $\rmk_1=E I_1$ and $\rmk_2=E I_2$, with $I_1=l_1^3 l_2/12$ and $I_2=l_1 l_2^3/12$ the principal moments of inertia \cite{LandauBook, AudolyBook}. Hence, the bending energy of an anisotropic rod is quadratic in the material curvatures, \cite{LandauBook, AudolyBook, Marko1994, Tu2008, GuvValVaz2014}:
\begin{equation} \label{def:anisenden}
\calF_{AB} = \sum_{I=1}^2 \frac{\rmk_I}{2} \, \left(\kappa_I - c_I\right)^2 \,.
\end{equation}
If the material frame is obtained by a constant rotation of the FS frame, so the twist angle is constant, $\theta =\theta_0$ (the twist reduces to the torsion $\kappa_3 = \tau$), the two bending moduli are equal $\rmk_1= \rmk_2$, and the spontaneous normal curvatures are given by $c_1 = c_{0} \sin \theta_0$, $c_2 = c_{0} \cos \theta_0$, then $\calF_{AB}$ reduces to $\calF_{IB}$.
\\
Cross terms in the material curvatures are usually disregarded in the energy of rods. For instance, a term proportional to the product of the normal curvatures, $\kappa_1 \kappa_2$, can be neglected by choosing the normal vectors along the principal axes of inertia \cite{LandauBook}. However, such term is relevant in certain models, for instance, it might be introduced to account for the interaction between the molecules of a membrane's edge,  where one of the normals is fixed by the normal direction of the membrane \cite{Asgari2015}. With the other cross terms we can define the following energy density representing a coupling between the bending and twisting of a rod
\begin{equation} \label{Fchiral}
 \calF_{BT} = \rmk_{13} \kappa_1 \kappa_3 + \rmk_{23} \kappa_2 \kappa_3\,,
\end{equation}
where $\rmk_{i3}$, $i=1,2$, are constants quantifying the bend-twist coupling. Such terms have been added to the energy of semiflexible polymers to describe the asymmetry in the chiral structure of the DNA \cite{Marko1994, Haijun1998}.
\\
Furthermore, one could go beyond in the anharmonic regime and consider higher order terms in the curvatures, which can be constructed from the derivatives of the tangent vector or by simply considering a Taylor series expansion of the energy in the material curvatures  \cite{Helfrich1991, LandauBook, Marko1994, Kamien2002}. For instance, from Eq. (\ref{eq:e3pce3pp}) we obtain a cubic term in the curvature and torsion given by $\bfe_3 \cdot \bfe_3' \times \bfe_3''= \kappa^2 \tau$, which has its  counterpart involving the twist $\kappa^2 \kappa_3$. Besides, a cubic term in the twist can considered, so we have the energy density for isotropic rods
\begin{equation}
 \calF_{IC} = \rmk_C \kappa^2 \kappa_3 + \rmk_{C3} \kappa_3^3\,. \quad (\rmk_{C}, \;
 \rmk_{C3} \; \mbox{constants})
\end{equation}
Both terms could be positive or negative depending on the sign of $\kappa_3$, feature that may also account for the chirality of the filament or rod \cite{Marko1994, Eslami2009} (property not shared by a term $\kappa^3$ which is always positive). In order to keep the total energy bounded from below quartic terms could be included, such as
\begin{equation}
 \calF_{IQ} = \rmk_Q \kappa^4 + \rmk_{Q3} \kappa^2 \kappa_3^2 + \rmk_{Q33}\kappa_3^4\,.  \quad (\rmk_{Q}, \; \rmk_{Q3}, \; \rmk_{Q33} \; \mbox{constants})
\end{equation}
Likewise, the corresponding terms for anisotropic rods in the anharmonic regime could be \cite{Marko1994, Helfrich1991, Kamien2002, Eslami2009}
\begin{subequations} \label{def:anisocubquaren}
\begin{eqnarray}
 \calF_{AC} &=& \rmk_{C1} \kappa_1^3 + \rmk_{C2} \kappa_2^3 +( \rmk_{C13} \kappa_1^2 + \rmk_{C23} \kappa_2^2) \kappa_3 + \rmk_{C3} \kappa_3^3\,, \quad (\rmk_{CI}, \; \rmk_{CI3}, \; \rmk_{C3} \; \mbox{constants})\\
 \calF_{AQ} &=& \rmk_{Q1} \kappa_1^4 +\rmk_{Q2} \kappa_2^4 +(\rmk_{Q13} \kappa_1^2+ \rmk_{Q23} \kappa_2^2) \kappa_3^2 + \rmk_{Q33} \kappa_3^4\,. \quad (\rmk_{QI}, \; \rmk_{QI3}, \; \rmk_{Q33} \; \mbox{constants})
 \end{eqnarray}
\end{subequations}
Some of these terms might vanish on account of symmetry properties of the curve. For instance requiring the energy of a segment of the curve to be invariant under a rotation about $\bfe_1$ by an angle of $\pi$ implies the vanishing of odds terms in $\kappa_1$, such as those  $\propto \kappa_1 \kappa_3, \kappa_1^3$, so that $\rmk_{13}=0$ and $\rmk_{C1}=0$ \cite{Marko1994}.
\\
The equilibrium configurations of the rod are determined by minimizing the energy $E[\bfY]$ defined in Eq. (\ref{def:H}). The usual manner involves the determination of the variation of the curvatures in terms of the variation of the embedding functions, which are presented in Appendix \ref{App:altderELeqs}. However, the variational principle can be greatly simplified by introducing Lagrange multipliers implementing the relations between the different geometric quantities. Three auxiliary fields are required: a three dimensional vector $\bfF$, enforcing the identification of the derivatives of the embedding functions as the tangent vector; a symmetric tensor $\lambda_{ij}$ (with six independent  components) ensuring the orthonormality of the material frame, and the components $\rmS_i$ of a three-dimensional vector, implementing the definition of the material curvatures. Therefore, we consider the  functional
\begin{equation} \label{def:HE}
E_C[\bfY, \bfe_i, \kappa_i, \kappa_i', \bfF, \lambda_{ij}, \Lambda_{i}] = E[\kappa_i,\kappa_i'] + \int \left[ \bfF \cdot \left(\bfY' - \bfe_3\right) + \frac{1}{2} \lambda_{ij} \left( \bfe_i \cdot \bfe_j - \delta_{ij} \right) + \rmS_i \left(\frac{1}{2} \varepsilon_{ijk} \, \bfe'_j \cdot \bfe_k - \kappa_i \right) \right] \rmd s \,.
\end{equation}
Since the definitions of the curvatures and vectors in terms of the derivatives of the embedding functions have been specified explicitly, their variations can be considered independently. The benefit is that in this manner the total energy $E$ is now regarded as a functional depending on $\kappa_i$, rather than on derivatives of $\bfY$. In full, $E_C$ has a functional dependence on the 3 components of the embedding functions, the 3 material curvatures and the 9 components of the material basis, as well as on the 12 components of the Lagrange multipliers. Thus, as shown in the following, the variations of $E_C$ with respect to $\bfY$, $\kappa_i$ and $\bfe_i$ yield 15 equations, which determine the 12 components of $\bfF$, $\lambda_{ij}$ and $\Lambda_i$, whereas the 3 remaining equations correspond to the EL equations governing the equilibria of the rods.
\\
First, from the response of $E_C$ under a variation of the embedding functions $\delta \bfY$  follows that the vector $\bfF$ is conserved in equilibrium \cite{LandauBook, CapoChryssGuv2002}: 
\begin{equation} \label{eq:deltaHY}
\frac{\delta E_C}{\delta \bfY} = \mathbf{0}  \quad \Rightarrow  \quad \bfF'=\mathbf{0}\,.
\end{equation}
In Sec. (\ref{sec:FMEucInv}),  this vector will be identified as the constant force along the rod. Expressing the vector in the material basis as $\bfF = \rmF_I \bfe_I + \rmF \bfe_3$, its arc-length derivative can be spanned likewise as
\begin{equation}
 \nabla_s \bfF := \bfF' = \mathcal{E}_I \bfe_I + \mathcal{E}_3 \bfe_3\,,
\end{equation}
where the components are given by
\begin{equation} \label{ELderI3}
\mathcal{E}_I = \nabla_s \rmF_I + \varepsilon_{IJ} \kappa_J \rmF_3  \,, \quad \mathcal{E}_3 = \nabla_s \rmF_3 + \varepsilon_{IJ} \kappa_I \rmF_J\, .
\end{equation}
The equilibrium equations for the rod are given by the vanishing of the normal projections, $\mathcal{E}_I=0$.
\\
From the variations of $E_C$ with respect to the material curvatures, we readily identify the Lagrange multipliers $\rmS_i$ as the EL derivatives of the energy density with respect to the material curvatures,
\begin{equation} \label{def:Si}
\frac{\delta E_C}{\delta \kappa_i} = 0  \quad \Rightarrow  \quad \rmS_i := \frac{\delta E}{\delta \kappa_i} = \frac{\partial \calF }{\partial \kappa_i} - \left(\frac{ \partial \calF }{\partial \kappa_i'}\right)'\,.
\end{equation} 
Furthermore, stationarity of $E_C$ under variations with respect to the material frame implies that
\begin{equation} \label{eq:deltaHei}
\frac{\delta E_C}{\delta \bfe_i} = \mathbf{0}  \quad \Rightarrow \quad \lambda_{ij}  \, \bfe_j + \frac{1}{2} \, \epsilon_{ijk} \, \rmS_j' \, \bfe_k + \rmS_j \left(\kappa_j \, \bfe_i - \kappa_i \, \bfe_j\right) = \delta_{i3} \bfF \,.
\end{equation}
The linear independence of the material frame $\bfe_i$ in the equations of the normal variations ($i=1,2$) allows us to determine the following components
\begin{subequations} \label{eq:ELder12}
\begin{align} 
\lambda_{11} & =- \kappa_2 \, \rmS_2 - \kappa_3 \rmS_3  \,, & \lambda_{12} &=  \frac{1}{2} \, \rmS_3' + \kappa_1 \, \rmS_2\,,&  \lambda_{13} &= - \frac{1}{2} \rmS_2' + \kappa_1 \rmS_3 \,,& \\
\lambda_{21} & =- \frac{1}{2} \, \rmS_3' + \kappa_2 \, \rmS_1 \,, & \lambda_{22} & = - \kappa_1 \, \rmS_1  - \kappa_3 \rmS_3 \,,&  \lambda_{23} &=  \frac{1}{2} \, \rmS_1' + \kappa_2 \rmS_3\,.&
\end{align}
\end{subequations}
The identity $\lambda_{12} = \lambda_{21}$ determines the arc length derivative of the tangential EL derivative of the energy
\begin{equation} \label{eq:sym12}
\rmS_3' = \kappa_2 \, \rmS_1 - \kappa_1 \, \rmS_2\,, \quad \mbox{or} \quad \nabla_s \rmS_3 = -\varepsilon_{IJ} \kappa_I \rmS_J\,. 
\end{equation}
As we show in Sec. (\ref{sec:FMEucInv}), this relation reflects the fact that the torque per unit length along the curve does not have a tangential component, so the three material curvatures are not independent. Another derivation of this relation in terms of $\theta$ is provided in Appendix \ref{App:FStoMF}, where the connection between the material curvatures is made explicit.
\\
Using the symmetry of the $\lambda_{ij}$, we have that $\lambda_{31}=\lambda_{13}$ and $\lambda_{32}=\lambda_{23}$. Thus, substituting Eqs. (\ref{eq:ELder12}) in Eq. (\ref{eq:deltaHei}) with $i=3$, we obtain that the components of $\bfF$ are
\begin{equation} \label{def:matFrvcF}
\rmF_1 = - \rmS'_2 - \kappa_3 \, \rmS_1 + \kappa_1 \, \rmS_3 \,, 
\quad
\rmF_2 = \rmS'_1 - \kappa_3 \, \rmS_2 + \kappa_2 \, \rmS_3 \,, \quad
\rmF_3 = \lambda_{33} + \kappa_1 \, \rmS_1 + \kappa_2 \, \rmS_2 \,.
\end{equation}
We determine the remaining component $\lambda_{33}$ from the tangential projection of the conservation law of $\bfF$. Substituting Eqs. (\ref{def:matFrvcF}) in $\mathcal{E}_3$ given in Eq. (\ref{ELderI3}),  and using Eq. (\ref{eq:sym12}), we get
\begin{equation} \label{eq:Fpdote1}
\mathcal{E}_3 = \left(\lambda_{33} + \kappa_1 \rmS_1 + \kappa_2 \, \rmS_2 \right)' + \kappa_1 \, \rmS_1' + \kappa_2 \, \rmS_2' + \kappa_3 \rmS_3' = 0\,.
\end{equation}
The last three terms can be recast as a total derivative. We have that the total derivative of the energy density is
\begin{equation} \label{eq:calFp}
 \calF' = \frac{\partial \calF}{\partial \kappa_i}\kappa_i' + \frac{\partial \calF}{\partial \kappa_i'} \kappa_i'' = \rmS_i \kappa_i' + \left( \frac{\partial \calF}{\partial \kappa_i'} \kappa_i '\right)' = -\kappa_i \rmS_i' + \left(\kappa_i \rmS_i + \frac{\partial \calF}{\partial \kappa_i'} \kappa_i '\right)' \,.
\end{equation}
Taking into account that the momenta conjugate to the material curvature and the Hamiltonian associated to $\calF$ are defined by
\begin{equation} \label{def:piHam}
 p_i = \frac{\partial \calF}{\partial \kappa_i'} \,, \quad \calH = \kappa_i' p_i -\calF\,,
\end{equation}
we can rewrite Eq. (\ref{eq:calFp}) as
\begin{equation}
 \kappa_i \rmS_i' = \left(\kappa_i \rmS_i + \calH \right)' \,.
\end{equation}
Substituting this result in Eq. (\ref{eq:Fpdote1}) we have that $\calE_3$ is a total derivative (a consequence of the reparametrization invariance of the energy \cite{CapoChryssGuv2002}), which enables us to determine $\lambda_{33}$ as
\begin{equation} \label{eq:lambda33}
\lambda_{33} = - 2 \left(\kappa_1 \, \rmS_1 + \kappa_2\, \rmS_2 \right) - \kappa_3 \, \rmS_3  - \calH  + \mu\,,
\end{equation} 
where the constant of integration $\mu$ can be regarded either as a Lagrange multiplier fixing total length, or as an intrinsic linear tension along the rod.
\\
Substituting Eq. (\ref{eq:lambda33}) back in Eq. (\ref{def:matFrvcF}) we determine the tangential component
\begin{equation} \label{eq:F3}
\rmF_3 = - \left(\kappa_1 \, \rmS_1 + \kappa_2 \, \rmS_2 + \kappa_3 \, \rmS_3 + \calH - \mu \right)\,.
\end{equation}
The components of the vector $\bfF$ can be written concisely as 
\begin{equation} \label{eq:FIF3}
\rmF_I = -\varepsilon_{IJ} \nabla_s \rmS_J + \kappa_I \rmS_3 \,,  
\quad 
\rmF_3 = - \left( \kappa_i \, \rmS_i + \calH - \mu \right) \,.
\end{equation}
Substituting these components in the two normal projections of the conservation law of $\bfF$, $\mathcal{E}_I$ given in Eq. (\ref{ELderI3}), we obtain the EL equations governing the equilibria of the rod
\begin{equation}
\mathcal{E}_I = - \varepsilon_{IJ} \left( \nabla_s^2 \rmS_J + \kappa_J \left( \kappa_i \rmS_i + \calH -\mu \right) \right) + \nabla_s (\kappa_I \rmS_3)=0\,, \quad I=1,2\,.
\end{equation}
The first terms of these EL equations are antisymmetric under the permutation $1 \leftrightarrow 2$. This contrasts with their counterparts in the FS framework, where the gauge symmetry on the description of the cross section is broken and the EL equations look dissimilar, see Appendix \ref{App:FStoMF}.
\\
In full, the vector $\bfF$ reads
\begin{equation} \label{eq:vecF}
\bfF = - \left(\nabla_s \rmS_2 - \kappa_1 \, \rmS_3 \right) \, \bfe_1 
+ \left(\nabla_s \rmS_1 + \kappa_2 \, \rmS_3\right) \, \bfe_2 
- \left(\kappa_1 \, \rmS_1 + \kappa_2 \, \rmS_2 + \kappa_3 \, \rmS_3 + \calH - \mu \right)\, \bfe_3\,.
\end{equation}
whereas the EL equations are given by
\begin{subequations} \label{eq:EL12}
\begin{eqnarray}
\mathcal{E}_1 &=& - \rmS_2'' - \rmS_1 \kappa_3' - 2 \kappa_3 \rmS_1' + \left(\kappa_1 \rmS_3 \right)' - \kappa_2 \, \left(\kappa_1 \, \rmS_1 + \kappa_2 \, \rmS_2 + 2 \kappa_3 \, \rmS_3 + \calH - \mu \right) + \kappa_3^2 \, \rmS_2 = 0 \,, \\
\mathcal{E}_2 &=& \rmS_1'' - \rmS_2 \kappa_3' -2 \kappa_3 \rmS_2' + \left(\kappa_2 \rmS_3\right)' + \kappa_1 \, \left( \kappa_1 \, \rmS_1 + \kappa_2 \, \rmS_2 + 2 \kappa_3 \, \rmS_3 + \calH - \mu \right) - \kappa_3^2 \, \rmS_1 = 0 \,. 
\end{eqnarray}
\end{subequations}
Eq. (\ref{eq:sym12}) can be used to eliminate $\rmS'_3$ in favor of $\rmS_1$ and $\rmS_2$ in the EL equations, obtaining
\begin{subequations} \label{eq:EL12simp}
\begin{eqnarray}
\mathcal{E}_1 &=& - \rmS_2'' - \rmS_1 \kappa_3' -2 \kappa_3 \rmS_1' +\kappa_1' \rmS_3 - \kappa_2 \, \left(2 \kappa_3 \rmS_3 + \calH - \mu \right) - \left(\kappa^2 - \kappa_3^2 \right) \, \rmS_2  = 0 \,, \\
\mathcal{E}_2 &=& \rmS_1'' - \rmS_2 \kappa_3' - 2 \kappa_3 \rmS_2' + \kappa_2' \rmS_3 + \kappa_1 \, \left(2 \kappa_3 \, \rmS_3 + \calH - \mu \right) + \left(\kappa^2 - \kappa_3^2 \right) \, \rmS_1 = 0 \,. 
\end{eqnarray}
\end{subequations}
If the rod lies on a surface, one of the normal vectors is aligned with the surface normal $\mathbf{n}$, say $\bfe_2 = \mathbf{n}$ (so the other normal vector will be the conormal, $\bfe_1 = \mathbf{l}$), the material curvatures coincide with the normal and geodesic curvatures, $\kappa_1 = \kappa_n$ and $\kappa_2 = \kappa_g$, whereas the twist agrees with the geodesic torsion $\kappa_3 = \tau_g$. In this case, the vanishing of the EL derivative $\mathcal{E}_1$ corresponds to the only EL equation, accounting for the only mode of deformation. In turn, the EL derivative $\mathcal{E}_2$ represents the normal force exerted by the curve on the surface \cite{GuvVaz2012, GuvValVaz2014}. Moreover, the confinement constriction severely restricts the conformation of the curve. In particular, restraining the twisting of the curve, so identity (\ref{eq:sym12}) no longer holds; it will be balanced by a tangential confining force. Therefore, only Eqs. (\ref{eq:EL12}) are appropriate in order  to analyze the equilibrium configurations and force of confined curves, since it would be illegitimate to use Eqs. (\ref{eq:EL12simp}) in such case. 
\\
Since in the normal components of the vector $\bfF$ enter only the first order derivatives $\rmS_i'$, the EL Eqs. (\ref{eq:EL12}) involve the second order derivatives $\rmS_i''$. By contrast, their FS counterparts are of second and third order in arc length derivatives of the EL derivatives of the energy with respect to the torsion, see Appendix \ref{App:FStoMF}. Even though the material EL equations have this advantage, instead of solving them, as we show in Sec. (\ref{sec:1stint}), we can exploit the Euclidean invariance of the energy to identify their first integrals, which makes easier the integration of the material curvatures.  

\section{Force and torque vectors: conservation laws from Euclidean symmetries} \label{sec:FMEucInv}

The change of the energy as a response under a deformation of the rod is given by
\begin{equation} \label{eq:varH}
\delta E =  \int \left(- \bfF' \cdot \delta \bfY + \delta Q' \right) \rmd s \,, 
\quad
\delta Q = \bfF \cdot \delta \bfY -\frac{1}{2} \varepsilon_{ijk} \rmS_i \bfe_j \cdot \delta \bfe_k + p_i \, \delta \kappa_i\,.
\end{equation}
The first term represents the variation of the energy in the bulk, given by $\bfF' = \mathcal{E}_I \, \bfe_I$, $I=1,2$, which vanishes in equilibrium. The second term is a total derivatives arising from integration by parts, so it represents the variation of the energy at the boundaries.
\\
Let us consider a constant infinitesimal deformation of the rod. In general, such deformation of the embedding functions will be a composition of a translation by an infinitesimal constant vector $\delta \mathbf{a}$ and a rotation by an infinitesimal constant vector $\delta \bm{\Omega}$, given by $\delta \bfY = \delta \mathbf{a} + \delta \bm{\Omega} \times \bfY$. Under this deformation the material frame changes as $\delta \bfe_i = \delta \bm{\Omega} \times \bfe_i$, and the material curvatures do not change, $\delta \kappa_i=0$, so from Eq. (\ref{eq:varH}) it follows that the change in the energy of the rod is
\begin{equation} \label{eq:deltaHEucMot}
\delta E = \delta \mathbf{a} \cdot \int \bfF' \rmd s + \delta \bm{\Omega} \cdot \int \bfM' \rmd s\,,
\end{equation}
where the vector $\bfM$ is defined by
\begin{equation} \label{def:MvecMF}
\mathbf{M} = \bfY \times \bfF + \bfS \,, \quad \bfS =  \rmS_i \, \bfe_i \,.
\end{equation}
The vector $\bfS$ is independent of the embedding functions. Its components $\rmS_i$, given by the EL derivatives of the energy with respect to the material curvatures defined in Eq. (\ref{def:Si}), provide the generalized constitutive relations. Unlike their FS counterparts, these components do not involve arc length derivatives of the $\rmS_i$, see Appendix \ref{App:FStoMF}.
\\
The Euclidean invariance of the energy entails not only the conservation law of $\bfF$, but also the conservation of $\bfM$: since $\delta \mathbf{a}$ and $\delta \bm{\Omega}$ are arbitrary in Eq. (\ref{eq:deltaHEucMot}), $\delta E = 0 \Rightarrow \bfF'= \bf{0}$ and $\mathbf{M}'=\bf{0}$.
Differentiation of the expression of $\bfM$, given in (\ref{def:MvecMF}), yields  $\bfM' - \bfY \times \bfF' - \bfe_3 \times \bfF - \bfS' = 0 $. Since the first two terms vanish on account of the conservation laws of $\bfF$ and $\bfM$, we get the equation
\begin{equation} \label{id:Sp}
 \bfS' = \bfF \times \bfe_3 \,.
\end{equation}
The projections of this equation onto the normal vectors reproduce the normal components of the force vector, $\rmF_I$, given in Eq. (\ref{eq:FIF3}). Since $\bfS'$ is orthogonal to the tangent vector, we have $\bfS'\cdot \bfe_3 = (\bfS \cdot \bfe_3)'- \bfS \cdot \bfe_3'=\rmS_3' + \kappa_2 \rmS_1 - \kappa_1 \rmS_2=0$, which reproduces Eq. (\ref{eq:sym12}). From this relation follows that for any energy whose dependence on the material curvatures $\kappa_1$ and $\kappa_2$ occurs through the FS curvature $\kappa$, $\rmS_I=(\kappa_I/\kappa) \partial \calF/\partial \kappa$,  $I=1,2$, so the cross term will vanish and the tangential component of the intrinsic torque, $\rmS_3$, will be constant. 
\\
Also, $\bfS'$ is orthogonal to $\bfF$, so projecting it along the direction of force vector, $\hat{\bfF} = \bfF/\rmF$, and using its conservation follows that $\bfS' \cdot \hat{\bfF} = \left(\bfS \cdot \hat{\bfF} \right)' =0$, i.e. $\rmJ :=\bfS \cdot \hat{\bfF}$ is constant.
\\
Projecting $\bfS'$ onto the Darboux vector, we have
\begin{equation} \label{eq:SpdotD}
\bfS' \cdot \bfD + \bfF \cdot \left(\bfD \times  \bfe_3\right) =(\bfS \cdot \bfD)' -\bfS \cdot \bfD' + \bfF \cdot \bfe_3' =0\,.
\end{equation}
Taking into account that $\bfF$ is conserved and using the identity $\bfS \cdot \bfD' = \kappa_i ' \rmS_i = - \calH '$, where $\calH$ is the Hamiltonian defined in Eq. (\ref{def:piHam}), we can write Eq. (\ref{eq:SpdotD}) as a total derivative 
\begin{equation}
\left(\bfS \cdot \bfD + \calH + \rmF_3 \right)' =0\,,
\end{equation}
which upon integration, reproduces Eq. (\ref{eq:F3}) for the tangential component of the force. Thus, the Hamiltonian $\calH$ corresponding to $\calF$ is not conserved. On the other hand, it can be checked that the constant $\mu$ is the Hamiltonian associated to the effective energy density, and it is conserved because $E_c$ does not depend explicitly on arc length.
\\
We now focus on a constant infinitesimal Euclidean motion of a boundary $b$ of the rod.
The energy of the surrounding region does not change at first order, so the change in the energy reads\footnote{The subindex represents evaluation at the boundary $b$.}
\begin{equation} \label{eq:deltaHEucMotbnd}
\delta E = \delta Q|_b = \delta \mathbf{a} \cdot \bfF|_b + \delta \bm{\Omega} \cdot \bfM|_b \,,
\end{equation}
which represents the virtual work done on the curve as a response to the boundary change. The two contributions are given by the product of the displacement and rotation of the boundary with vectors $\bfF$ and $\bfM$ respectively, so they are identified as the force and torque exerted on the rod at the boundary. Furthermore, since $\bfF$ and $\bfM$ are conserved along the rod, they represent the force and torque exerted on the point at $s$ by its neighboring segment at $s +\rmd s$.\footnote{This is the sign convention used by Landau \cite{LandauBook}, in which the description is in terms of stresses exerted on the rod, rather than by it.} In particular, the component $\rmF_3$ represents the force exerted on a segment of the rod by the neighboring segment with grater arc-length along the tangential direction: regions of the rod with $\rmF_3 > 0$ ($\rmF_3 < 0$) are under tension (compression). Furthermore a positive (negative) $\mu$ introduces tension (compression) along the rod.

\section{First integrals} \label{sec:1stint}

The conservation laws of $\bfF$ and $\bfM$ also permits us to identify two conserved quantities. The first one is given by the squared magnitude of the force vector, $\rmF^2 = \bfF \cdot \bfF$, whereas the second one is the projection of the torque vector along the direction of the force vector, $\rmF \rmJ = \bfM \cdot \bfF=\bfS \cdot \bfF$ \cite{CapoChryssGuv2002}. From them, the two normal components of the force vector, involving first order derivatives of $\rmS_I$, $I=1,2$, can be determined in terms of the tangential component and of the component of the intrinsic torque 
\begin{subequations} \label{eq:H1pH2p}
\begin{eqnarray} 
\rmF_1 &=& -\rmS_2' -\kappa_3 \, \rmS_1 + \kappa_1 \, \rmS_3 = \frac{\rmS_1 A - \rmS_2 B}{\rmS_1^2 + \rmS_2^2}   \,, \\
\rmF_2 &=& \rmS_1' - \kappa_3 \,\rmS_2 + \kappa_2 \,\rmS_3 = \frac{\rmS_2 A + \rmS_1 B}{\rmS_1^2 + \rmS_2^2}  \,,
\end{eqnarray}
\end{subequations}
where $A$ and $B$ are functions of the $\rmS_i$ and $\rmF_3$, defined by
\begin{subequations} \label{eq:defAB}
\begin{eqnarray} 
A & = & \rmF \rmJ - \rmF_3 \rmS_3\,,\\
B & = & \pm \sqrt{\left(\rmS_1^2 + \rmS_2^2\right) \left(\rmF^2 - \rmF_3^2\right) - A^2}\,, \label{Bsqrt}
\end{eqnarray}
\end{subequations}
and with $\rmF_3$ defined by Eq. (\ref{eq:F3}). Equations (\ref{eq:H1pH2p}) allow us to determine $\rmS_1'$ and $\rmS_2'$ in terms of the components $\rmS_i$, $i=1,2,3$.  Hence, they provide first integrals of the EL eqs. (\ref{eq:EL12}), which involve $\rmS_1''$ and $\rmS_2''$.\footnote{It can be checked that Eqs. (\ref{eq:H1pH2p}), altogether with Eq. (\ref{eq:sym12}), indeed satisfy the EL equations (\ref{eq:EL12}).}
\\
We can express the functions $A$ and $B$ in terms of the normal components of $\bfF$ as
\begin{subequations} \label{eq:altdefAB}
\begin{eqnarray} 
A & = & \rmF_1 \rmS_1 + \rmF_2 \, \rmS_2 = \rmS_2 \nabla_s \rmS_1 - \rmS_1 \nabla_s \rmS_2 + \rmS_3 \mathbf{S} \cdot \mathbf{D}_\perp \label{AScvdrS} \,,\\
B & = & \frac{1}{2} (\rmS^2)' = \bfS \cdot \bfF \times \bfe_3 \,, \quad \rmS^2 = \bfS \cdot \bfS = \rmS_1^2 + \rmS_2^2 + \rmS_3^2 \,. \label{BS2}
\end{eqnarray}
\end{subequations}
Thus, if $\bfS$ is constant or if it is orthogonal to $\bfF$ and $\bfe_3$, then $B=0$. It follows from Eq. (\ref{BS2}) that the negative solution in Eq. (\ref{Bsqrt}) only amounts to a change of orientation, $s \rightarrow -s$, that is, traversing the curve in the opposite sense, so in the following we consider only the positive solution. Also, $B^2 \geq 0$, implies that the following condition must be fulfilled
\begin{equation}
| \rmF \rmJ -\rmF_3  \rmS_3 | \leq \sqrt{(\rmS_1^2 + \rmS_2^2)(\rmF^2-\rmF_3^2)}\,.
\end{equation}
Since $\calF=\calF(\kappa_i, \kappa_i')$, the components $\rmS_i$ depend on the second order derivatives of $\kappa_i$, so these equations, along with Eq. (\ref{eq:sym12}), provide a system of three equations of third order in the derivatives of the material curvatures. However, if $\calF$ depends at most linearly in the derivatives $\kappa_i'$, such that the derivatives $\partial \calF/\partial \kappa_i'$ are constant, then the components $\rmS_i$ depend only on $\kappa_i$. In consequence, Eqs. (\ref{eq:H1pH2p}) are of first order in the derivatives of the material curvatures, providing a system of three first order differential equations of the material curvatures $\kappa_i$, which are easier to solve than the second order EL Eqs. (\ref{eq:EL12}), like in a Hamiltonian formalism. In general this procedure to obtain quadratures is not possible in the FS framework \cite{CapoChryssGuv2002}, because the component of $\bfF$ along the binormal depends on the second order derivatives of the EL derivatives of the energy with respect to the torsion (see Appendix \ref{App:FStoMF}). Thus, even if the energy density does not depend on the derivatives of the curvature and torsion but depends on the torsion at least quadratically, the binormal component of the force would involve the second order arc length derivative of the torsion.
\\
In order to solve these equations and to determine the three constants of integration, $\rmF$, $\rmJ$ and $\mu$, six appropriate boundary conditions should be provided. 
Like for the classical Euler elastica, to obtain non-trivial configurations of open rods, their boundaries should be subject to external forces or torques, so $\rmF \neq 0$ or $\rmJ \neq 0$ \cite{LandauBook}.\footnote{For an open rod with free ends, there are no external forces or torques on the boundaries, and since since they are conserved along the rod, the force and torque vectors vanish everywhere, $\bfF =\mathbf{0}$, and $\bfM = \mathbf{0}$. These conditions imply the vanishing of the EL derivatives of the energy density with respect to the material curvatures, $\rmS_i = 0$, $i=1,2,3$, so the shape of the rod will be dictated by the spontaneous material curvatures or will be a straight line for null spontaneous curvatures.} One possibility to supply the required six boundary conditions could be the specification of the material basis at the ends of the rod, for which suffices to fix one normal and the tangent vector,  i.e. $\delta \bfe_1|_b = \mathbf{0}$ and $\delta \bfe_3|_b = \mathbf{0}$ (the other normal is then also fixed, as well as the material curvatures, $\delta \kappa_i=0$, so $\delta Q$ vanishes identically), which amounts to six boundary conditions (see Appendix \ref{App:altderELeqs}). Another possibility would be to consider the opposite case where the tangent and the normal vectors are free at the boundaries. In such case Eq. (\ref{eq:varH}) implies $\rmS_i|_b = 0$, $i=1,2,3$, which also corresponds to six boundary conditions. For closed curves, periodicity of the coordinate functions impose three boundary conditions. Also, if the energy depends on the derivatives of the material curvatures, the stationarity of the energy at the boundary requires that $p_i=0$ at the boundary, which adds three boundary conditions.

\section{Reconstruction of the centerline of the rod} \label{sec:Reconstruction}

The embedding functions of rod can be determined with respect to a cylindrical coordinate system \cite{LandauBook, Langer1996, SingerSantiago2008}. The $Z$ axis is chosen to be aligned with the constant force, so $\bfF = \rmF \hat{\bf z}$. Moreover, after an appropriate constant translation the torque vector $\bfM$ can be aligned with the force vector $\bfF$,\footnote{Under a translation by a constant vector $\mathbf{a}$, $\bfY \rightarrow \bfY + \mathbf{a}$, the intrinsic torque vector $\bfS$ does not change, so the torque vector changes as $\bfM \rightarrow \tilde{\bfM} = \bfM + \bfF \times \mathbf{a}$. Setting $\mathbf{a}=\hat{\bfF} \times \bfM/\rmF$ and defining $\rmJ = \bfM \cdot \hat{\bfF}$, the new torque vector becomes $\tilde{\bfM} = \rmJ \hat{\bfF}$.}such that $\mathbf{M} = \rmJ \hat{\bfF}= \bfY \times \bfF + \bfS$, and $\rmJ = \bfM \cdot \hat{\bfF} = \bfS \cdot \hat{\bfF}$. The vector $\bfF \times \bfY = \bfS - \rmJ \hat{\bfF}$ defines a rotational vector field along the rod. By expressing the curve's embedding functions in a cylindrical coordinate system as $\bfY = \rho \hat{\bm \rho} + z \hat{\bf z}$, this rotational field is along the azimuthal direction $\bfF \times \bfY = \rmF \rho \hat{\bm{\varphi}}$. In consequence the intrinsic torque vector has no radial component, $\bfS = \rmF \rho \hat{\bm{\varphi}} + \rmJ \hat{\bf z}$. By taking norms we obtain the radial coordinate
\begin{equation} \label{eq:rhocyl}
|\rmF| \rho = \sqrt{\rmS^2 -  \rmJ^2} \,.
\end{equation} 
Hence, if the magnitude of $\bfS$ is constant ($B=0$), so is $\rho$. \\
The unit azimuthal vector is given by
\begin{equation} \label{def:uvphi}
\hat{\bm \varphi} = \frac{1}{\sqrt{\rmS^2 - \rmJ^2}}\left(\bfS - \rmJ \hat{\bfF}\right) \,.
\end{equation}
We obtain the unit radial vector from the cross product $\hat{\bm \varphi} \times \hat{\bf z}$, which reads
\begin{equation} \label{def:uvrho}
\hat{\bm{\rho}} = \frac{1}{\sqrt{\rmS^2 - \rmJ^2}} \, \bfS \times \hat{\bfF}.
\end{equation}
Recall that the tangent vector in the cylindrical basis is given by $\bfe_3 = \rho' \hat{\bm \rho} + \rho \, \varphi' \hat{\bm \varphi} + z' \hat{\bf z}$. Using identity (\ref{id:Sp}) in the projection of $\bfe_3$ onto $\hat{\bm{\rho}}$ and integrating reproduces Eq. (\ref{eq:rhocyl}). Likewise, from the projections of $\bfe_3$ onto $\hat{\bm{\varphi}}$ and $\hat{\mathbf{z}}$ we get the derivatives of the azimuthal and height coordinates 
\begin{equation} \label{eq:phipzp}
\varphi' = \frac{\rmF \, \rmS_3 - \rmJ \, \rmF_3}{\rmS^2 - \rmJ^2} \,, \quad z' = \frac{\rmF_3}{\rmF} \,.
\end{equation}
In the first equation we have used Eq. (\ref{eq:rhocyl}) to substitute $\rho$. For closed curves, there are three periodicity conditions
\begin{equation} \label{eq:closure}
\Delta \rho = 0\,, \quad \Delta \varphi = 2 \pi n\,, \; n \in \mathbbm{N}\,, \quad \Delta z = 0\,.
\end{equation}
Thus, once the material curvatures have been determined, upon integration of these equations the reconstruction of the rod is completed.

\section{Isotropic Kirchhoff elastic rods} \label{Sec:Isorods}

Here, we apply this framework to the case of isotropic Kirchhoff rods, whose associated energy is given by the sum of the bending and twisting energies, defined in Eq. (\ref{def:TBEnden}),
\begin{equation} \label{eq:BendEnDens}
\calF = \calF_B + \calF_{Tw} = \frac{\rmk}{2} \left(\kappa - c_0\right)^2 + \frac{\rmk_3}{2} \left(\kappa_3 - c_3\right)^2 \,,
\end{equation}
To reduce the number of parameters we rescale all quantities by the inverse of $\rmk$. The rescaled components of the intrinsic torque are 
\begin{subequations}
\begin{eqnarray}
\rms_I &=& \frac{\rmS_I}{\rmk} = \left(1 - \frac{c_0}{\kappa}\right) \kappa_I \,, \quad I=1,2\,, \\
\rms_3 &=& \frac{\rmS_3}{\rmk} = \chi_3 \left(\kappa_3 - c_3\right)\,, \quad \chi_3 = \frac{\rmk_3}{\rmk}.
\end{eqnarray}
\end{subequations}
Using Eq. (\ref{def:Dperp}), we can write the normal part of $\mathbf{s}$ in terms of the tangent vector as $\mathbf{s}_\perp = (1-c_0/\kappa) \bfe_3 \times \bfe_3'$. The norm of this normal part is $|\mathbf{s}_\perp|^2= \rms_1^2 + \rms_2^2 = \left(\kappa - c_0\right)^2$. Due to the symmetric dependence of $\calF$ on $\kappa$, we have $\kappa_2 \rms_1= \kappa_1 \rms_2$, and Eq. (\ref{eq:sym12}) implies that $\rms_3$ and $\kappa_3$ are constant along the rod \cite{LoveBook, Langer1996, AudolyBook}:
\begin{equation} \label{eq:H3pIso}
\rms_3' = \chi_3 \, \kappa_3' = 0 \,, \quad \Rightarrow \quad \kappa_3 = \kappa_{3c}\,.
\end{equation}
Hence, the sum of the derivative of the twist angle and the torsion remains constant along the rod $\theta' + \tau = \kappa_{3c}$, so $\theta$ can be determined if $\tau$ is known. Furthermore, the total twist is proportional to the total length of the curve $L$, $\calT w = \kappa_{3c} L/(2 \pi)$.
\\
The derivatives of $\rms_1$ and $\rms_2$ are
\begin{subequations}\label{eq:H2pH3pIsoFil}
\begin{eqnarray}
\rms_1' &=& \left(1 - \frac{c_0}{\kappa^3} \, \kappa_2^2\right) \kappa_1' + \frac{c_0}{\kappa^3} \kappa_1 \, \kappa_2 \kappa_2' = \kappa_1' - c_0 \frac{\kappa_2}{\kappa} (\kappa_3-\tau)\,,\\
\rms_2' &=& \frac{c_0}{\kappa^3} \kappa_2 \, \kappa_1 \kappa_1' + \left(1 - \frac{c_0}{\kappa^3} \, \kappa_1^2\right) \kappa_2'  = \kappa_2' + c_0 \frac{\kappa_1}{\kappa} (\kappa_3-\tau) \,,
\end{eqnarray}
\end{subequations}
where we have used Eq. (\ref{def:nrmcovderomega}) in the latter expressions. The rescaled components of the force, given by Eqs. (\ref{eq:vecF}) are
\begin{subequations} \label{eq:FMquaden}
\begin{eqnarray} 
\rmf_1 &:=& \frac{\rmF_1}{\rmk} = -\nabla_s \kappa_2 + \left( c_0 \frac{\tau}{\kappa} + \rms_{3} \right) \, \kappa_1   \,, \\
\rmf_2&:=&\frac{\rmF_2}{\rmk} = \nabla_s \kappa_1 + \left( c_0 \frac{\tau}{\kappa} + \rms_{3} \right) \kappa_2  \,, \\
\rmf_3&:=&\frac{\rmF_3}{\rmk} = - \frac{1}{2} \left(\kappa^2 - c_0^2 \right) + \zeta  \,, 
\end{eqnarray}
\end{subequations}
where the constant $\zeta$ is defined by
\begin{equation}
\zeta := - \frac{\chi_3}{2 } \,  \left(\kappa_{3c}^2 -c_3^2\right)  + \frac{\mu}{\rmk}\,.
\end{equation}
Thus, the difference between the twist $\kappa_3$ and the spontaneous twist $c_3$, along with the spontaneous curvature $c_0$ renormalizes the scaled intrinsic tension $\mu/\rmk$.
\\
From the first integrals, given by Eqs. (\ref{eq:H1pH2p}), we obtain the derivatives of the material normal curvatures 
\begin{subequations} \label{eq:k1pk2piso}
\begin{eqnarray}
 \nabla_s \kappa_1 &=& a_\mathrm{I}(\kappa) \kappa_2  + b_\mathrm{I}(\kappa) \frac{\kappa_1}{\kappa}\,, \label{eq:k1pisorod}\\
 \nabla_s \kappa_2 &=& - a_\mathrm{I}(\kappa)  \kappa_1 + b_\mathrm{I}(\kappa) \frac{\kappa_2}{\kappa}\,, \label{eq:k2pisorod}
\end{eqnarray}
 \end{subequations}
where we have defined the following rescaled quantities
\begin{subequations} \label{def:alphabetaisoKR}
\begin{eqnarray} 
a_\mathrm{I}(\kappa) &=&\frac{A}{(\rmk(\kappa-c_0))^2}-\frac{\rms_3 \kappa}{\kappa-c_0} 
= \frac{\upsilon}{(\kappa - c_0)^2} - \frac{\rms_3}{2}  \,,\\
b_\mathrm{I}(\kappa)^2 &=&  \left(\frac{B}{\rmk^2(\kappa-c_0)}\right)^2 = \rmf^2 -\left( \frac{1}{2} \left(\kappa^2-c_0^2\right) - \zeta \right)^2 - \left(  \frac{\upsilon}{\kappa -c_0} + \frac{\rms_3}{2} \left( \kappa+c_0\right) \right)^2 \,,
\label{betaisoKR} \\
\upsilon &:=& \rmf \, \rmj - \rms_3 \zeta \,, \quad \rmf := \frac{\rmF}{\rmk}\,, \quad \rmj := \frac{\rmJ}{\rmk} \,.
\end{eqnarray}
\end{subequations}
This provides a system two first order differential equations with four parameters, $\kappa_{3c}$, $\chi_3$ and the two spontaneous curvatures $\kappa_{0}$ and $c_3$. The three constants of integration $\rmf$, $\rmj$ and $\zeta$ are to be determined from boundary or periodicity conditions.
 \\
From Eqs. (\ref{eq:k1pk2piso}) we can obtain two equivalent equations in terms of the curvature and torsion. The first one is obtained by multiplying Eq. (\ref{eq:k1pisorod}) by $\kappa_2$ and Eq. (\ref{eq:k2pisorod}) by $\kappa_1$, taking their difference, and using Eq. (\ref{def:nrmcovderomega}), which allows us to determine the torsion as a function of the curvature
\begin{equation} \label{eq:tauIsoKR}
\tau = a_\mathrm{I}=- \frac{\upsilon}{\left(\kappa - c_0\right)^2} + \frac{\rms_3}{2} \,.
\end{equation}
A twist difference $\rms_3$ introduces a constant torsion. As mentioned above constant $\theta$ renders $\tau$ constant, which also implies that $\kappa$ is also constant, so the corresponding curves are helices.
\\
From the other linear combination, multiplying Eq. (\ref{eq:k1pisorod}) by $\kappa_1$ and Eq. (\ref{eq:k2pisorod}) by $\kappa_2$, and summing them provides a quadrature for the curvature $\kappa' = b $, which can be recast in the form of a particle in a potential
\begin{equation}  \label{eq:Fstintkappaisorod}
 \kappa'{}^2 + V(\kappa) = E\,,
\end{equation}
where
\begin{subequations}
 \begin{eqnarray}
V(\kappa) &=& \frac{\kappa^4}{4}
- \left(\frac{c_0^2}{2} +\zeta -\frac{\rms_3^2}{4} \right) \kappa^2+\frac{\rms_3^2}{2} c_0 \kappa + \frac{\upsilon^2}{(\kappa - c_0)^2} + \frac{2 \rms_3 c_0 \upsilon}{\kappa -c_0}  \,,\\
E&=& \rmf^2 - \left(\frac{c_0^2}{2}  + \zeta \right)^2 - \frac{\rms_3^2}{4} c_0^2 - \rms_3 \upsilon \,.
\end{eqnarray}
\end{subequations}
Equations (\ref{eq:tauIsoKR}) and (\ref{eq:Fstintkappaisorod}), which could also have been obtained directly from Eqs. (\ref{eq:altdefAB}), correspond to the first integrals of the FS EL equations presented in Appendix \ref{App:FStoMF}. Under the redefinition $\tau \rightarrow \tilde{\tau}= \tau - \rms_3/2$, these equations reduce to the equations corresponding to the Euler elastica with spontaneous curvature. For vanishing spontaneous curvature, the same equations are obtained if the dependence on the torsion is linear, as for the functional involving the first three conserved geometric quantities of the LIE hierarchy \cite{Langer1996}, because in such case the EL derivatives of the torsion with respect to the material normal curvatures vanish
\begin{equation} \label{eq:ELtaukI}
 \frac{\delta \tau}{\delta \kappa_I} = \frac{\partial \tau}{\partial \kappa_I} - \left(\frac{\partial \tau}{\partial \kappa_I'}\right)'=0\,, \quad I=1,2\,,
\end{equation}
so $\rmS_1$ and $\rmS_2$ are not changed, whereas $\rmS_3$ is also constant (alternatively, in the FS approach the EL derivative of the energy with respect to the torsion, $\delta \calF/\delta \tau$, is constant).
\\
In the case of vanishing spontaneous curvature, $c_0=0$, the quadrature (\ref{eq:Fstintkappaisorod}) can be solved in terms of Jacobi elliptic functions. For $c_0 \neq 0$, the simplest solutions correspond to helices.

\subsection{Helices} \label{sec:helices}

\begin{figure}[htb]
 \centering
\subfigure[$\alpha=0$]{\includegraphics[scale=0.5]{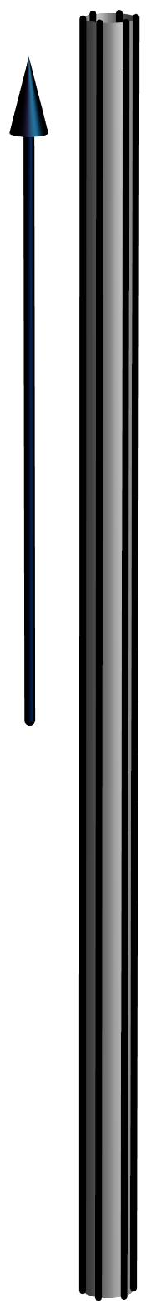}}
\hfil
\subfigure[$\alpha=\frac{\pi}{8}$]{\includegraphics[scale=0.5]{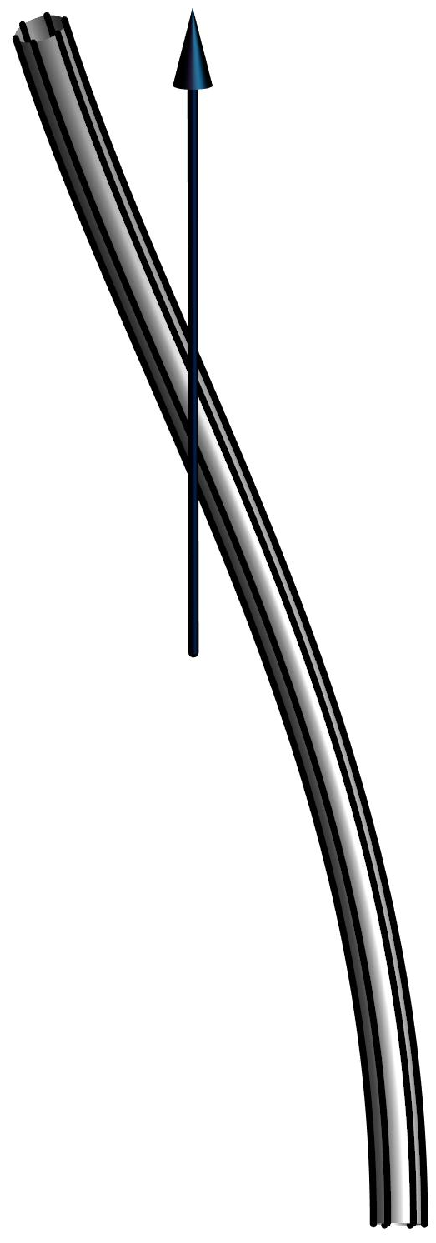}}
\hfil
\subfigure[$\alpha=\frac{\pi}{4}$]{\includegraphics[scale=0.5]{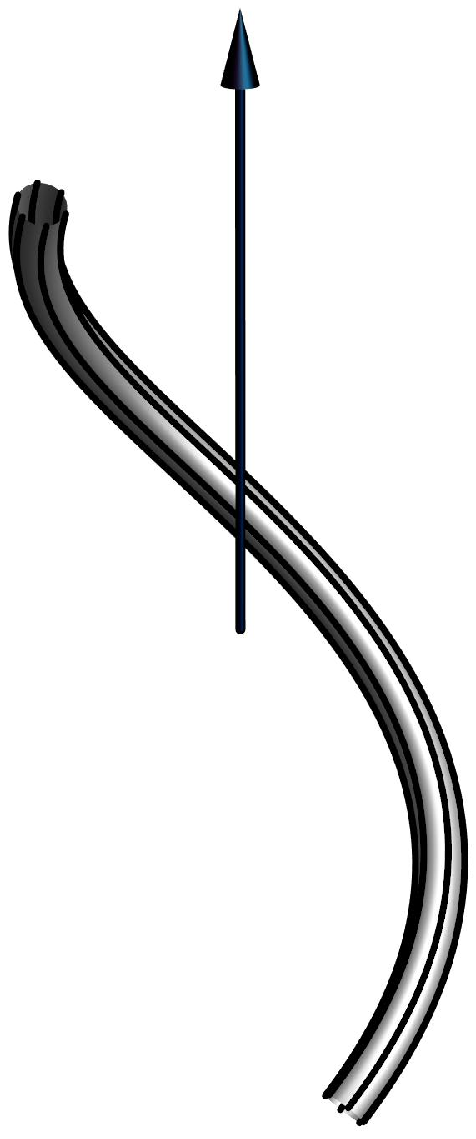}}
\hfil
\subfigure[$\alpha=\frac{3\pi}{8}$]{\includegraphics[scale=0.5]{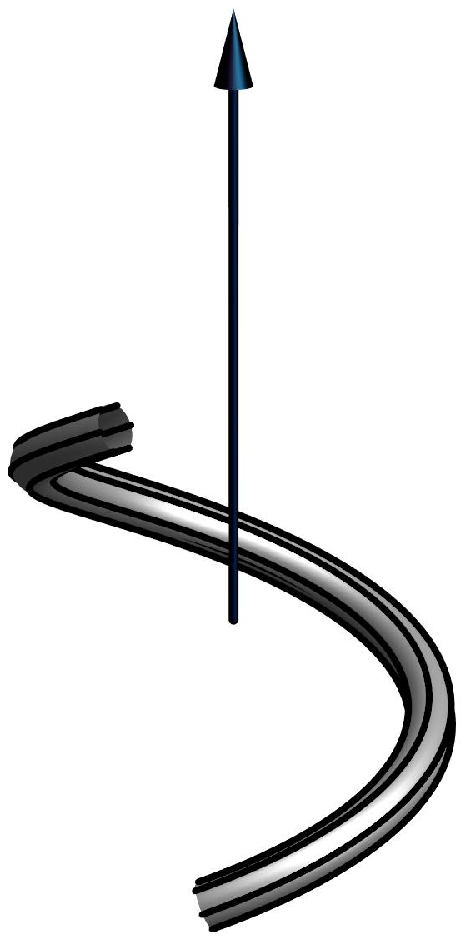}}
\hfil
\subfigure[$\alpha=\frac{\pi}{2}$]{\includegraphics[scale=0.5]{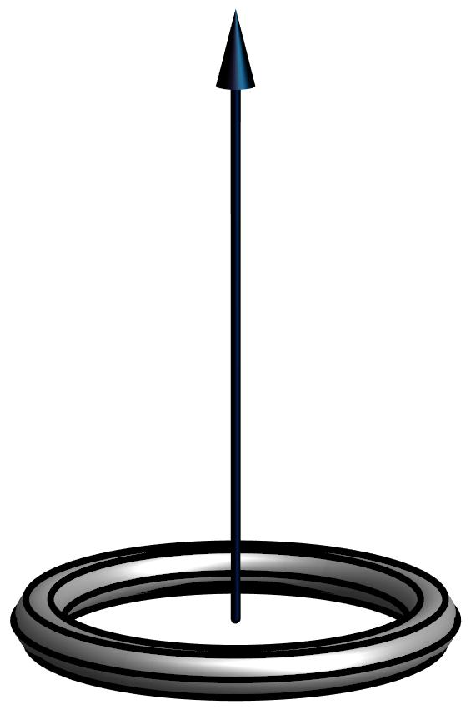}}
\caption{Helices of length $L=2\pi \rho$ with different tilts. The arrow represents the helical axis.}
 \label{fig:2}
\end{figure}

\begin{figure}[htb]
 \centering
 \subfigure[]{\includegraphics[scale=0.45]{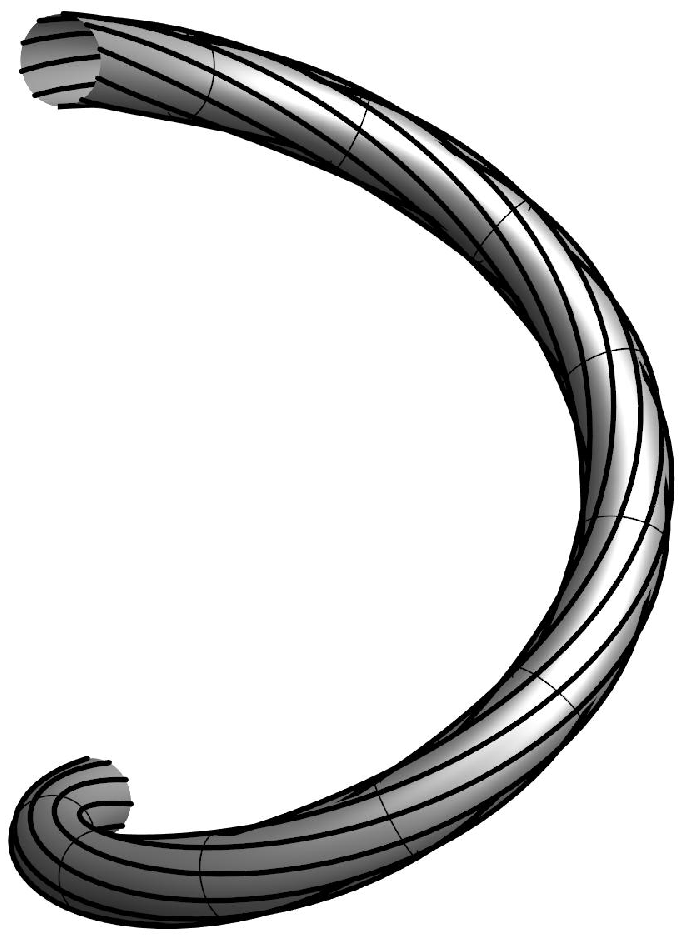}}
 \hfil
 \subfigure[]{\includegraphics[scale=0.45]{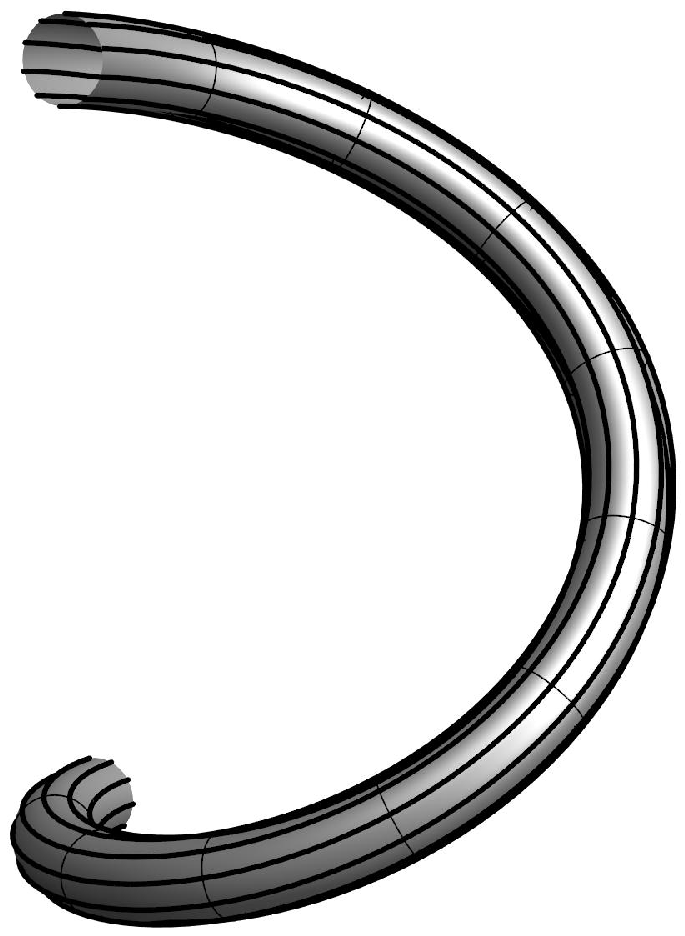}}
 \hfil
 \subfigure[]{\includegraphics[scale=0.45]{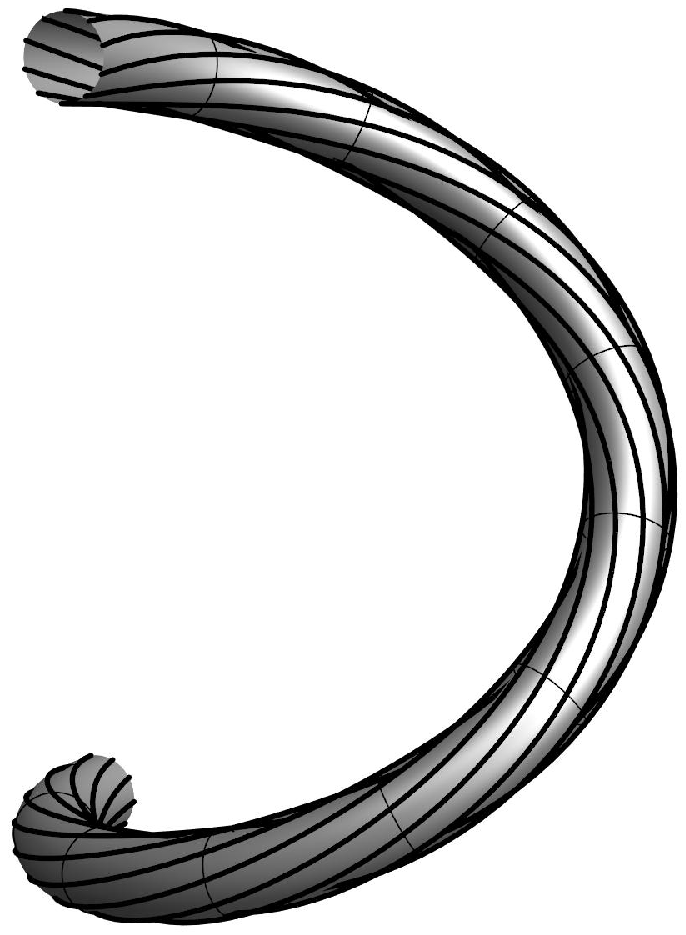}}
 \caption{Helices with a) overtwist, b) null twist and c) undertwist.} \label{fig:3}
\end{figure}

 According to Lancret's theorem, helices are characterized by a constant ratio of FS torsion to FS curvature $\tau/\kappa=\cot \alpha$, with $\alpha$ the constant angle that the tangent makes with the helical axis.
\\
A circular helix of radius $\rho$ and pitch $2 \pi p$ has constant curvature and torsion, given by $\kappa = \rho/\ell^2$ and $\tau = p/\ell^2$, where $\ell^2 = \rho^2+ p^2$. The radius $\rho$ and $p$ are related to $\alpha$ by $\rho /\ell =  \sin \alpha$ and $p/\ell = \cos \alpha$, so $p=\rho \cot \alpha$. Since for constant $\rho$, $p \rightarrow \infty$ as $\alpha \rightarrow 0$, it is more convenient to parametrize the helix in terms of $\rho$ and $\alpha$, for instance $\kappa$ and $\tau$ can be expressed as
\begin{equation} \label{ktISOhlx}
\kappa = \frac{\sin^2 \alpha}{\rho} \,,  
\quad 
\tau = \frac{\sin 2 \alpha}{2 \rho}\,. 
\end{equation}
The parameters $\rmf$, $\rmj$ and $\zeta$ of the helix can be related to $\rho$, $\alpha$ and $\rms_3$. On account of the glide rotational symmetry of the helices, the azimuthal and height are related, $z=p \phi$, where $\phi=s/\ell$. In terms of $\rho$ and $\alpha$ these coordinates are given by 
\begin{equation}
\phi=\frac{\sin \alpha \, }{\rho}s\,, 
\quad 
z=\cos \alpha \, s\,.
\end{equation}
For $\alpha =0$ we have $\kappa=0$, $\tau=0$, $\phi =0$ and $z=s$, so the helix degenerates into a straight line parallel to the $Z$ axis, passing through the $X$ axis at a distance $\rho$, whereas for $\alpha = \pi/2$, $\kappa=1/\rho$, $\tau=0$, $\phi=s/\rho$ and $z=0$, so the helix becomes a circle of radius $\rho$ on the $X-Y$ plane.
Helices of length $L=2\pi \rho$ and different values of $\alpha$ are shown in Fig. \ref{fig:2}. The midpoint of the helix is chosen as $s=0$, so the boundaries are located at $s=\pm L/2= \pm \pi \rho$.
\vskip0pc \noindent
For circular helices Eqs. (\ref{eq:rhocyl}) and (\ref{eq:phipzp}) for the radial coordinate and the derivatives of the azimuthal and height coordinates read
\begin{subequations}
\begin{eqnarray}
\rmf^2 \rho^2 &=& (\kappa-c_0)^2 +\rms_3^2 - \rmj^2 \,, \\ 
\phi'&=&\frac{\sin \alpha}{\rho} = \frac{\rmf \rms_3  - \rmj \rmf_3 }{\rmf^2 \rho^2}\,, \\
z'&=& \cos \alpha =\frac{\rmf_3}{\rmf} \,.
\end{eqnarray}
\end{subequations}
where $\rmf_3$ is given by Eq. (\ref{eq:FMquaden}). Solving for $\rmf$, $\rmj$ and $\zeta$, we obtain \cite{LandauBook, Thompson1996},\footnote{The force and torque agrees with the results of Ref.  \cite{LandauBook} with the change $\alpha \rightarrow \pi/2-\alpha$.}
\begin{subequations} \label{fjzetaisohlx}
\begin{eqnarray}
\rmf &=& \frac{1}{\rho} \left(-(\kappa - c_0) \cos \alpha  + \rms_3 \sin \alpha \right) \,,\\
\rmj &=&(\kappa - c_0) \sin \alpha + \rms_3 \cos \alpha  \,,\\
\zeta &=& \frac{1}{2} (\kappa^2-c_0^2) - \left( \kappa - c_0\right) \frac{\cos^2 \alpha}{\rho}+ \tau \rms_3 \,.
\end{eqnarray}
\end{subequations}
The signature of these solutions is chosen according to the convention that under compression (tension), $\rmf>0$ ($\rmf<0$), the curvature of the helix gets smaller (larger) that its spontaneous curvature $c_0>\kappa$ ($c_0<\kappa$) or equivalently the radius of curvature $R=1/\kappa$ gets larger (smaller) than the spontaneous radius of curvature $R_0=1/c_0$, $R>R_0$ ($R<R_0$). The same applies for the twist difference $\rms_3$. Moreover, $\rmf$, $\rmj$ and $\zeta$ vanish for $\kappa=c_0$ and $\rms_3 =0$ ($\kappa_{3c}=c_3$).
\\
Substituting expression (\ref{fjzetaisohlx}) in Eq. (\ref{betaisoKR}) yield  $b_\mathrm{I}=\kappa'=0$, and recall that $a_\mathrm{I} = -\tau$, so Eqs. (\ref{eq:k1pk2piso}) reduce to $\kappa_I' - \theta' \epsilon_{IJ} \kappa_J=0$, with constant $\theta' =\kappa_{3c} - \tau:=q$, whose solutions are 
\begin{equation} \label{kIsolsISOKR}
\kappa_1 = \kappa \sin \theta\,, \quad \kappa_2 = \kappa \cos \theta\,, \quad \theta = q(s-s_0)\,.
\end{equation}
Overtwisted (undertwisted) helices corresponds to $q>0$ ($q<0$), whereas if $q=0$, $\theta$ is constant, so the material frame coincides with the FS frame, and torsion becomes the only relevant quantity \cite{Nizette1999}. Examples of these three kind of helices are shown in Fig. \ref{fig:3}.
\vskip0pc \noindent
Using these results in Eqs. (\ref{eq:FMquaden}), we found that the normal components are $\rmf_I = \kappa_I \left(\rms_3 -\tau (1-c_0/\kappa) \right) $, $I=1,2$. Therefore, the scaled components of the force along the principal normal and binormal are $\rmf_\bfN = (\kappa_2 \rmf_1 - \kappa_1 \rmf_2)/\kappa= -\kappa' =0$, and $\rmf_\bfB  = (\kappa_1 \rmf_1 + \kappa_2 \rmf_2)/\kappa= -\tau (\kappa-c_0)+\rms_3 \kappa$. In consequence, there is no azimuthal force $\rmf_{\bm{\phi}}= \sin \alpha \rmf_3 -\cos \alpha \rmf_\bfB =0$, and the constant force is entirely along the helical axis, chosen as the $Z$ axis,  $\bm{\rmf}=\rmf \hat{\mathbf{z}}$. Helices with vanishing force, $\rmf=0$, require a twist difference $\rms_3 = \tau (1-c_0/\kappa)$, as well as a torque along the helical axis of magnitude $j= (\kappa-c_0) \csc \alpha$. Conversely, for a helix with vanishing torque, $\rms_3= - (\kappa-c_0) \tan \alpha$ and it is subject to a force of magnitude $\rmf = - (\kappa-c_0)/(\rho \cos \alpha)$. In this case, the limit curves are  the line with $\alpha=0$, $c_0=0$ (which satisfies $\rmf=0$, $\rmj=\rms_3$ and $\zeta=0$); and the circles with $\alpha=\pi/2$, $\rmf=\rms_3/\rho$, $\rmj=1/\rho -c_0$ and $\zeta=(1/\rho^2 -c_0^2)/2$.

\subsection{Twisting instabilities of isotropic helices} \label{subsec:persoliso}

Here we perform a perturbative analysis of circular helices which are already subject to  forces and torques, discussed above, which provides useful information about their twisting instabilities. Expanding the curvature, torsion, twist and the constants as\footnote{One natural choice of small parameter for the expansion of these quantities would be the thickness of the rod, which enters through the bending modulus \cite{Audoly2007}, but since we scaled all quantities with the inverse of $\rmk$, we cannot use it as the small parameter. Although we left unspecified the small parameter of the expansion at this point,  as shown below, once the EL equation is solved and the curve is reconstructed, we can identify the appropriate small parameter for each kind of deformation.}
\begin{subequations}
\begin{align}
\kappa &= \kappa_{(0)} + \kappa_{(1)} + \kappa_{(2)} + \dots \,
&\quad 
\tau &= \tau_{(0)} +\tau_{(1)}+ \tau_{(2)} +\dots \,, 
&\quad 
\kappa_3 &= \kappa_{3(0)} + \kappa_{3(1)} + \kappa_{3(2)} +\dots \,,\\
\rmf &= \rmf_{(0)}+ \rmf_{(1)}+\rmf_{(2)} + \dots \,, 
&\quad 
\rmj &= \rmj_{(0)} +\rmj_{(1)}+\rmj_{(2)}  + \dots\,, 
&\quad 
\zeta &= \zeta_{(0)} + \zeta_{(1)} + \zeta_{(2)} +\dots \,, 
\end{align}
\end{subequations}
where the zeroth order of the FS curvatures and the parameters are given in Eqs. (\ref{ktISOhlx}), (\ref{kIsolsISOKR}) and (\ref{fjzetaisohlx}). The twist difference is also expanded as $\rms_3 =\rms_{3(0)} + \rms_{3(1)} +\rms_{3(2)}  \dots$, where $\rms_{3(0)} = \chi_3 (\kappa_{3(0)} -c_3)$, $\rms_{3(n)} = \chi_3 \kappa_{3(n)}$, $n \in  \mathbbm{N}$. For the purpose of extending these results to anisotropic rods, we consider the case of vanishing spontaneous curvatures. In this case, the zeroth order of  parameters $\rmf$, $\rmj$ and $\zeta$  can be expressed as (recall $\ell=(\kappa_{(0)}^2+\tau_{(0)}^2)^{-1/2}$)
\begin{equation} \label{eq:f0j0zeta0}
\rmf_{(0)} = \frac{\rms_{3(0)}-\tau_{(0)}}{\ell}\,, 
\quad
\rmj_{(0)}= \ell \left(\kappa_{(0)}^2 + \rms_{3(0)} \tau_{(0)} \right)\,,
\quad 
\zeta_{(0)} = \frac{\kappa_{(0)}^2}{2}+\tau_{(0)} (\rms_{3(0)} - \tau_{(0)}) \,.
\end{equation}
To zeroth order, Eqs. (\ref{eq:tauIsoKR}) and (\ref{eq:Fstintkappaisorod}) are satisfied.
To first order these equations read\footnote{In order to avoid the singular behavior when $\kappa \rightarrow 0$ in the limit curves, these equations are multiplied by $\kappa^2$.}
\begin{subequations}
\begin{eqnarray}
\ell \tau_{(0)} \rmf_{(1)} + \frac{\rmj_{(1)}}{\ell} - \tau_{(0)} \rms_{3(1)} - \zeta_{(1)}
&=& 0 \,, \\
\kappa_{(0)} \left(\left(2 \tau_{(0)} - \rms_{3(0)}\right) \kappa_{(1)}(s) + \kappa_{(0)} \tau_{(1)}(s)\right) + \rmj_{(0)} \rmf_{(1)} + \rmf_{(0)} \rmj_{(1)}-\rms_{31} \left(\frac{\kappa_{(0)}^2}{2}+\zeta _{(0)}\right) - \rms_{3(0)} \zeta_{(1)}
 &=&0\,.
\end{eqnarray}
\end{subequations}
We can determine $\zeta_{(1)}$ from the first equations, whereas for $\kappa_{(0)} \neq 0$ the latter allows us to determine $\tau_{(1)}$.
\\
To second order, Eq. (\ref{eq:Fstintkappaisorod}) reads
\begin{equation}
\kappa_{(1)}{}'(s)^2 + c_{\kappa(0)} \kappa_{(1)}(s)^2 + c_{\kappa(1)} \kappa_{(1)}(s) = c_{\kappa(2)} \,, 
\end{equation}
where
\begin{subequations}
\begin{eqnarray}
 c_{\kappa(0)} &=& \kappa_{(0)}^2 + \left(2 \tau_{(0)} - \rms_{3(0)}\right){}^2\,, \label{ckappa0} \\ 
 c_{\kappa(1)} &=& 2 \rho \left(\tau_{(0)}-\rms_{30}\right) \rmf_{(1)}-\frac{2 \left(\kappa_{(0)}^2+\tau_{(0)} \left(2 \tau_{(0)}-s_{3(0)}\right)\right)}{\rho} \rmj_{(1)} + \frac{ \left(2 \rms_{3(0)} \left(\kappa_{(0)}^2-\tau_{(0)}^2\right)+4 \tau_{(0)}^3\right)}{\kappa_{(0)}} \rms_{3(1)}\,,\\
 c_{\kappa(2)} &=&\left(\tau_{(0)}-\rms_{3(0)}\right){}^2 \left(-p\, \rmf_{(2)} + \zeta_{(2)}-\frac{\rmj_{(2)}}{\ell} + \tau_{(0)} \rms_{3(2)} \right) +2 \left(\tau_{(0)} -\rms_{3(0)}\right) \left( \rmj_{(1)} - \, \ell \, \tau_{(0)} \,\rms_{3(1)} \right) \rmf_{(1)}\nonumber \\
 &-&\frac{\left(\kappa_{(0)}^2+\tau_{(0)}^2\right)}{\rho^2}\rmj_{(1)}^2 + \frac{2  \left(\kappa_{(0)}^2 \rms_{3(0)}+\tau_{(0)}^3\right)}{\kappa_{(0)} \rho} \rms_{3(1)} \rmj_{(1)} -  \tau_{(0)} \left(\frac{\tau_{(0)}^3}{\kappa_{(0)}^2}-\tau_{(0)}+2 \rms_{3(0)}\right) \rms_{3(1)}^2\,.
\end{eqnarray}
\end{subequations}
Thus, the general solution is
\begin{equation}
\kappa_{(1)}(s) = \kappa_{M(1)} \sin Q (s-s_0) + \kappa_{c(1)}  \,,
\quad Q^2 = c_{\kappa(0)}\,, \quad \kappa_{M(1)} = \sqrt{\frac{c_{\kappa(2)}}{c_{\kappa(0)}} - \frac{c_{\kappa(1)}^2}{4 c_{\kappa(0)}^2}} \,,
\quad \kappa_{c(1)} = \frac{c_{\kappa(1)}}{2 c_{\kappa(0)}} \,.
\end{equation}
From Eqs. (\ref{eq:rhocyl}) and (\ref{eq:phipzp}), we find that the first order perturbations of $\rho$, and the derivatives of $\phi$ and $z$ are proportional to $\kappa_{(1)}$
\begin{subequations} \label{crhophiz}
\begin{eqnarray}
\rho_{(1)} &=& c_{\rho(0)} \kappa_{(1)}(s) + c_{\rho{(1)}}\,,\\
\phi_{(1)}' &=& c_{\phi(0)}\kappa_{(1)}(s) + c_{\phi(1)} \\
z_{(1)}' &=&c_{z(0)} \kappa_{(1)}(s) +c_{z(1)} \,,
 \end{eqnarray}
\end{subequations}
where
\begin{subequations} 
\begin{eqnarray}
 c_{\rho(0)} &= &\frac{\kappa_{(0)}}{\rmf_0^2 \rho_{(0)}} \,, \quad c_{\rho{(1)}} = -\frac{1}{\rmf_{(0)}}\left(\rho_{(0)} \rmf_{(1)}+\frac{1}{\rmf_{(0)} \rho_{(0)}} \left(  \rmj_{(0)} \rmj_{(1)} + \rms_{3(0)} \rms_{3(1)} \right) \right)\,,\\
 c_{\phi(0)} &=& \frac{\kappa_{(0)}}{\rmf_{(0)}^3 \rho_{(0)}^3} \left( \left(\frac{2 \zeta_{(0)}-\kappa_{(0)}^2}{\rmf_{(0)} \rho_{(0)}} + \rmf_{(0)} \rho_{(0)} \right) \rmj_{(0)}-\frac{2 \rms_{3(0)}}{\rho_{(0)}}\right)\,,\nonumber\\
 c_{\phi(1)} &=&  \frac{\rms_{3(0)}\rmf_{(1)}}{\rmf_{(0)}^2 \rho_{(0)}^2}  + \left( \left( \frac{\rmj_{(0)} \left(\kappa_{(0)}^2-2 \zeta_{(0)}\right)}{\rmf_{(0)} } + 2 \rms_{3(0)} \right) \frac{ \rmj_{(0)} }{\rmf_{(0)}\rho_{(0)}^2}+\frac{\kappa _{(0)}^2}{2}-\zeta_{(0)}\right) \frac{\rmj_{(1)}}{\rmf_{(0)}^2 \rho _{(0)}^2} \nonumber\\
&+& \left( \left( \frac{\rmj_{(0)} \left(2 \zeta _0-\kappa _0^2\right)}{\rmf_{(0)}} -2 s_{3{(0)}} \right) \frac{\rms_{3(0)}}{\rmf_{(0)}^2 \rho_{(0)}^2}+ 1 \right)  \frac{\rms_{3(1)}}{\rmf_{(0)} \rho_{(0)}^2}  -\frac{\rmj_{(0)}}{\rmf_{(0)}^2 \rho_{(0)}^2}\zeta_{(1)}\,, \\
c_{z(0)}&=& -\frac{\kappa_{(0)}}{\rmf_{(0)}} \,, \quad c_{z(1)} = - \frac{1}{\rmf_{(0)}} \left( \left(\frac{\kappa_{(0)}^2-2 \zeta _{(0)}}{2 \rmf_{(0)}} + p\right)\rmf_{(1)}-\frac{\rmj_{(1)}}{\ell} + \tau_{(0)}\rms_{3(1)} \right)\,.
\end{eqnarray}
\end{subequations}
Integration of Eqs. (\ref{crhophiz}) yields
\begin{subequations}
 \begin{eqnarray}
 \phi_{(1)} &=& -\frac{c_{\phi(0)} \kappa_{M(1)}}{Q} \cos Q \left(s-s_0\right) +  \left(c_{\phi(1)} + c_{\phi(0)} \kappa _{c(1)} \right) s \,, \\
 z_{(1)}&=& -\frac{c_{z(0)} \kappa_{M(1)}}{Q} \cos Q \left(s-s_0\right) + \left(c_{z(1)} + c_{z(0)} \kappa _{c(1)}\right)s\,.
\end{eqnarray}
\end{subequations}
We expand the embedding functions and the tangent vectors as $\bfY = \bfY_{(0)} + \bfY_{(1)} + \dots $ and $\bfe_3 = \bfe_{3(0)} + \bfe_{3(1)} + \dots $, where
\begin{subequations}
\begin{align}
\bfY_{(0)} &= \rho_{(0)} \hat{\bm{\rho}} + z_{(0)} \hat{\mathbf{z}}  \,, 
&\quad  \bfY_{(1)} &= \rho_{(1)} \hat{\bm{\rho}}+ \rho_{(0)} \phi_{(1)} \hat{\bm{\phi}} + z_{(1)} \hat{\mathbf{z}}\,,\\
\bfe_{3(0)} &= \rho_{(0)} \phi_{(0)}' \hat{\bm{\phi}} + z_{(0)}' \hat{\mathbf{z}} \,, 
&\quad 
\bfe_{3(1)} &= \left( \rho_{(1)}' - \rho_{(0)} \phi_{(0)}' \phi_{(1)} \right) \hat{\bm{\rho}} + \left(\rho_{(0)} \phi_{(1)}' + \rho_{(1)} \phi_{(0)}'\right) \hat{\bm{\phi}} + z_{(1)}' \hat{\mathbf{z}}\,.  \label{def:e31}
\end{align}
\end{subequations}
The first order change in the arc-length element is $\rmd s_{(1)} = \bfe_{3(0)} \cdot \bfe_{3(1)} \rmd s$, so the first order change in total length $L_{(1)} = L -L_{(0)}$ is given by
\begin{equation} \label{eq:L1const}
\frac{L_{(1)}}{\rho_{(0)}} = \left(\sin \alpha \, \phi_{(1)}  + \cos \alpha   \frac{z_{(1)}}{\rho_{(0)}} \right)\Big{|}_{-L/2}^{L/2} + \frac{1}{\ell^2} \int_{-L/2}^{L/2} \rho_{(1)} \rmd s \,.
\end{equation}
There are three families of solutions satisfying this fixed length condition along with a combination of fixed coordinates at the boundaries\footnote{$\phi$ and $z$ have a similar behavior as $\rho_{(1)}'$.}
\begin{subequations}
\begin{eqnarray}
\rho_{(1)}(\pm L/2)&=&0 \,, \label{bc:rho} \\
 \phi_{(1)} (\pm L/2) &=&0, \quad z_{(1)} (\pm L/2) =0\,. \label{bc:phiz}
 \end{eqnarray}
\end{subequations}

\begin{figure}[htb]
 \centering
\subfigure[]{\includegraphics[width=0.475\textwidth]{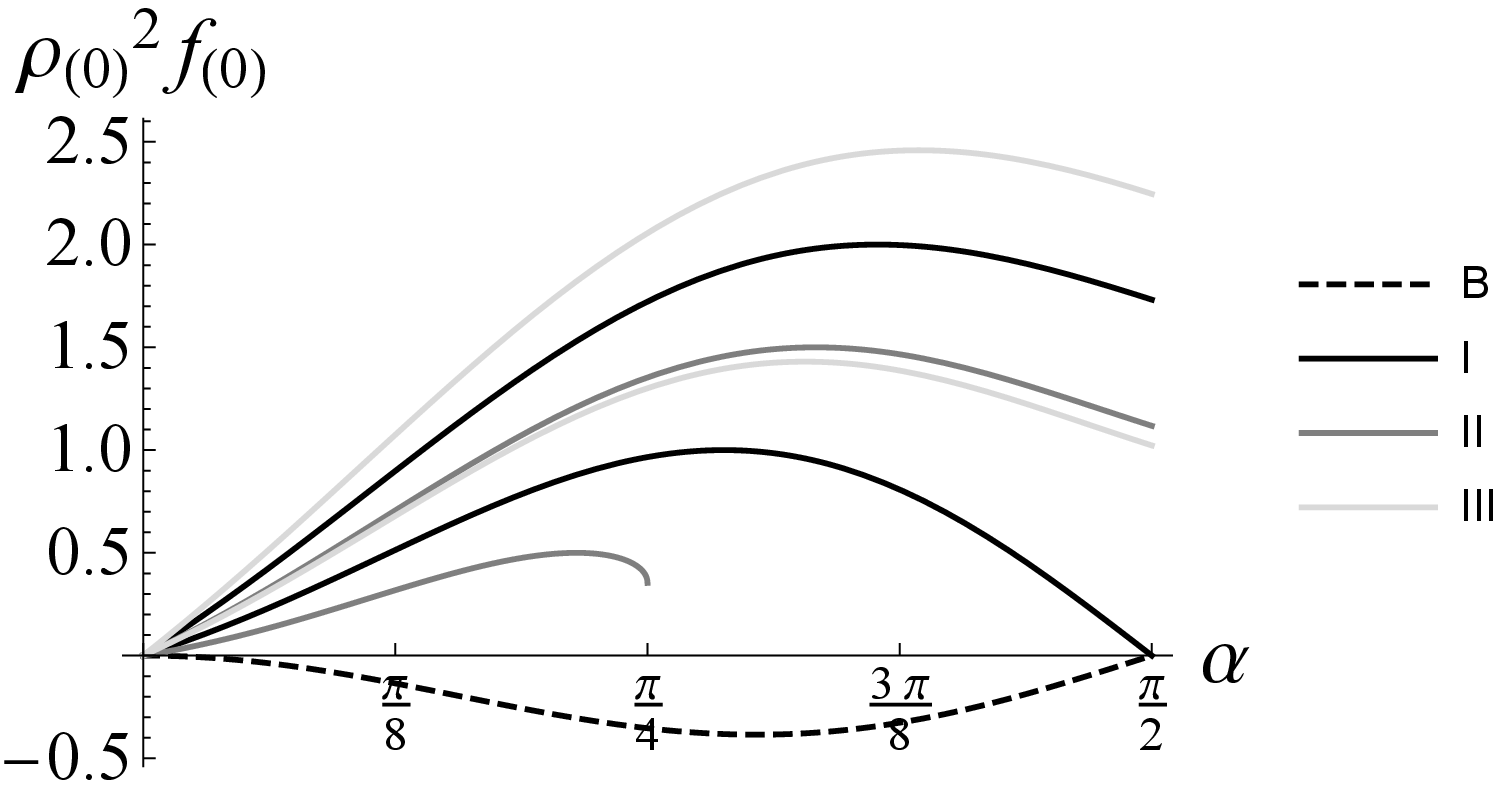}}
\hfil
\subfigure[]{\includegraphics[width=0.475\textwidth]{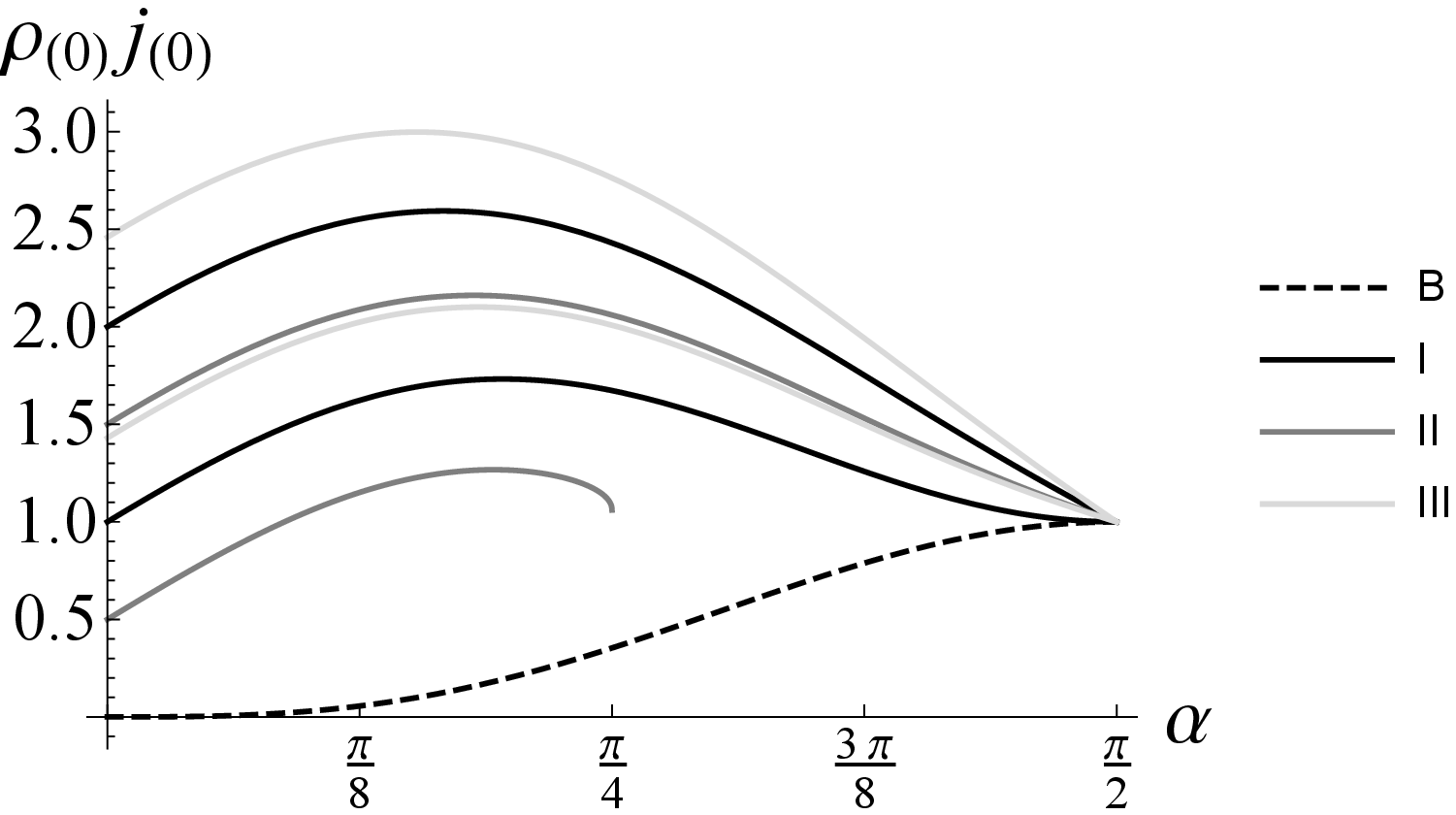}}
\\
\subfigure[]{\includegraphics[width=0.475\textwidth]{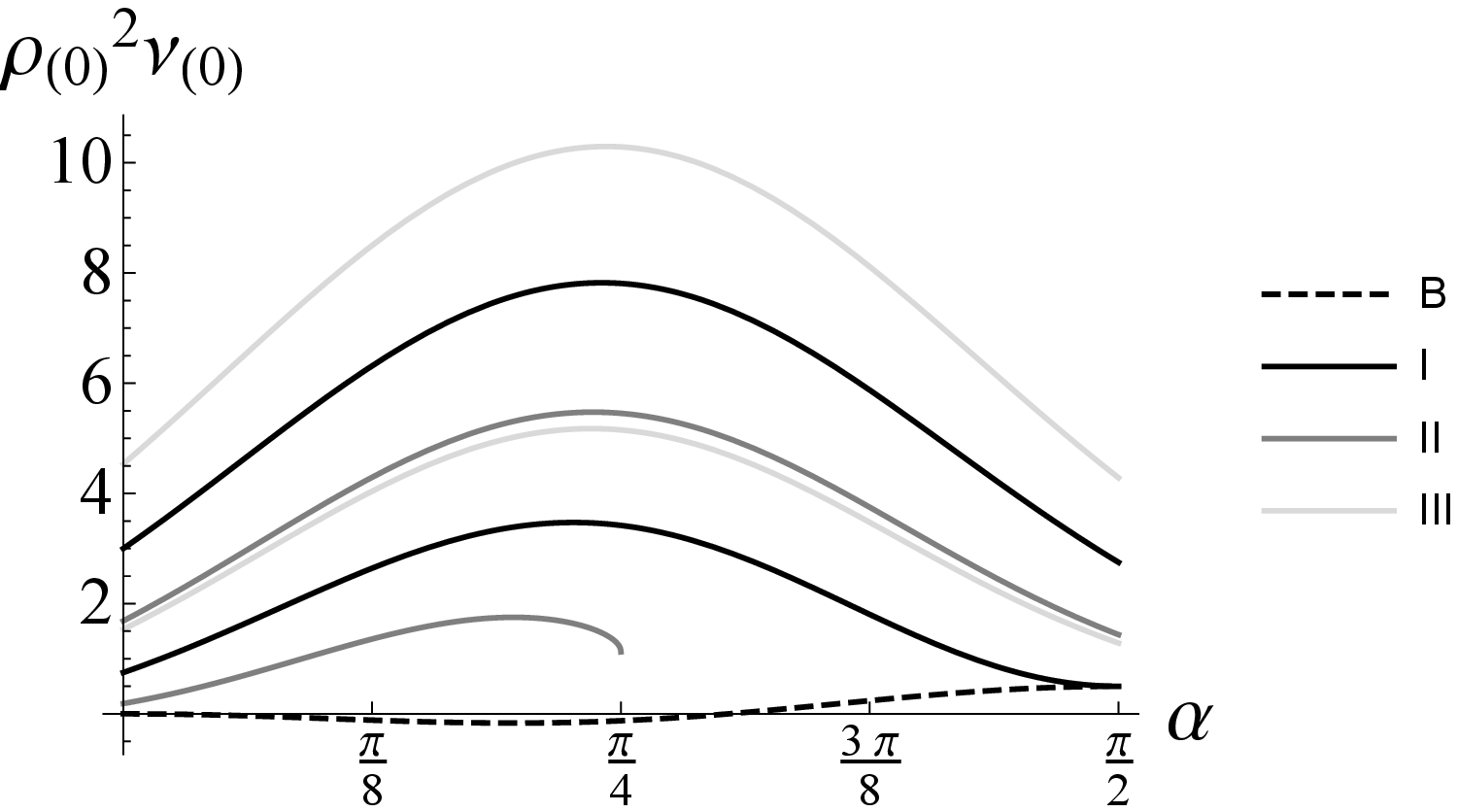}}
\hfil
\subfigure[]{\includegraphics[width=0.475\textwidth]{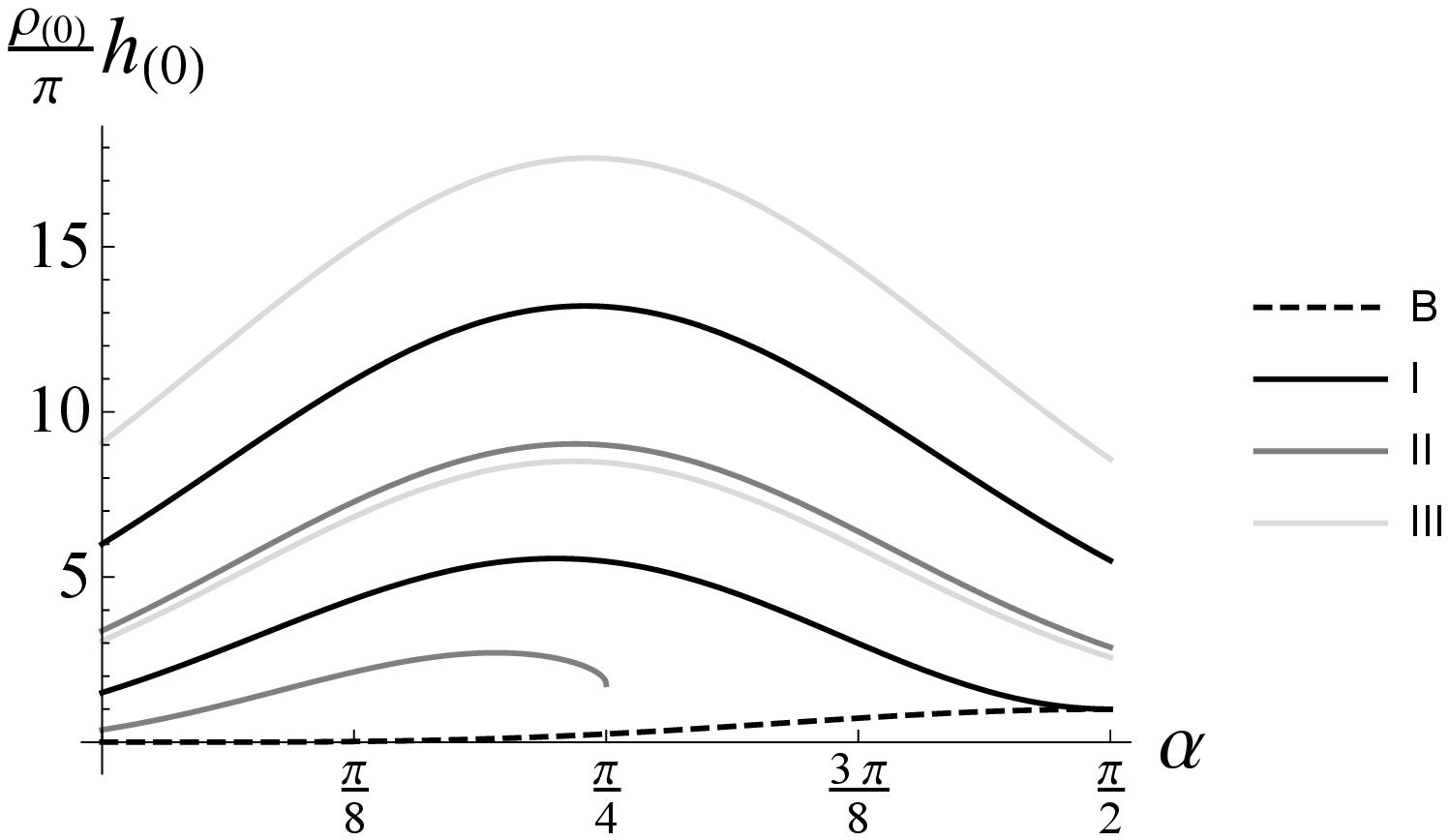}}
\caption{Lowest order of the scaled force, torque, intrinsic tension and total energy for the three families of deformed isotropic helices of length $L= 2 \pi \rho_{(0)}$ and $\chi_3 =2/3$. $\alpha=0,\pi/2$ correspond to the line and circle respectively. The dashed line represents the case of pure bending. The black lines represent values for the first family with $Q \rho_{(0)}=1,2$ (lower and upper lines respectively). Gray lines represent values for the second family with $Q \rho_{(0)}=1/2,3/2$, (line ending at $\pi/4$ and complete one respectively, for $n=2$ it agrees with the first curve of the first family). The light gray lines represent values for the third family with $Q \rho_{(0)}=1.43, 2.46$ (lower and upper lines respectively).}
 \label{fig:7}
\end{figure}

\begin{enumerate}[I.]

\item 
\textbf{Fixed radial coordinate at the boundaries}. Setting $s_0=0$ and imposing Eqs. (\ref{eq:L1const}) and (\ref{bc:rho}) lead to $Q L = 2 \pi n$, $n \in \mathbb{N}$, or $Q \rho_{(0)} =n$. Combining this result with Eq. (\ref{ckappa0}) allow us to determine $\rms_{3(0)}$
\begin{equation}
\rho_{(0)} \rms_{3(0)}= \sqrt{ n^2-(\rho_{(0)} \kappa_{(0)})^2} + 2 \rho_{(0)} \tau_{(0)} \,.
\end{equation}
Since $\rho_{(0)} \kappa_{(0)}= \sin^2 \alpha$, for $0 \leq \alpha < \pi/2$ ($\alpha=0$ corresponding to the line), the lowest deformation mode has one period, $n=1$. For the circle with $\alpha = \pi/2$ the lowest deformation mode has two periods, $n=2$, known as the Michell's instability \cite{Goriely2006}. 
\\
The first order corrections to the cylindrical coordinates are given by
\begin{equation} \label{eq:rhophiz12}
\rho_{(1)}(s) = c_{\rho(0)} \kappa_{M(1)} \sin Q s  \,, \quad
\phi_{(1)}(s) = -\frac{c_{\phi(0)}}{Q} \kappa_{M(1)} \cos Q s  \,,   \quad
z_{(1)}(s) =  -\frac{c_{z(0)}}{Q} \kappa_{M(1)} \cos Q s  \,.
\end{equation}
\\
This family of solutions does not satisfy condition (\ref{bc:phiz}), so the azimuthal and height coordinates of the curve change at the boundaries. Moreover, from Eq. (\ref{def:e31}) it follows that the tangent vector changes in the radial direction at the boundaries. From the last equation we can express the maximum curvature perturbation in terms of the first order change in the height of the boundaries
\begin{equation}
 \kappa_{M(1)}= (-1)^{n} \, n \csc^2 \alpha \, \rmf_{(0)} \, z_{b(1)}\,, \quad z_{b(1)}:=z_{(1)}(\pm L/2)\,.
\end{equation}
The lowest admissible modes of this kind of perturbed helices are shown in Fig. \ref{fig:4}
\begin{figure}[htb]
 \centering
\subfigure[$\alpha=\frac{\pi}{64}$]{\includegraphics[scale=0.6]{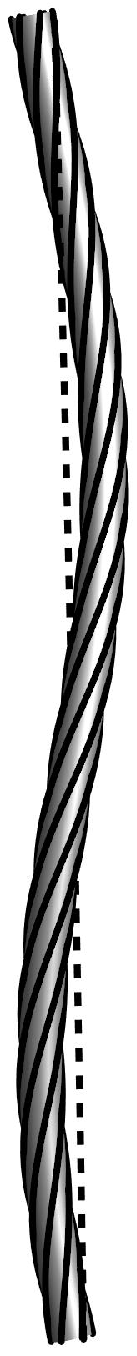}}
\hfil
\subfigure[$\alpha=\frac{\pi}{8}$]{\includegraphics[scale=0.6]{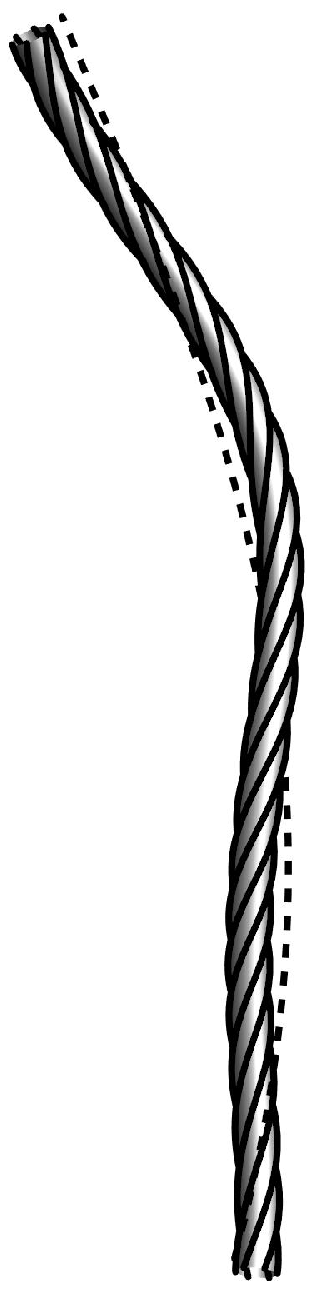}}
\hfil
\subfigure[$\alpha=\frac{\pi}{4}$]{\includegraphics[scale=0.6]{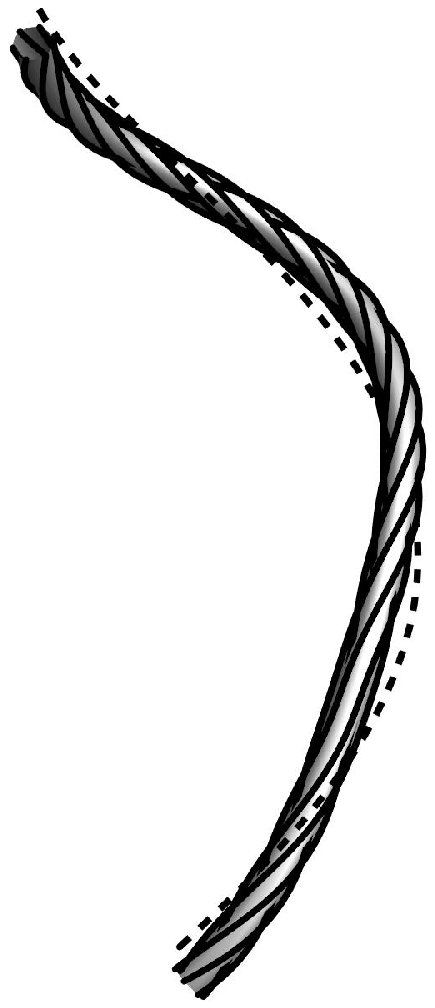}}
\hfil
\subfigure[$\alpha=\frac{3\pi}{8}$]{\includegraphics[scale=0.6]{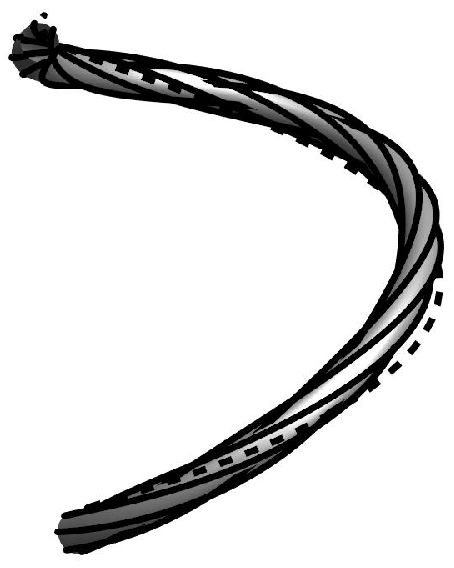}}
\hfil
\subfigure[$\alpha=\frac{\pi}{2}$]{\includegraphics[scale=0.62]{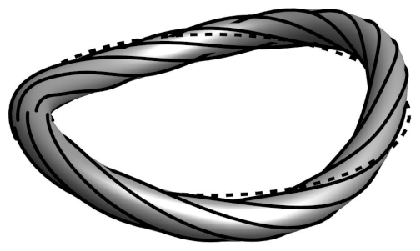}}
\caption{Deformation of helices of length $L=2\pi \rho_{(0)}$, $\chi_3=2/3$ and wavenumber $Q=n/\rho_{(0)}$ without boundary changes of the radial coordinate. For (a)-(d), $n=1$ and for (e), $n=2$. The magnitude of the perturbation has been exaggerated for visualization purposes. The original helices are shown with dashed lines.}
 \label{fig:4}
\end{figure}

\item 
\textbf{Fixed azimuthal and height coordinates at the boundaries}. Requiring condition (\ref{bc:phiz}) to be satisfied along with condition (\ref{eq:L1const}), leads to $Q s_0 = -\mathrm{mod}(n-1,2)\pi/2$,\footnote{$\mathrm{mod}(a,b)$ stands for $a$ modulo $b$} and $QL = \pi n$, $n \in \mathbb{N}$. Thus $Q \rho_{(0)} = n/2$ and in this case $\rms_{3(0)}$ is given by
\begin{equation}
\rho_{(0)} \rms_{3(0)}= \sqrt{\left(\frac{n}{2}\right)^2-(\rho_{(0)} \kappa_{(0)})^2} + 2 \rho_{(0)} \tau_{(0)} \,.
\end{equation}
For even $n$, say $n=2m$, these helices have the same value of $\rms_{3(0)}$ as those of case $\mathrm{I}$ with $m$.
\\
For $0 \leq \alpha < \pi/4$ the helices, close to the line, admit $n \geq 1$, for $\pi/4 \leq \alpha < \pi/2$, helices must have $n \geq 2 $, whereas for the circle with $\alpha=\pi/2$, the admissible modes are $n \geq 3$. These solutions do not satisfy condition (\ref{bc:rho}), so the curve is displaced along the radial direction at the boundaries. Moreover, Eq. (\ref{def:e31}) implies that the tangent vector changes along the azimuthal and radial directions.
\\
The first order corrections to the cylindrical coordinates are given by
\begin{subequations} \label{eq:rhophizII}
\begin{align} 
&\mbox{Odd} \; n, &\quad  \rho_{(1)} &= c_{\rho(0)} \kappa_{M(1)} \sin Q s  \,, 
&\quad
\phi_{(1)} &= -\frac{c_{\phi(0)}}{Q} \kappa_{M(1)} \cos Q s  \,,  
& \quad
z_{(1)} &=  -\frac{c_{z(0)}}{Q} \kappa_{M(1)} \cos Q s  \,,\\
&\mbox{Even} \; n, &\quad \rho_{(1)} &= c_{\rho(0)} \kappa_{M(1)} \cos Q s  \,, 
&\quad
\phi_{(1)} &= \frac{c_{\phi(0)}}{Q} \kappa_{M(1)} \sin Q s  \,,  
& \quad
z_{(1)} &= \frac{c_{z(0)}}{Q} \kappa_{M(1)} \sin Q s  \,,
\end{align}
\end{subequations}
We can determine the maximum curvature perturbation in terms of the first order change of the radial coordinate at the boundaries
\begin{subequations}
\begin{align}
\mbox{Odd} \; n, &\qquad  \kappa_{M(1)}= (-1)^{(n-1)/2} \, \csc^2 \alpha \, \rho^2_{(0)} \rmf^2_{(0)} \, \rho_{b(1)}\,, \quad \pm \rho_{b(1)}:=\rho_{(1)}(\pm L/2) \,, \\
\mbox{Even} \; n, &\qquad  \kappa_{M(1)} = (-1)^{n/2} \, \csc^2 \alpha \, \rho^2_{(0)} \rmf^2_{(0)} \, \rho_{b(1)}\,, \quad \rho_{b(1)}:=\rho_{(1)}(\pm L/2) \,.
\end{align}
\end{subequations}
The lowest admissible modes of this family of perturbed helices are shown in Fig. \ref{fig:5}.
\begin{figure}[htb]
 \centering
\subfigure[$\alpha=\frac{\pi}{64}$]{\includegraphics[scale=0.6]{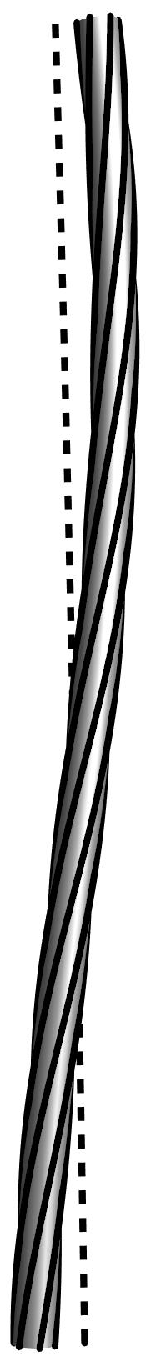}}
\hfil
\subfigure[$\alpha=\frac{\pi}{8}$]{\includegraphics[scale=0.6]{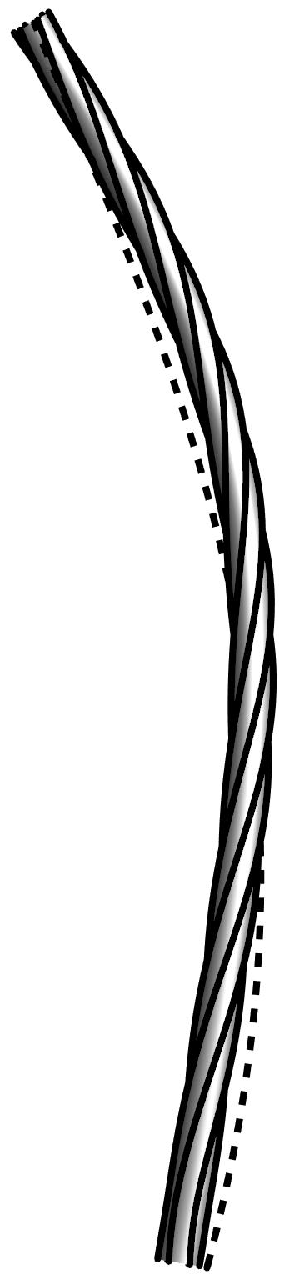}}
\hfil
\subfigure[$\alpha=\frac{\pi}{4}$]{\includegraphics[scale=0.6]{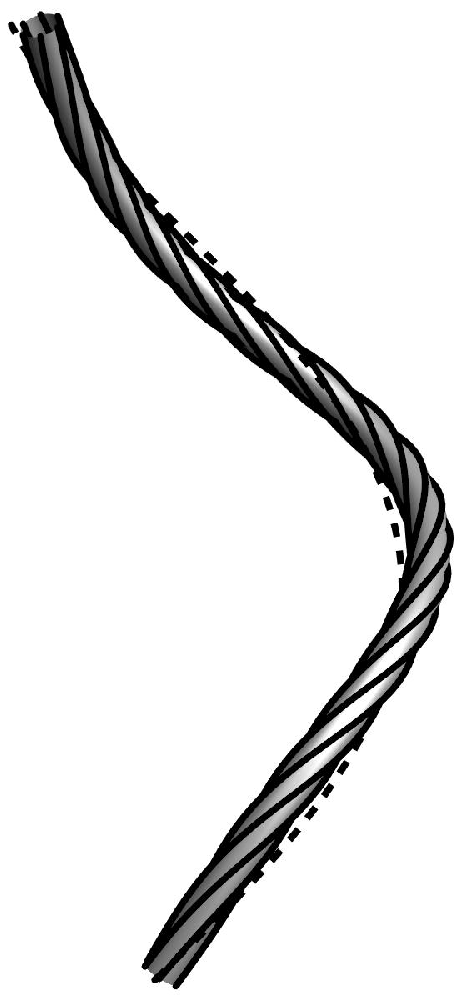}}
\hfil
\subfigure[$\alpha=\frac{3\pi}{8}$]{\includegraphics[scale=0.6]{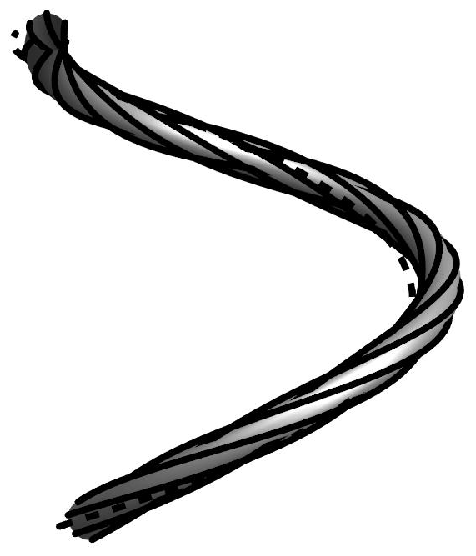}}
\hfil
\subfigure[$\alpha=\frac{\pi}{2}$]{\includegraphics[scale=0.62]{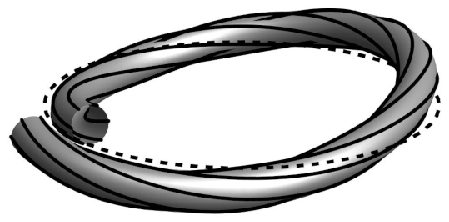}}
\caption{Deformation of helices of length $L=2\pi \rho_{(0)}$, $\chi_3=2/3$ and wavenumber $Q=n/(2 \rho_{(0)})$ without change of the azimuthal and height coordinates at the boundaries. For (a) and (b), $n=1$, (c) and (d), $n=2$ and for (e), $n=3$. The magnitude of the perturbation has been exaggerated for visualization purposes. The original helices are shown with dashed lines.}
 \label{fig:5}
\end{figure}

\item
\textbf{Fixed boundaries}. For $Q s_0 = -\pi/2$, both conditions (\ref{bc:rho}) and (\ref{bc:phiz}) can be satisfied, so the boundaries remain fixed. In this case the wavenumber $Q$ is determined from the transcendental equation
\begin{equation}
\tan \frac{Q L}{2} = \frac{Q L}{2}\,.
\end{equation}
The first three solutions to this equation are $Q L/2=4.493, 7.725, 10.904$ \cite{LandauBook}, so $Q \rho_{(0)}= 1.43, 2.459, 3.471$. For the first wavenumber, $\rms_{3(0)}$ is given by
\begin{equation}
\rho_{(0)} \rms_{3(0)}= \sqrt{2.046-(\rho_{(0)} \kappa_{(0)})^2} + 2 \rho_{(0)} \tau_{(0)} \,.
\end{equation}
In this case the first order corrections to the cylindrical coordinates of the curve are given by
\begin{subequations}
\begin{eqnarray}
\rho_{(1)}(s) &=& c_{\rho(0)} \kappa_{M(1)} \left( \cos Q s -\cos \frac{Q L}{2}   \right) \,,\\
\phi_{(1)}(s) &=& \frac{c_{\phi(0)}}{Q} \kappa_{M(1)} \left( \sin Q s - 2 \sin \frac{Q L}{2} \left(\frac{s}{L} \right)   \right)  \,,  \\
z_{(1)}(s) & = & \frac{c_{z(0)}}{Q} \kappa_{M(1)} \left( \sin Q s - 2 \sin \frac{Q L}{2} \left(\frac{s}{L} \right)  \right)  \,.
\end{eqnarray}
\end{subequations}
The derivative of the radial coordinate does not vanish at the boundaries $\rho_{(1)}'(\pm L/2) \propto \sin Q L/2 \neq 0$, so the tangent vector changes along the radial direction at the boundaries. The maximum curvature perturbation can be expressed in terms of the first order change in the radial coordinate at the midpoint of the rod
\begin{equation}
\kappa_{M(1)} = \frac{ \csc^2 \alpha \, \rho^2_{(0)} \rmf^2_{(0)}}{1 -\cos \frac{Q L}{2}} \, \rho_{0(1)}\,, \quad \rho_{0(1)}:=\rho_{(1)}(0)\,.
\end{equation}
Deformed helices with the lowest wavenumber are plotted in Fig. \ref{fig:6}.
\begin{figure}[htb]
 \centering
\subfigure[$\alpha=\frac{\pi}{64}$]{\includegraphics[scale=0.6]{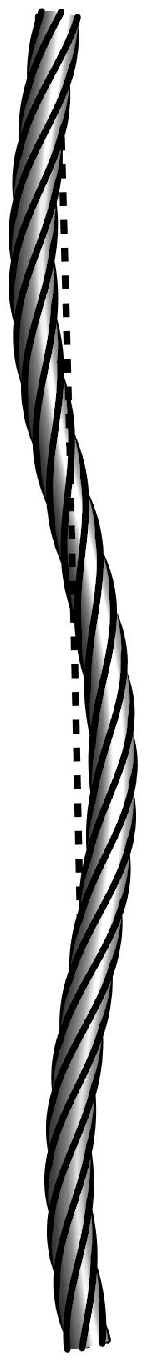}}
\hfil
\subfigure[$\alpha=\frac{\pi}{8}$]{\includegraphics[scale=0.6]{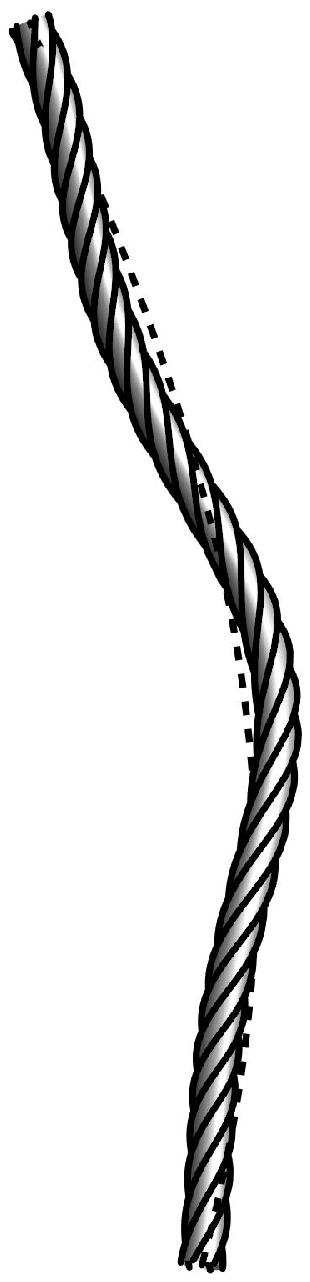}}
\hfil
\subfigure[$\alpha=\frac{\pi}{4}$]{\includegraphics[scale=0.6]{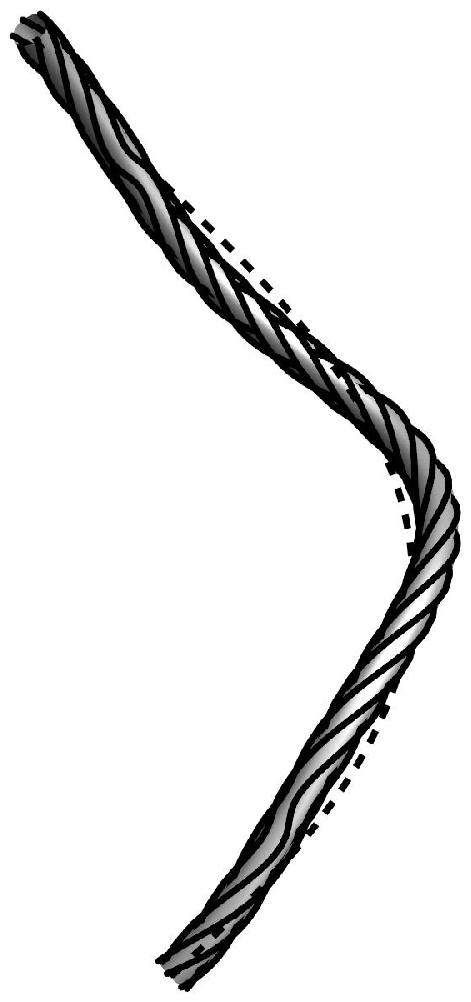}}
\hfil
\subfigure[$\alpha=\frac{3\pi}{8}$]{\includegraphics[scale=0.62]{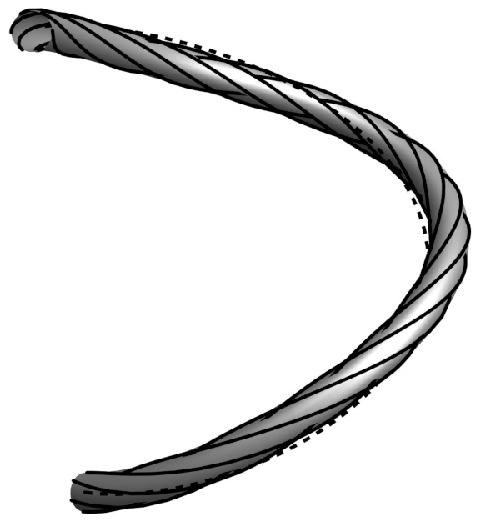}}
\hfil
\subfigure[$\alpha=\frac{\pi}{2}$]{\includegraphics[scale=0.62]{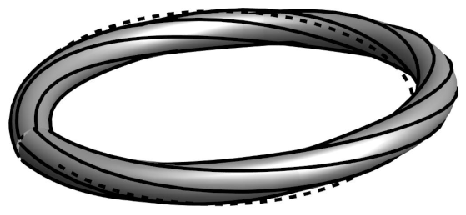}}
\caption{Deformation of helices of length $L= 2 \pi \rho_{(0)}$, $\chi_3=2/3$ and wavenumber $Q= 1.43/\rho_{(0)}$. The magnitude of the perturbation has been exaggerated for visualization purposes. The original helices are shown with dashed lines.}
 \label{fig:6}
\end{figure}
\end{enumerate}
Having determined $\rms_{3(0)}$ the zeroth order of the force, torque and intrinsic tension $\nu_{(0)}= \zeta_{(0)} + \rms_{3(0)}/(2 \chi_3)$, given by Eqs. (\ref{eq:f0j0zeta0}), are completely determined. Besides, the zeroth order of the total energy of the helices, given by
\begin{equation} \label{eq:istoten}
 h_{(0)}= \frac{H_{(0)}}{\rmk} = \pi \rho_{(0)} \left( \kappa_{(0)}^2 + \frac{\rms_{3(0)}^2}{\chi_3} \right)\,. 
\end{equation}
is also determined. Their scaled values are plotted in Figs. \ref{fig:7}(a)-(d) as a function of $\alpha$. In these plots, values for the first family are shown with black lines (lower one with $n=1$ and upper one with $n=2$); for the second family are shown with gray lines ($n=1$ for the one ending at $\pi/4$, $n=2$ coincides with $n=1$ of the first family, and $n=3$ for the complete one); for the third family are shown with light gray lines (lower one with $Q \rho_{(0)} = 1.43$ and upper one with $Q \rho_{(0)} = 2.459$ respectively). For the purpose of comparison, also the case of pure bending ($\rms_{3(0)}=0$) is shown with dashed lines.
\\
The scaled force $\rho_{(0)}^2 \rmf_{(0)}$, shown in Fig. \ref{fig:7}(a), vanishes for the line with $\alpha=0$ in all cases. Intermediate helices are under tension ($\rmf_{(0)} >0$), which initially increases monotonically, reaching a maximum and as the helices tend to the circle it decreases towards the value $\sqrt{(\rho_{(0)}Q)^2-1}$, vanishing only for the state $n=1$ of first family. By contrast in the case of pure bending, the helices are always under compression ($\rmf_{(0)} <0$). The discontinuities in the force, either for the first family for $\alpha = \pi/2$ to deform the circle into a two-fold configuration, or the second family as $\alpha = \pi/4$ to pass from $n=1$ to $n=2$, are analogous to the Euler buckling instability \cite{LandauBook, AudolyBook}.
\\
The scaled torque $\rho_{(0)} \rmj_{(0)}$ is plotted in Fig. \ref{fig:7}(b). For the line it vanishes in the pure bending case, whereas for the three families its value is given by $ \rho_{(0)}Q$. In all cases, its value increases, reaching a maximum and then decreases as the helices approach the circle, for which it achieves a unit value, $\rho_{(0)} \rmj_{(0)} = 1$.
\\
We observe in Fig. \ref{fig:7}(c), that except for the lowest case of the second family terminating at $\alpha=\pi/4$, in all other cases the scaled intrinsic tension $\rho_{(0)}^2 \nu_{0}$ has a value $(\rho_{(0)}Q)^2/(2 \chi_3)$, first increases and then decreases towards the circle with a value of $1/2(1-1/\chi_3+(\rho_{(0)}Q)^2/\chi_3)$. For the pure bending case it vanishes for the line and starts with negative values, later becoming positive as the helices tend to the circle with a value of $1/2$.
\\
The scaled total energy, shown in Fig. \ref{fig:7}(d), begins with a value $(\rho_{(0)}Q)^2/\chi_3$ for the line, then increases and decreases as the circle is approached, achieving a value $1-1/\chi_3+(\rho_{(0)}Q)^2/\chi_3$. In the pure bending case, it vanishes for the line and increases towards a unit value for the circle.
 
\section{Anisotropic Kirchhoff elastic rods} \label{Sec:Anisorods}

Now we consider anisotropic rods whose energy density, defined in Eq. (\ref{def:anisenden}), is given in full  by
\begin{equation} \label{eq:ABendEnDens}
\calF = \calF_{AB} + \calF_{Tw} = \frac{\rmk_1}{2} \left(\kappa_1 - c_1\right)^2 + \frac{\rmk_2}{2} \left(\kappa_2 - c_2\right)^2 + \frac{\rmk_3}{2} \left(\kappa_3 - c_3\right)^2 \,.
\end{equation}
Like before, we rescale all quantities by the inverse of $\rmk_1$, for instance the rescaled components of the intrinsic torque are 
\begin{equation}
\rms_1 = \frac{\rmS_1}{\rmk_1}=\kappa_1 - c_1, 
\quad 
\rms_2 = \frac{\rmS_2}{\rmk_1} = \chi_2 \left(\kappa_2 - c_2\right) \,,
\quad 
\rms_3 = \frac{\rmS_3}{\rmk_1} = \chi_3 \left(\kappa_3 - c_3\right) \,,
\quad 
\chi_i = \frac{\rmk_i}{\rmk_1}\,, \quad i=2,3 \,.
\end{equation}
The squared norm of the normal part no longer involves $\kappa^2$, $\rms_1^2 +\rms_2^2 = \left(\kappa_1 - c_1\right)^2 + \chi_2^2 \left(\kappa_2 - c_2\right)^2$. 
\\
In this case Eq. (\ref{eq:sym12}) reads
\begin{equation} \label{eq:k3paniso}
\chi_3 \kappa_3' =  \kappa_2 \left(\kappa_1 -c_1 \right) - \chi_2 \kappa_1 \left( \kappa_2 - c_2\right) \,,
\end{equation}
so the asymmetry in the bending energy prevents the conservation of the twist along the rod. Only  for $\chi_2=1$ and null spontaneous curvatures $c_1=c_2=0$ the twist $\kappa_3$ becomes constant.
\\
The components of the force are given by
\begin{subequations} \label{eq:F123aniso}
\begin{eqnarray} 
\rmf_1 &:=& \frac{\rmF_1}{\rmk_1} = - \chi_2 \kappa_2' + \left( \chi_3-1 \right) \, \kappa_1 \kappa_3  + c_1 \kappa_3 - \chi_3 c_3 \kappa_1 \,, \nonumber \\
\rmf_2&:=&\frac{\rmF_2}{\rmk_1} = \kappa_1' + \left( \chi_3 - \chi_2 \right) \kappa_2 \kappa_3 + \chi_2 c_2 \kappa_3 - \chi_3 c_3 \kappa_2 \,, \\
\rmf_3&:=&\frac{\rmF_3}{\rmk_1} = - \frac{1}{2} \left(\kappa_1^2 - c_1^2 \right) - \frac{\chi_2}{2} \left(\kappa_2^2 - c_2^2 \right) - \frac{\chi_3}{2} \left(\kappa_3^2 - c_3^2 \right) + \nu  \,, \nonumber 
\end{eqnarray}
\end{subequations}
where we have defined $\nu= \mu/\rmk_1$.
\\
From the first integrals, given by Eqs. (\ref{eq:H1pH2p}), we obtain the derivatives of the normal curvatures
\begin{subequations} \label{eq:k1pk2panisorod}
\begin{eqnarray}
 \kappa_1' &=& (\kappa_1 -c_1) b_{\textrm{A}}(\kappa_i) + \chi_2 \left(\kappa_2-c_2\right) a_\textrm{A}(\kappa_i) - \chi_{3} \, \left(\kappa_3-c_3\right) \, \kappa_2  \,, \\
 \chi_2 \kappa_2 ' &=& - \left(\kappa_1-c_1\right)  a_\textrm{A}(\kappa_i) + \chi_2 (\kappa_2 -c_2) b_\textrm{A}(\kappa_i) + \chi_{3} \, \left(\kappa_3-c_3\right) \, \kappa_1  \,, 
\end{eqnarray}
\end{subequations}
where we have defined the following rescaled quantities
\begin{subequations} \label{abaniso}
\begin{eqnarray} 
a_{\textrm{A}}(\kappa_i) &:=& \frac{A}{\rmk_1^2(\rms_1^2+\rms_2^2)} + \kappa_3 = \frac{\rmf \rmj -\rms_3 \rmf_3}{\rms_1^2 + \rms_2^2 } + \kappa_3\,, \quad \rmf := \frac{\rmF}{\rmk_1}\,, \quad \rmj := \frac{\rmJ}{\rmk_1} \,, \\
b_\textrm{A}(\kappa_i)^2 &:=& \left(\frac{B}{\rmk_1^2 (\rms_1^2+\rms_2^2)}\right)^2 =  \frac{\rmf^2 - \rmf_3^2}{\rms_1^2 + \rms_2^2} - \left( \frac{\rmf \rmj -\rms_3 \rmf_3}{\rms_1^2 + \rms_2^2 } \right)^2\,. 
\end{eqnarray}
\end{subequations}
Equations (\ref{eq:k3paniso}) and (\ref{eq:k1pk2panisorod}) constitute a system of three first order differential equations with five parameters, two ratios of bending and twisting moduli $\chi_2$ and $\chi_3$, and three spontaneous curvatures $c_i$, $i=1,2,3$. Like before, the three constants of integration $\rmf$, $\rmj$ and $\nu$ are to be determined from boundary conditions.
\\
If we consider the chiral energy density (\ref{Fchiral}) the components of the intrinsic torque and force have the additional contributions 
\begin{subequations} \label{eq:sifichiral}
\begin{eqnarray}
\rms_I &\rightarrow& \rms_I + \chi_{I3} \kappa_3\,, \quad \rms_3 \rightarrow \rms_3 + \chi_{13} \kappa_1 + \chi_{23} \kappa_2\,, \quad \chi_{I3}=\rmk_{I3}/\rmk_1 \\
\rmf_1 & \rightarrow & \rmf_1 + \chi_{13} \left(\kappa_1^2-\kappa_3^2\right)-\chi_{23} \left(\kappa_3' - \kappa_1\kappa_2\right)\,, \\
\rmf_2 & \rightarrow & \rmf_2 + \chi_{13} \left(\kappa _3'+\kappa _1 \kappa _2\right)+\chi_{23} \left(\kappa _2^2-\kappa _3^2\right) \,,\\
\rmf_3 & \rightarrow & \rmf_3 - \left(\chi_{13} \kappa _1 + \chi_{23} \kappa _2 \right)\kappa_3 \,.
\end{eqnarray}
\end{subequations}
We observe an additional asymmetry of the normal components of the force.
The first integrals are now given by
\begin{subequations}
 \begin{eqnarray}
\xi \kappa_1'&=&  \left(\chi_{13} \chi_{23} \rms_1 + \left(\chi_{23}^2 - \chi_2 \chi_3\right)\rms_2 \right)a_A(\kappa_i) + \left(\rms_1 \left(\chi_{23}^2- \chi_2 \chi_3\right) - \chi_{13} \chi_{23} \rms_2\right)b_A(\kappa_i) \nonumber  \\
&& + \chi_{13} \chi_2 (\kappa_2 \rms_1-  \kappa_1 \rms_2) +s_3 \left(\kappa _2 \chi_2 \chi_3-k_{23} \left(\kappa _1 k_{13}+\kappa _2 k_{23}\right)\right) \,, \\
\xi \kappa_2' &=& -\left( \left(\chi _{13}^2 - \chi_3\right) \rms_1 + \chi_{13} \chi_{23}  \rms_2 \right) a(\kappa_i) + \left(- \chi_{13} \chi_{23} \rms_1 + \left(\chi _{13}^2-\chi_3\right)\rms_2 \right) b(\kappa_i) \nonumber \\
&&+ \left(\kappa _2 \rms_1 -\kappa _1 \rms_2\right) \chi_{23} + \left( \left(\chi_{13}^2-\chi_3\right)\kappa_1 + \chi_{13} \chi_{23} \kappa_2\right)\rms_3 \,, \\
\xi \kappa_3' &=& \left( \chi_2 \chi_{13} \rms_2 - \chi _{23} \rms_1 \right)a_A(\kappa_i) + \left( \chi_{13} \chi_2 \rms_1 +  \chi _{23} \rms_2 \right) b_A(\kappa_i) + \left(\kappa _1 \chi _{23}-\kappa _2 \chi _{13} \chi_2\right)\rms_3 + \chi_2 \left(\kappa_2 \rms_1 - \kappa_1 \rms_2\right)\,.
\end{eqnarray}
\end{subequations}
where we have defined the constant $\xi=(\chi _{13}^2 - \chi_3)\chi_2+\chi_{23}^2$; 
$a_A(\kappa_i)$ and $b_A(\kappa_i)$ are defined as in Eqs. (\ref{abaniso}), but with the $\rms_i$ and $\rmf_3$ given as in Eqs. (\ref{eq:sifichiral}). In this case the three differential equations for the three curvatures have the same structure. It is noteworthy to mention that this procedure could be employed to address also the cubic and quartic energy densities, such as the ones given above in Eqs. (\ref{def:anisocubquaren}). Although, the resulting expressions may seem daunting, they might be solved numerically. For the sake of concreteness and simplicity, in the following we do not consider such additional terms in the energy. 

\subsection{Twisting instabilities of anisotropic helices}

Now we turn to analyze the instabilities due to torsion on anisotropic helical rods. To establish a relation with the isotropic cases presented before, we consider the case of vanishing spontaneous curvatures and twist, $c_i=0$.  Furthermore, if $\chi_2 = 1$, Eqs. (\ref{eq:k1pk2panisorod}) reduce to Eqs (\ref{eq:k1pk2piso}) with $c_0=0$, whereas Eq. (\ref{eq:k3paniso}) also implies that the twist is constant. Therefore, we consider a curve whose normal curvatures differ slightly, so we can expand the equations about the circular helices, presented in Sec. \ref{sec:helices}. To this end we expand the material curvatures and constants as
\begin{subequations}
\begin{align}
\kappa_1 &= \kappa_{1(0)} + \kappa_{1(1)} + \kappa_{1(2)} + \dots \,
&\quad 
\kappa_2 &= \kappa_{2(0)} +\kappa_{2(1)}+ \kappa_{2(2)} +\dots \,, 
&\quad 
\kappa_3 &= \kappa_{3(0)} + \kappa_{3(1)} + \kappa_{3(2)} +\dots \,,\\
\rmf &= \rmf_{(0)}+ \rmf_{(1)}+\rmf_{(2)} + \dots \,, 
&\quad 
\rmj &= \rmj_{(0)} +\rmj_{(1)}+\rmj_{(2)}  + \dots\,, 
&\quad 
\nu &= \nu_{(0)} + \nu_{(1)} + \nu_{(2)} +\dots \,, 
\end{align}
\end{subequations}
where 
\begin{equation} \label{anisok1k2thetaqzero}
\kappa_{1(0)}= \kappa_{(0)} \sin \theta_{(0)}\,, \quad \kappa_{2(0)}= \kappa_{(0)} \cos \theta_{(0)}\,, \quad \theta_{(0)}= q (s-s_0)\,, \quad q= \kappa_{3(0)}- \tau_{(0)}\,,
\end{equation}
and $\nu_{(0)} = \zeta_{(0)} + \rms_{3(0)}^2/(2 \chi_3)$, with $\rmf_{(0)}$, $\rmj_{(0)}$ and $\zeta_{(0)}$ given by Eqs. (\ref{eq:f0j0zeta0}). Moreover, we expand the ratio as $\chi_{2} = 1 + \chi_{2(1)} + \dots$, so $\chi_{2(1)}$ captures the anisotropy of the rod. We consider a finite ratio $\chi_{3}$, so $\rms_3 = \rms_{3(0)} + \rms_{3(1)} + \dots $, with $\rms_{3(0)}= \chi_{3} \kappa_{3(0)}$, $\rms_{3(1)}= \chi_{3} \kappa_{3(1)}$. 
\\
To zeroth order Eqs. (\ref{eq:k1pk2panisorod}) are satisfied and (\ref{eq:k3paniso}) is satisfied for constant $\rms_{3(0)}$.
\\
In this case, instead of expanding the first integrals (\ref{eq:k1pk2panisorod}), we consider the EL Eqs. given by 
\begin{subequations} \label{ELeqsaniso}
 \begin{eqnarray}
  \mathcal{E}_1 &=& - \chi_2 \kappa_2'' + \left( \left( \chi_ 3 -2 \right) \kappa_3 - \chi_3 c_3 \right) \kappa_1 ' + \left( \left( \chi_ 3 - 1 \right) \kappa_1 + c_1 \right) \kappa_3 '  \nonumber \\
  &-& \kappa_2 \left(\frac{1}{2} \left(\kappa_1^2-c_1^2 \right) +\frac{\chi_2}{2} \left(\kappa_2^2-c_2^2 \right) + \frac{\chi_3}{2} \left(3 \kappa_3 ^2 -2 c_3 \kappa_3  - c_3^2 \right)  - \nu \right) + \chi_2 \kappa_3^2 (\kappa_2-c_2)  \,, \\
  \mathcal{E}_2 &=& \kappa_1'' + \left( \left( \chi_ 3 -2 \chi_2 \right) \kappa_3 - \chi_3 c_3 \right) \kappa_2 ' + \left( \left( \chi_ 3 -\chi_2 \right) \kappa_2 + \chi_2 c_2 \right) \kappa_3 '  \nonumber \\
  &+& \kappa_1 \left(\frac{1}{2} \left(\kappa_1^2-c_1^2 \right) +\frac{\chi_2}{2} \left(\kappa_2^2-c_2^2 \right) + \frac{\chi_3}{2} \left(3 \kappa_3 ^2 -2 c_3 \kappa_3  - c_3^2 \right)  - \nu \right) -\kappa_3^2 (\kappa_1-c_1) \,.
 \end{eqnarray}
\end{subequations}
In this case, to first order the EL eqs. (\ref{ELeqsaniso}) and Eq. (\ref{eq:k3paniso}) read
\begin{subequations} \label{ELeqsanisoord1}
\begin{eqnarray}
\kappa_{1(1)}'' +  (\chi_3-2) \kappa_{3(0)} \kappa_{2(1)}' + (\chi_3-1) \kappa_{2(0)} \kappa_{3(1)}'+ \left(\kappa_{1(0)}^2+q (\kappa_{3(0)} (\chi_3-1)-\tau_{(0)})\right)\kappa_{1(1)} && \nonumber \\ 
+\kappa_{1(0)} \left[ \kappa_{2(0)} \kappa_{2(1)}- ( (2-\chi_3)\tau_{(0)} - 2 \chi_3 \kappa_{3(0)} )\kappa_{3(1)} + \frac{1}{2} \chi_{2(1)}  \left(\kappa_{2(0)}^2+4q \kappa_{3(0)} \right) -\nu_{(1)} \right]  &=&0 \label{EL1eqsanisoord1} \,,\\
\kappa_{2(1)}'' - (\chi_3-2) \kappa_{3(0)} \kappa_{1(1)}' - (\chi_3-1) \kappa_{1(0)} \kappa_{3(1)}' + \left(\kappa_{2(0)}^2+q (\kappa_{3(0)} (\chi_3-1)-\tau _{(0)})\right)\kappa_{2(1)} && \nonumber \\
+\kappa_{2(0)} \left[ \kappa_{1(0)} \kappa_{1(1)} - ( (2-\chi_3)\tau _{(0)}-2 \kappa_{3(0)} \chi_3)\kappa_{3(1)} + \frac{1}{2} \chi_{2(1)}  \left(\kappa_{2(0)}^2 - 2 \left(\kappa_{3(0)}^2+q^2\right)\right) -\nu_{(1)} \right] &=& 0\,, \label{EL2eqsanisoord1}\\
 \rms_{3(1)}' +  \chi_{2(1)} \kappa _{1(0)} \kappa _{2 (0)} &=& 0\,; 
\end{eqnarray}
\end{subequations}
Integrating the last equation for $q \neq 0$,\footnote{For $q=0$ ($\kappa_{1(0)}=\kappa_{(0)}$ and $\kappa_{2(0)}=0$) or $\chi_{2(1)}=0$, $\rms_{3(1)}'=0$, so $\rms_{3(1)}$ is constant.}, we get that the first order correction to the twist has twice the period of the normal curvatures
\begin{equation}
\rms_{3 (1)} = \frac{\kappa_{(0)}^2 \chi_{2(1)}}{4 q} \cos 2 q (s-s_0)\,.
\end{equation}
Substituting this result in Eqs. (\ref{EL1eqsanisoord1}) and (\ref{EL2eqsanisoord1}), along with the ansatz
\begin{equation}
\kappa_{1(1)} = \sum_{j=1}^{3} A_{j(1)} \sin j q (s-s_0)  \,, \quad
 \kappa_{2(1)} = \sum_{j=1}^{3} B_{j(1)} \cos j q (s-s_0) \,, 
\end{equation}
we find that they are satisfied for the following values of the coefficients
\begin{subequations}
 \begin{eqnarray}
  A_{1(1)} &=& \frac{\nu_{(1)}}{\kappa_{(0)}} + \mathrm{A}_{(0)} \chi_{2(1)}\,, \quad B_{1(1)} = \frac{\nu_{(1)}}{\kappa_{(0)}} + \mathrm{B}_{(0)} \chi_{2(1)}\,, \\
  A_{2(1)} &=&B_{2(1)}=0  \,, \\
  A_{3(1)} &=& B_{3(1)} := C_{3(1)} = \mathrm{C}_{(0)} \chi_{2(1)}\,, 
 \end{eqnarray}
\end{subequations}
where $\mathrm{A}_{(0)}$, $\mathrm{B}_{(0)}$ and $\mathrm{C}_{(0)}$ are constants defined by the following relations
\begin{subequations}
\begin{eqnarray}
\frac{\mathrm{B}_{(0)}-\mathrm{A}_{(0)}}{2} &=& \frac{\kappa_{(0)} \left(-\kappa_{(0)}^2 (s_{3(0)}+2 q) + (2 q+\tau_{(0)})^2 (2 (q-\tau_{(0)}) + s_{3(0)})\right)}{4 q \left(\kappa_{(0)}^2-4 q^2+(\rms_{3(0)}-2 \tau_{(0)})^2 \right)} - \mathrm{C}_{(0)}  \,,\\
\frac{\mathrm{A}_{(0)} + \mathrm{B}_{(0)}}{2} &=& \frac{\tau_{(0)}^2}{2 \kappa_{(0)}}-\frac{\kappa_{(0)}}{4} \\
\mathrm{C}_{(0)}&=&\frac{\kappa_{(0)}^3 \left(\left(1-\frac{1}{\chi_3}\right) \left(-\kappa_{(0)}^2-4 \tau_{(0)}^2+4 q^2\right)+(q+3 \tau_{(0)}) (\rms_{3(0)}-\tau_{(0)})-3 q \tau_{(0)}\right)}{16 q^2 \left(\kappa_{(0)}^2-4 q^2+(\rms_{3(0)}-2 \tau_{(0)})^2\right)} 
\end{eqnarray}
\end{subequations}
This is consistent with the fact that four constants of integration are required. We can obtain the first order correction of the intrinsic tension, $\nu_{(1)}$, from the first order of the first integrals (\ref{eq:k1pk2panisorod})
\begin{equation}
\nu_{(1)} = \rmf_{(1)} \tau_{(0)} \ell + \frac{j_{(1)}}{\ell} - \frac{1}{4} \kappa_{(0)}^2 \chi_{2(1)}\,.
\end{equation}
Therefore, the first order corrections to the normal curvatures are
\begin{subequations}
\begin{eqnarray}
\kappa_{1(1)} &=& A_{1(1)} \sin q (s-s_0) + C_{3(1)} \sin 3 q (s-s_0)  \,, \label{kappa11} \\
 \kappa_{2(1)} &=& B_{1(1)} \cos q (s-s_0) + C_{3(1)} \cos 3 q (s-s_0) \,, \label{kappa21} 
\end{eqnarray}
\end{subequations}
involving only odd multiples of the wavenumber $q$. Thus, the anisotropy not only rescales the magnitude of zeroth order solution with a single wavenumber, but also introduces terms with thrice the wavenumber.
\\
The first order correction to the curvature, given by the relation $\kappa_{(0)}  \kappa_{(1)}= \kappa_{1(0)} \kappa_{1(1)}+ \kappa_{2(0)} \kappa_{2(1)}$, likewise has twice the wavenumber of the zeroth order normal curvatures
\begin{equation}
\kappa_{(1)}= \kappa_{M(1)} \cos 2 q (s-s_0) + \kappa_{c(1)} \,, 
\end{equation}
where
\begin{equation}
\kappa_{M(1)} = \frac{B_{1(1)}-A_{1(1)}}{2} +C_{3(1)}
\,, \quad
\kappa_{c(1)}= \frac{A_{1(1)}+B_{1(1)}}{2} \,.
\end{equation}
For the isotropic case with $\chi_{2(1)}=0$, $A_{1(1)}=B_{1(1)}$ and $C_{3(1)}=0$, so $\kappa_{(1)}$ becomes constant.
\\
Proceeding as in Sec. (\ref{subsec:persoliso}), we get that the first order corrections to the cylindrical coordinates are
\begin{subequations}
\begin{eqnarray}
\rho_{(1)} &=& a_{\rho(1)} \cos 2q(s-s_0) + b_{\rho{(1)}}\,,\\
\phi_{(1)}' &=& a_{\phi(1)} \cos 2q(s-s_0) + b_{\phi(1)} \,,  \\
z_{(1)}' &=& a_{z(1)} \cos 2q(s-s_0) + b_{z(1)} \,,
 \end{eqnarray}
\end{subequations}
where we have defined the following constants
\begin{subequations} \label{anisocrhophiz}
\begin{eqnarray}
 a_{\rho(1)} &= & \frac{\kappa_{(0)}  (\rms_{3(0)}+2 (q-\tau_{(0)}))}{4 q \left(\kappa_{(0)}^2-4 q^2+(\rms_{3(0)}-2 \tau_{(0)})^2\right)} \chi_{2(1)}\,, 
 \\
 b_{\rho{(1)}} &=& \frac{1}{\kappa_{(0)} (\rms_{3(0)}-\tau_{(0)})} \left(  \left(\frac{\tau _{(0)}}{\rms_{3(0)}-\tau_{(0)}}-\kappa_{(0)}^2 \ell^2\right) \ell \, \rmf_{(1)}- \tau _{(0)} \ell \rmj_{(1)} +\frac{\tau_{(0)}^2}{2 (\rms_{3(0)}-\tau_{(0)})} \chi_{2(1)} \right) \,,\\
 a_{\phi(1)} &=& \frac{ \left(2 q \tau_{(0)}-\kappa_{(0)}^2\right) (\rms_{3(0)}+2 q - \tau_{(0)})-2 q \tau_{(0)}^2 }{4 q  \ell \left(\kappa_{(0)}^2-4 q^2 + (\rms_{3(0)}-2 \tau_{(0)})^2\right)} \chi_{2(1)} \,, \\
 b_{\phi(1)} &=& \frac{1}{\kappa_{(0)} \rho_{0}  (\rms_{3(0)}-\tau_{(0)})} \left(\left(1-\frac{\tau_{(0)}}{\rms_{3(0)}- \tau_{(0)}}\right)\rmf_{(1)} + \tau_{(0)} \rmj_{(1)}-\frac{\tau_{(0)}^2 }{2} \left(\ell-\frac{1}{(\rms_{3(0)}-\tau_{(0)}) \ell}\right) \chi_{2(1)}\right) \,, \\
a_{z(1)}&=& \frac{\kappa_{(0)}^2 \ell \left(\kappa_{(0)}^2 - (2 q + \tau_{(0)}) (\rms_{3(0)}+2 (q - \tau_{(0)}))\right)}{4 q  \left(\kappa_{(0)}^2-4 q^2+(\rms_{3(0)}-2 \tau_{(0)})^2\right)} \chi_{2(1)} \,, \\
b_{z(1)} &=& -\frac{\ell \tau_{(0)}}{\rms_{3(0)}-\tau_{(0)}} \left(\ell \, \rmf_{(1)} +\frac{\tau_{(0)}}{2} \chi_{2(1)}\right)\,.
\end{eqnarray}
\end{subequations}
Integration of Eqs. (\ref{anisocrhophiz}) yields
\begin{subequations}
 \begin{eqnarray}
 \phi_{(1)} &=& \frac{a_{\phi(1)}}{2 q} \sin 2 q (s-s_0) + b_{\phi(1)} s \,, \\
 z_{(1)} &=& \frac{a_{z(1)}}{2q} \sin 2 q (s-s_0) + b_{z(1)} s\,.
\end{eqnarray}
\end{subequations}
Like for isotropic rods, we consider three families of deformations of helices with length $L=2 \pi \rho_{(0)}$, satisfying the fixed  length condition (\ref{eq:L1const}), as well as a combination of the boundary conditions (\ref{bc:rho}) and (\ref{bc:phiz}).

\begin{figure}[htb]
 \centering
\subfigure[]{\includegraphics[width=0.476\textwidth]{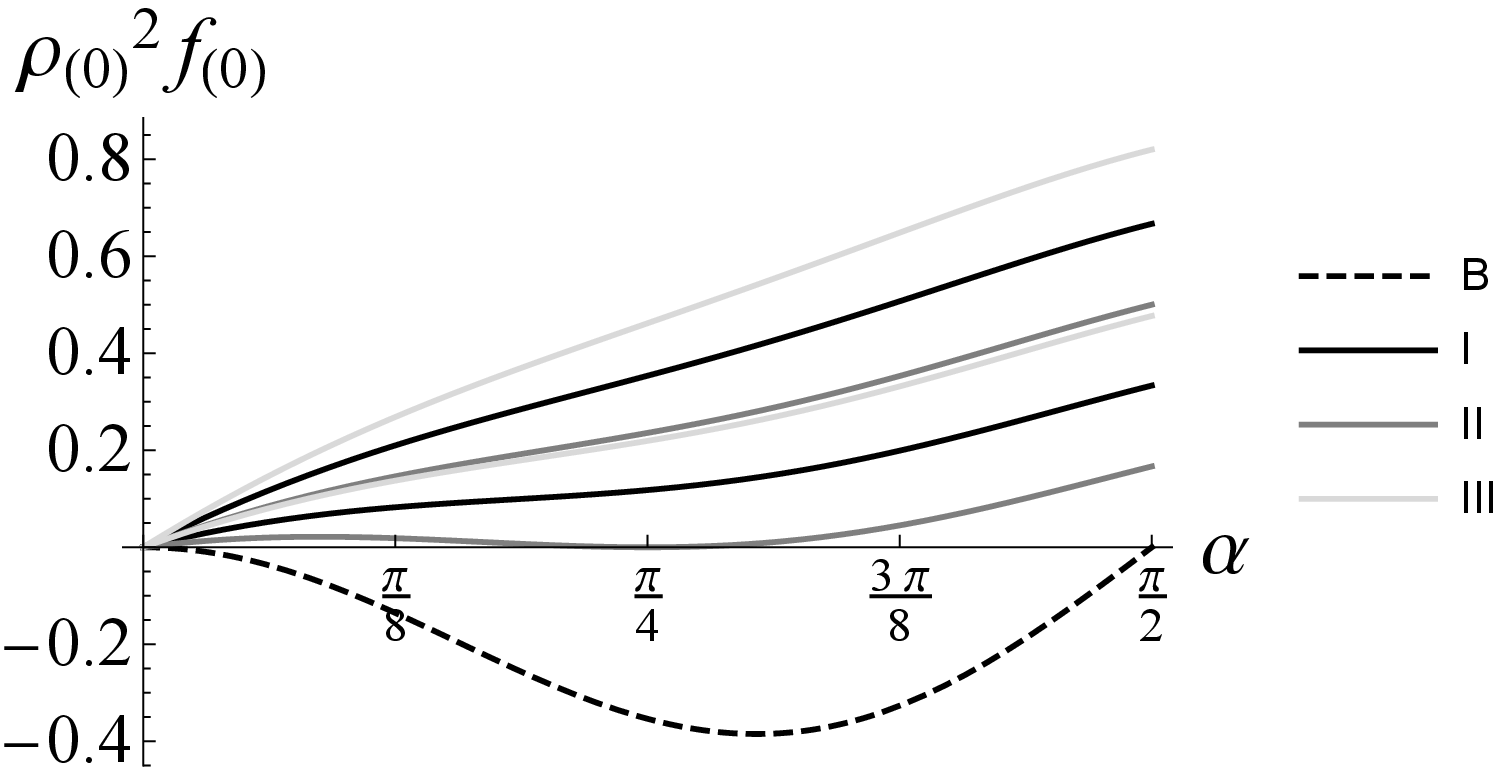}}
\hfil
\subfigure[]{\includegraphics[width=0.475\textwidth]{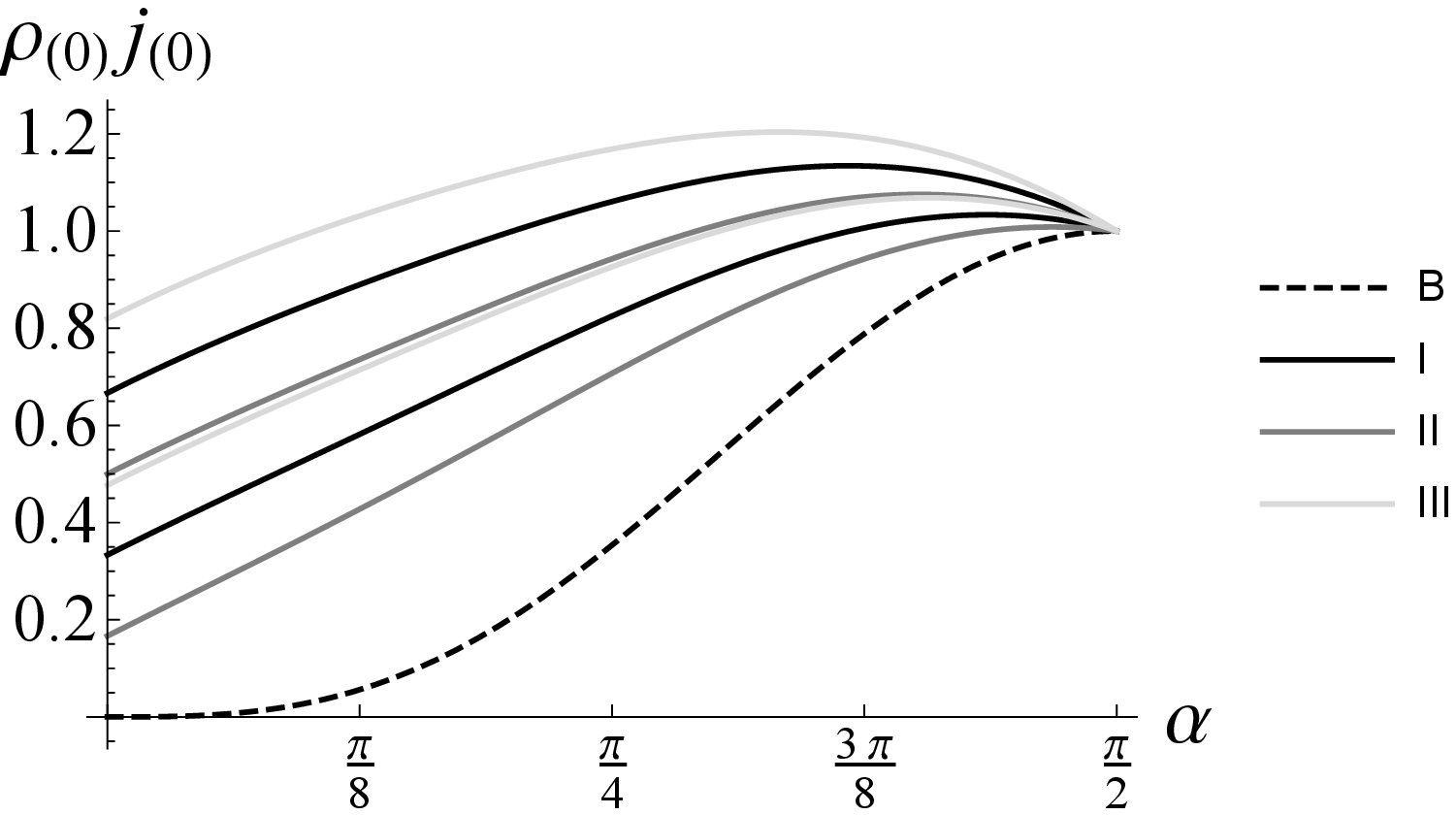}}
\\
\subfigure[]{\includegraphics[width=0.475\textwidth]{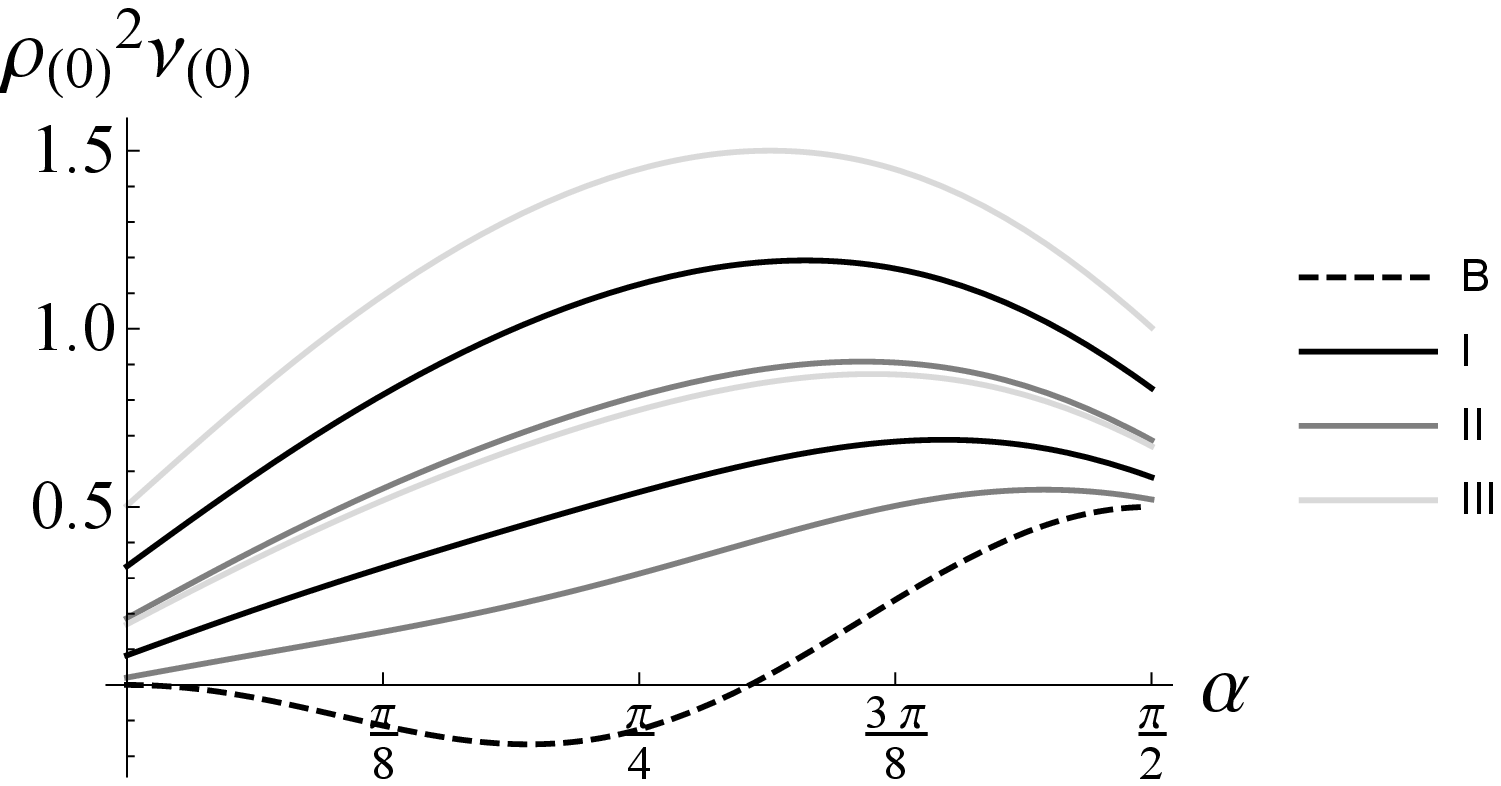}}
\hfil
\subfigure[]{\includegraphics[width=0.475\textwidth]{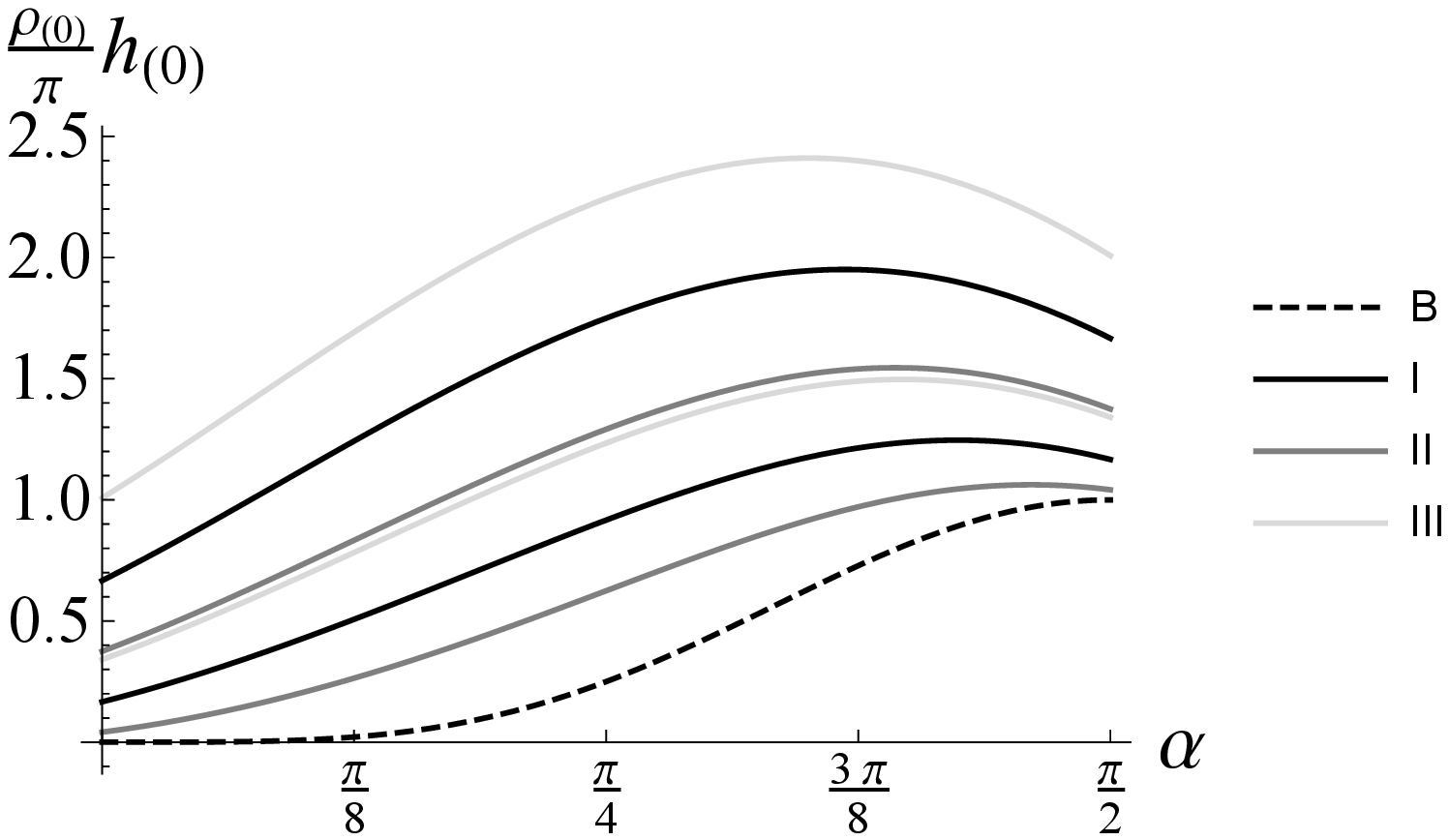}}
\caption{Lowest order of the scaled force, torque, intrinsic tension and total energy for the three families of deformed helices of length $L= 2 \pi \rho_{(0)}$ and $\chi_3 =2/3$. $\alpha=0,\pi/2$ correspond to the line and circle respectively. The dashed line represents the case of pure bending. The black lines represent values for the first family with $q \rho_{(0)}=1/2,1$ (lower ($n=1$) and upper ($n=2$) lines respectively). Gray lines represent values for the second family with $q \rho_{(0)}=1/4,3/4$, (lower ($n=1$) and upper ($n=3$) lines respectively, $n=2$ coincides with the first line of the first family). The light gray lines represent values for the third family with $q \rho_{(0)}=0.715, 1.23$ (lower and upper lines respectively).}
 \label{fig:11}
\end{figure}

\begin{enumerate}[I.]

\item 
\textbf{Fixed radial coordinate at the boundaries}. For $q s_0=\pi/4$, Eqs. (\ref{eq:L1const}) and (\ref{bc:rho}) imply that $q L = \pi n$, $n \in \mathbb{N}$, or $q \rho_{(0)} =n/2$. Substituting this result in the expression of $q$ given by Eq. (\ref{anisok1k2thetaqzero}), we get the zeroth order of the twist is determined only by the torsion
\begin{equation}
\rho_{(0)} \kappa_{3(0)} = \frac{n}{2} + \rho_{(0)} \tau_{(0)} \,.
\end{equation}
The lowest deformation mode for all  helices is one period $n=1$, which are shown in Fig. \ref{fig:8}. Since $\rho_{(0)} \tau_{(0)}= \sin 2 \alpha/2$, the line and the circle have the minimum twist, $\alpha =0, \pi/2$ ($\tau_{(0)}=0$) and is maximum for the intermediate helix with $\alpha= \pi/4$ ($\rho_{(0)}\tau_{(0)}=1/2$). The first order corrections to the cylindrical coordinates are given by
\begin{equation} \label{eq:anisorhophiz12}
\rho_{(1)} = a_{\rho(1)} \sin 2 q s  \,, \quad
\phi_{(1)} = -\frac{a_{\phi(1)}}{2q} \cos 2 q s  \,,   \quad
z_{(1)} =  -\frac{a_{z(1)}}{2 q}  \cos 2 q s  \,.
\end{equation}
These helices do not satisfy boundary condition (\ref{bc:phiz}), so at their boundaries the azimuthal and height coordinates change, whereas the tangent vector changes along the radial direction. The last equation permits us to determine $\chi_{2(1)}$ in terms of the first order change in the height of the boundaries
\begin{equation}
 \chi_{2(1)} =  \frac{8 (-1)^{n+1} q^2  \left(\kappa_{(0)}^2-4 q^2+(\rms_{3(0)}-2 \tau_{(0)})^2\right)}{\kappa_{(0)}^2 \ell \left(\kappa_{(0)}^2 - (2 q + \tau_{(0)}) (\rms_{3(0)}+2 q - 2 \tau_{(0)})\right)} \, z_{b(1)}\,, \quad z_{b(1)}:=z_{(1)}(\pm L/2)\,.
\end{equation}
\begin{figure}[htb]
 \centering
\subfigure[$\alpha=\frac{\pi}{64}$]{\includegraphics[scale=0.6]{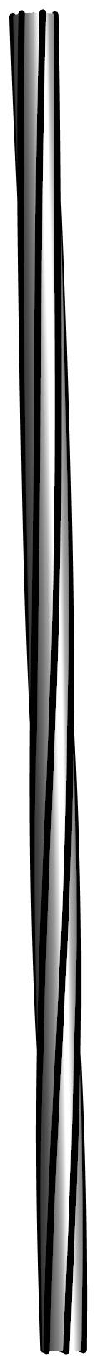}}
\hfil
\subfigure[$\alpha=\frac{\pi}{8}$]{\includegraphics[scale=0.6]{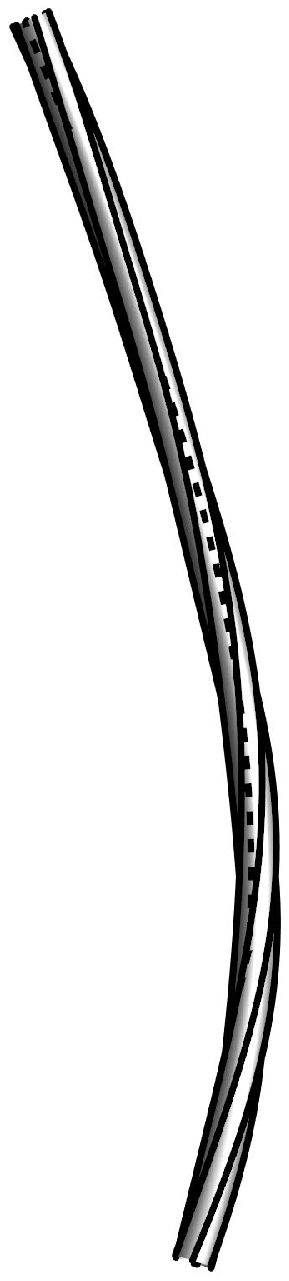}}
\hfil
\subfigure[$\alpha=\frac{\pi}{4}$]{\includegraphics[scale=0.6]{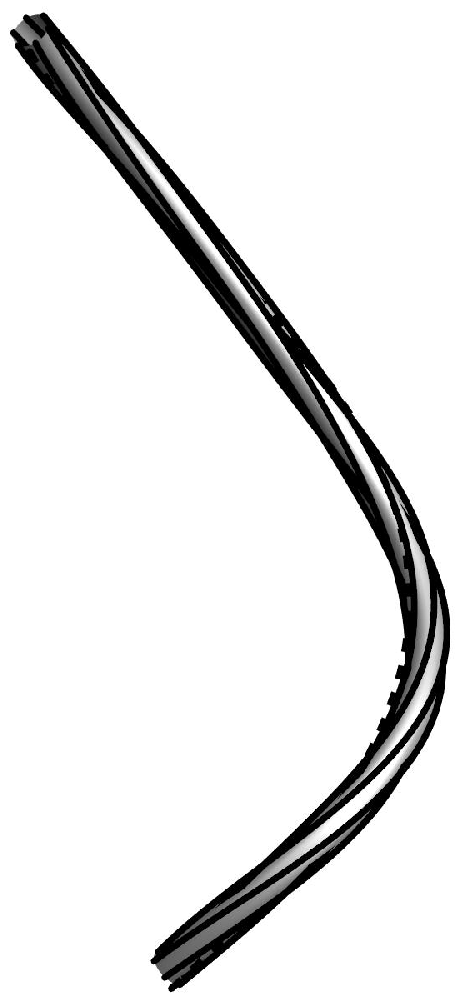}}
\hfil
\subfigure[$\alpha=\frac{3\pi}{8}$]{\includegraphics[scale=0.6]{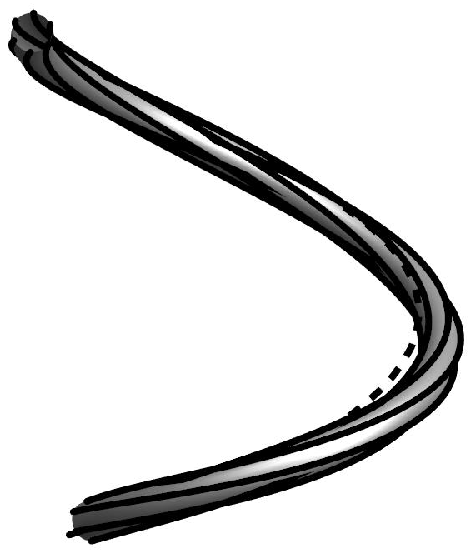}}
\hfil
\subfigure[$\alpha=\frac{\pi}{2}$]{\includegraphics[scale=0.6]{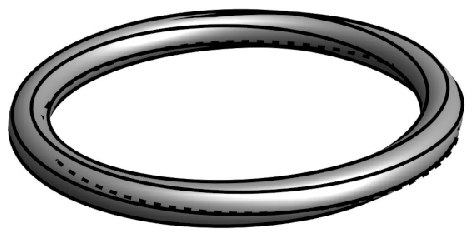}}
\caption{Deformation of slightly anisotropic helices of length $L=2\pi \rho_{(0)}$, $\chi_3=2/3$ and wavenumber $q=1/(2\rho_{(0)})$ ($n=1$) without boundary changes of the radial coordinate. The magnitude of the perturbation has been exaggerated for visualization purposes. The original helices are shown with dashed lines.}
 \label{fig:8}
\end{figure}

\item 
\textbf{Fixed azimuthal and height coordinates at the boundaries}. Conditions (\ref{bc:phiz}) and (\ref{eq:L1const}), imply $q s_0 = \mathrm{mod}(n,2)\pi/4$, and $qL = \pi n/2$, $n \in \mathbb{N}$ or $q \rho_{(0)} = n/4$, so the twist is given by
\begin{equation}
\rho_{(0)} \kappa_{3(0)} = \frac{n}{4} + \rho_{(0)} \tau_{(0)} \,.
\end{equation}
Like for isotropic rods, even deformation modes, say $n=2m$, have the same value of the twist $\kappa_{3(0)}$ as those of case $\mathrm{I}$ with $n=m$. Also the lowest deformation mode is $n =1$, which we show in Fig. \ref{fig:9}. These helices do not satisfy the boundary condition (\ref{bc:rho}), so they have a displacement along the radial direction at the boundaries, as well as a change of their tangent vectors along the azimuthal and radial directions.
\\
The first order corrections to the cylindrical coordinates are given by
\begin{subequations} \label{eq:anisorhophizII}
\begin{align} 
&\mbox{Odd} \; n, &\quad  \rho_{(1)} &= a_{\rho(1)} \sin 2 q s  \,, 
&\quad
\phi_{(1)} &= -\frac{a_{\phi(1)}}{2 q}  \cos 2 q s  \,,  
& \quad
z_{(1)} &=  -\frac{a_{z(1)}}{2q} \cos 2 q s  \,,\\
&\mbox{Even} \; n, &\quad \rho_{(1)} &= a_{\rho(1)}  \cos 2 q s  \,, 
&\quad
\phi_{(1)} &= \frac{a_{\phi(1)}}{2q}  \sin 2 q s  \,,  
& \quad
z_{(1)} &= \frac{a_{z(1)}}{2q} \sin 2q s  \,,
\end{align}
\end{subequations}
We can determine $\chi_{2(1)}$ in terms of the first order change of the radial coordinate at the boundaries
\begin{subequations}
\begin{align}
\mbox{Odd} \; n, &\quad \chi_{2(1)}= (-1)^{(n-1)/2} \, \frac{4 q \left(\kappa_{(0)}^2-4 q^2+(\rms_{3(0)}-2 \tau_{(0)})^2\right)}{\kappa_{(0)}  (\rms_{3(0)}+2 (q-\tau_{(0)}))} \rho_{b(1)} \,, \quad \pm \rho_{b(1)}:= \rho_{(1)}\left(\pm L/2\right) \,, \\
\mbox{Even} \; n, &\quad  \chi_{2(1)} = (-1)^{n/2} \, \frac{4 q \left(\kappa_{(0)}^2-4 q^2+(\rms_{3(0)}-2 \tau_{(0)})^2\right)}{\kappa_{(0)}  (\rms_{3(0)}+2 (q-\tau_{(0)}))} \rho_{b(1)}\,, \quad \rho_{b(1)}:=\rho_{(1)}(\pm L/2) \,.
\end{align}
\end{subequations}
\begin{figure}[htb]
 \centering
\subfigure[$\alpha=\frac{\pi}{64}$]{\includegraphics[scale=0.6]{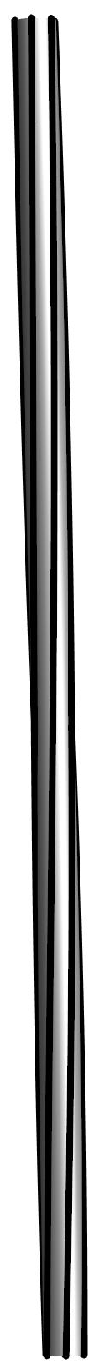}}
\hfil
\subfigure[$\alpha=\frac{\pi}{8}$]{\includegraphics[scale=0.6]{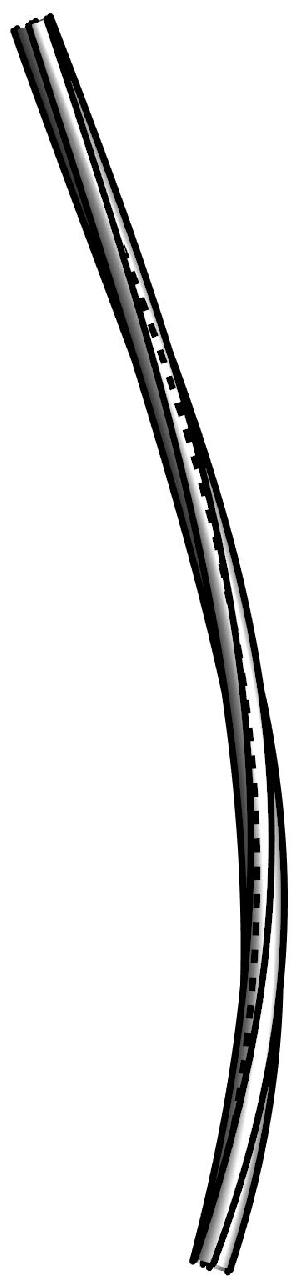}}
\hfil
\subfigure[$\alpha=\frac{\pi}{4}$]{\includegraphics[scale=0.6]{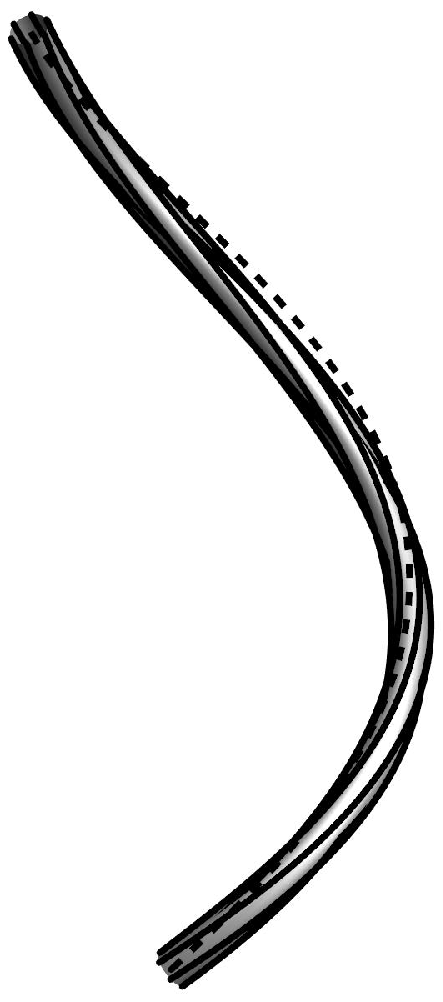}}
\hfil
\subfigure[$\alpha=\frac{3\pi}{8}$]{\includegraphics[scale=0.6]{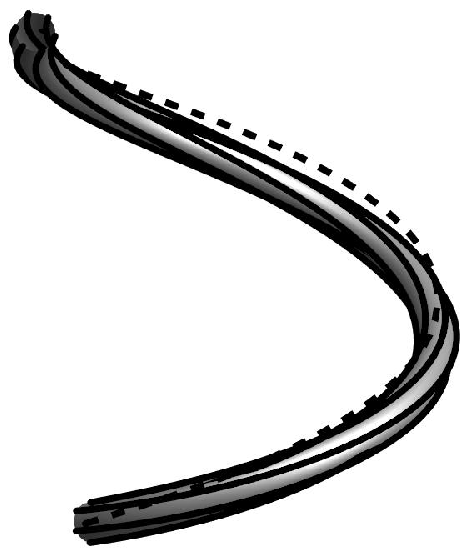}}
\hfil
\subfigure[$\alpha=\frac{\pi}{2}$]{\includegraphics[scale=0.62]{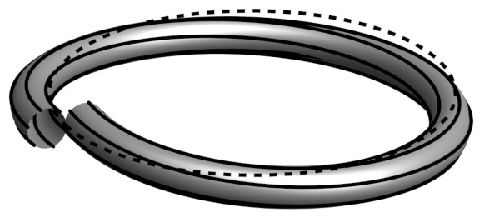}}
\caption{Deformation of slightly anisotropic helices of length $L=2\pi \rho_{(0)}$, $\chi_3=2/3$ and wavenumber $q=1/(4 \rho_{(0)})$ ($n=1$) without change of the azimuthal and height coordinates at the boundaries. The magnitude of the perturbation has been exaggerated for visualization purposes. The original helices are shown with dashed lines.}
 \label{fig:9}
\end{figure}

\item
\textbf{Fixed boundaries}. For $s_0 = 0$, boundary conditions (\ref{bc:rho}) and (\ref{bc:phiz}) are satisfied if the wavenumber $q$ fulfills the transcendental equation 
\begin{equation}
\tan q L = q L\,.
\end{equation}
The first three solutions to this equation are $q L =4.493, 7.725, 10.904$ \cite{LandauBook}, so $q \rho_{(0)}= 0.715, 1.23, 1.735$. For the first wavenumber the twist is given by
\begin{equation}
\rho_{(0)} \kappa_{3(0)} = 0.715 + \rho_{(0)} \tau_{(0)} \,.
\end{equation}
In this case the first order corrections to the cylindrical coordinates of the curve are given by
\begin{subequations}
\begin{eqnarray}
\rho_{(1)}(s) &=& a_{\rho(1)} \left( \cos 2q s -\cos q L   \right) \,,\\
\phi_{(1)}(s) &=& \frac{a_{\phi(1)}}{2 q} \left( \sin 2 q s - 2 \sin q L \left(\frac{s}{L} \right)   \right)  \,,  \\
z_{(1)}(s) & = & \frac{a_{z(1)}}{2q} \left( \sin 2 q s - 2 \sin q L \left(\frac{s}{L} \right)  \right)  \,.
\end{eqnarray}
\end{subequations}
Since, $\rho_{(1)}'(\pm L/2) \propto \sin q L \neq 0$, the tangent vector changes along the radial direction at the boundaries.
The parameter $\chi_{2(1)}$ can be expressed in terms of the first order change in the radial coordinate at the midpoint of the rod
\begin{equation}
\chi_{2(1)} =  \frac{4 q \left(\kappa_{(0)}^2-4 q^2+(\rms_{3(0)}-2 \tau_{(0)})^2\right)}{\kappa_{(0)}  (\rms_{3(0)}+2 (q-\tau_{(0)})) (1-\cos q L)} \, \rho_{0(1)}\,, \quad \rho_{0(1)}:=\rho_{(1)}(0)\,.
\end{equation}
Deformed helices with the lowest wavenumber are plotted in Fig. \ref{fig:10}.
\begin{figure}[htb]
 \centering
\subfigure[$\alpha=\frac{\pi}{64}$]{\includegraphics[scale=0.6]{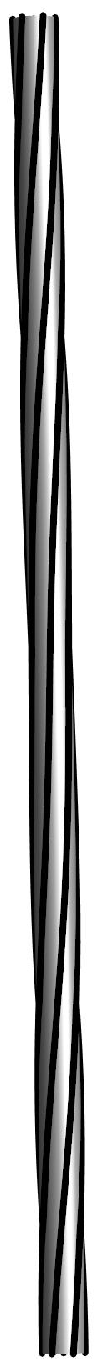}}
\hfil
\subfigure[$\alpha=\frac{\pi}{8}$]{\includegraphics[scale=0.6]{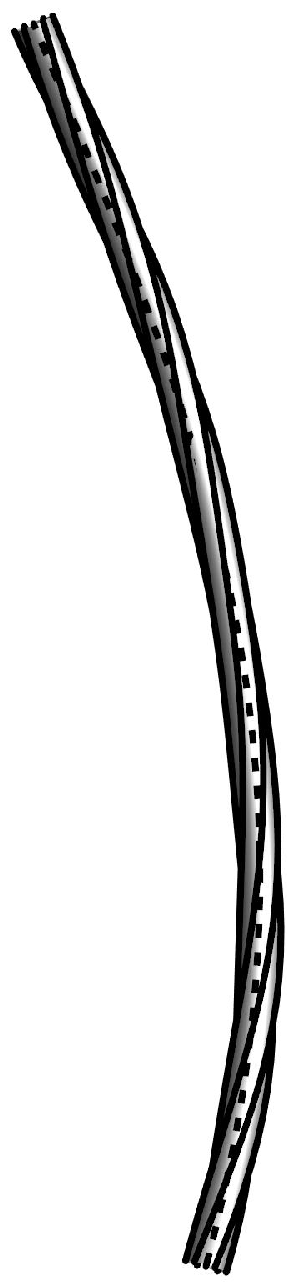}}
\hfil
\subfigure[$\alpha=\frac{\pi}{4}$]{\includegraphics[scale=0.6]{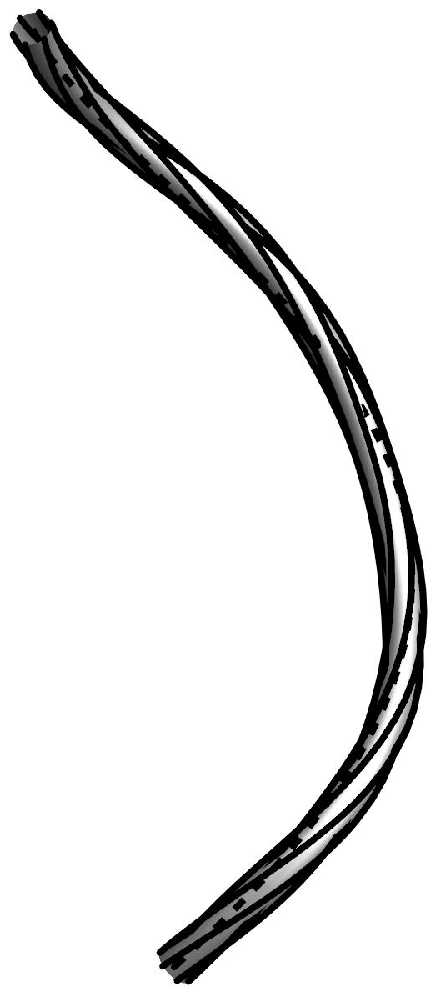}}
\hfil
\subfigure[$\alpha=\frac{3\pi}{8}$]{\includegraphics[scale=0.6]{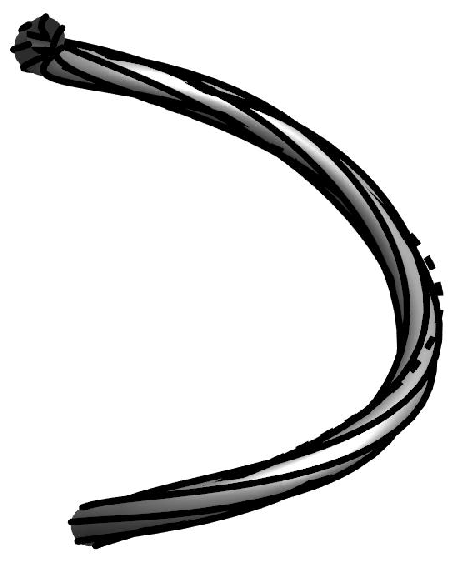}}
\hfil
\subfigure[$\alpha=\frac{\pi}{2}$]{\includegraphics[scale=0.6]{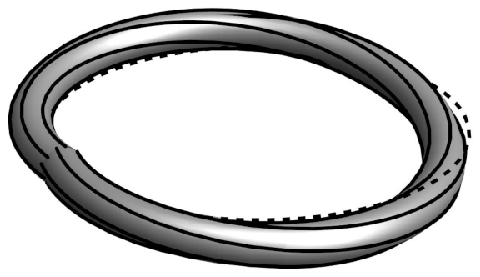}}
\caption{Deformation of slightly anisotropic helices of length $L= 2 \pi \rho_{(0)}$, $\chi_3=2/3$ and wavenumber $q= 0.715/\rho_{(0)}$. The magnitude of the perturbation has been exaggerated for visualization purposes. The original helices are shown with dashed lines.}
 \label{fig:10}
\end{figure}
\end{enumerate}
Since the anisotropy only enters to first order, the zeroth order of the scaled total energy of the anisotropic helices is also given by Eq. (\ref{eq:istoten}). The zeroth order of the force, torque, intrinsic tension and total energy of the helices are plotted in Figs. \ref{fig:11}(a)-(d) as a function of $\alpha$. The first family is shown with black lines (lower one with $n=1$ and upper one with $n=2$); for the second family are shown with gray lines (bottom line corresponds to $n=1$ and the top line to $n=3$, $n=2$ coincides with $n=1$ of the first family); for the third family they are shown with light gray lines (lower and upper ones with $q \rho_{(0)} = 0.714, 1.23$ respectively). The case of pure bending ($\rms_{3(0)}=0$) is shown with dashed lines.
\\
The scaled force $\rho_{(0)}^2 \rmf_{(0)}$, illustrated in Fig.  \ref{fig:11}(a), vanishes for the line with $\alpha=0$, but it is positive for the deformed helices ($\rmf_{(0)} >0$), so they are under tension, whereas helices under pure bending are always under compression ($\rmf_{(0)} <0$). Except for the lowest mode of deformation of the second family, which vanishes for $\alpha=\pi/4$ (and thus $\rms_{3(0)}=\tau_{(0)}$), the force on the deformed helices increases reaching a maximum value of $\rmf_{(0)}=\chi_{3} q/\rho_{(0)}$ for the circle with $\alpha=\pi/2$.
\\
The scaled torque $\rho_{(0)} \rmj_{(0)}$, plotted in Fig. \ref{fig:11}(b), vanishes initially for the line in the pure bending case, whereas for the three families it starts with a value $\rho_{(0)} \rmj_{(0)}= \chi_3 q$, and ends with a unit value $\rho_{(0)} \rmj_{(0)}= 1$ for the circle.
\\
The scaled intrinsic tension $\rho_{(0)}^2 \nu_{0}$, shown in Fig. \ref{fig:11}(c), begins with a value $\chi_3 (\rho_{(0)} q)^2/2$ for the line, increases and then decreases towards the circle with a value of $(1+\chi_3 (\rho_{(0)} q)^2)/2$.
\\
The scaled total energy $\rho_{(0)} h_{0}/\pi$, shown in Fig. \ref{fig:11}(d), begins with a value $\chi_3 (\rho_{(0)} q)^2$ for the line and ends with a value $1+\chi_3 (\rho_{(0)} q)^2$ for the circle.

\section{Discussion and conclusions} \label{Sec:conclusions}

We have developed a variational framework adapted to the material frame, which provides an alternative approach to the study of curves with geometric energies, not only in the harmonic approximation but also in the anharmonic approximation. We have derived the three equations governing the equilibrium states of such curves, two of them corresponding to the EL equations resulting from the normal projection of the conservation law of the force vector, and another one obtained from the tangential projection of the conservation law of the torque vector. So, in general they represent a system of three differential equations for the three material curvatures, whose initial conditions are provided by the vanishing of the boundary terms, being typically fixed or free ends and the specification of the material frame at the edges, unlike the usual Kirchhoff rods where the forces and torques at the boundaries are specified. The conserved quantities permitted us to identify the first integrals of the EL equations, which for energies not depending on second or higher order derivatives, provide directly quadratures for the material curvatures. By contrast, in general such quadratures cannot be obtained in the FS framework, because the component along the binormal depends the second order derivative of the EL derivative of the energy density with respect to the torsion \cite{CapoChryssGuv2002}.
\\
One of the benefits of working with the material frame, is that the twist is treated on the same footing as the normal curvatures, and not merely as a gauge degree of freedom like the twist angle in the FS framework. Moreover, one consequence of the fact that one arc-length derivative is encoded in the twist, is the reduction of the order of arc length derivatives in the force and torque vectors, as well as in the EL equations and their first integrals (rendering their expressions  more concise and symmetric), as compared with their counterparts expressed in the FS variables.
\\
Usually, the centerline of the rods is reconstructed employing Euler angles  \cite{vanderHeijden2000, AudolyBook, OReillyBook}. Here, we adopted the cylindrical coordinate system defined by the conserved force and torque vectors, which facilitates the integration of the centerline embedding functions, once the material curvatures have been integrated.  
\\
We have shown how our framework is appropriate for the description of elastic rods. The first integrals for isotropic rods reproduce the identification of an isotropic Kirchhoff rod as an Euler elastica with a renormalized torsion, along with the equation expressing the torsion as a function of the curvature \cite{Langer1996}. This reduction in number of equations is possible because for isotropic Kirchhoff rods the twist is conserved, a consequence of the symmetric dependence of the bending energy on the normal curvatures.
\\
For anisotropic rods the material curvatures are the natural choice to quantify the asymmetry of the bending energy, for they provide a direct decomposition of the curvature. In spite of the fact that the study of equilibrium states of anisotropic rods is considerably more involved because the twist is not conserved, it can be eased by working with the two first integrals and equation for the twist, because they provide a system of first order differential equations for the material curvatures which is suitable for a numerical analysis.
\\
To illustrate our framework, we analyzed the twisting instabilities on isotropic helices. We considered three deformations with different boundary conditions. The first one, with an integer wavenumber, correspond to deformation of helices with their boundaries fixed along the radial coordinate. The second one, with a half integer wavenumber, is obtained from perturbations changing the boundaries only along the radial direction. The third one, whose wavenumber is a real number corresponding the solution of a transcendental equation, results from deformations maintaining the boundaries. Likewise, we analyzed deformations of anisotropic helices with similar boundary conditions, but we found that the bending anisotropy introduces an additional perturbation whose wavenumber is the triple of the wavenumber of the initial twisting. It is conceivable that perturbations with higher wavenumbers enter in the higher corrections of the material curvatures. We have also discussed the forces and torques required to drive these three kinds of deformations of helical rods, along with their intrinsic tension and total energies.
\\
These perturbative results for isotropic and anisotropic rods, which describe the initial deformations of helical rods and generalize the twisting instabilities of linear and circular rods, provide a starting point to the analysis of equilibrium states in the non-linear regime. Besides, it would be interesting to analyze the stability of the deformed equilibrium states. To that end, it is required to calculate the second variation of the energy \cite{GelfandBook} and to analyze the spectrum of the associated differential operator \cite{Guven2011, GuvVaz2012, Vazquez2017}.
\\
There are several directions in which this framework could be useful. Although we focused our discussion only on energies depending quadratically on the material curvatures, as  pointed out for anisotropic rods, it can be applied to address models taking into account the asymmetry of the rods by introducing in the energy cubic and quartic terms in the material curvatures. Furthermore, as shown in the derivation, this framework can be also applied to energies depending on the derivatives of the material curvatures, such as that of Helfrich rods \cite{Helfrich1991, Tu2008, Starostin2009}, or energies depending on the torsion. The latter ones would involve first order derivatives of the normal material curvatures, so the momenta conjugate to the curvatures will be relevant. Examples of this kind of energies are the total curvature $\calF = \kappa^2 + \tau^2$, or its generalization given by the sum of the bending and torsional energies with spontaneous curvature and torsion, $\calF = (\rmk/2) \left(\kappa - c_0 \right)^2+  (\rmt/2) \left(\tau - \tau_0 \right)^2$, \cite{Starostin2008}. Furthermore, this framework could be generalized to include a dependence of the energy on the material basis, as in the case of paramagnetic filaments \cite{Biswal2003, Goubalt2003, Cebers2003, Vazquez2017}, or for inhomogeneous rods with variable bending or twisting moduli \cite{Guven2019, Palmer2020A, Palmer2020B}. Likewise, this framework could be extended to address Cosserat rods by relaxing the untretchability of the rods, as well as by taking into account the shear in their cross section \cite{Rubin2000, Gazzola2018}.

\section*{Acknowledgements}

D. A. S. was partially supported by UADY under Projects PFCE-2019-12 and P/PROFEXE-2020-31MSU0098J-13. P. V. M. acknowledges support by CONACYT under Grant Cátedra CONACYT 439-2018.

\begin{appendix}

\section{Energy and stresses in the FS frame} \label{App:FStoMF}

Throughout this work we have described  the force and torque vectors, as well as the equilibrium equations, with respect to the material frame. In this appendix we show how those expressions are related to their counterparts in the FS frame adapted to the central curve $\Gamma$, which is formed by the tangent $\mathbf{T} = \bfY'(s)$, the principal normal $\mathbf{N}$ and the binormal $\mathbf{B}$. The FS formulas, describing how this frame rotates along the curve, are 
\begin{equation} \label{def:FSeqs}
\mathbf{T}' = \mathbf{D} \times \mathbf{T}=  \kappa \, \mathbf{N}\,, \quad \mathbf{N}' = \mathbf{D} \times \mathbf{N} = -\kappa \, \mathbf{T} + \tau \mathbf{B} \,, \quad \mathbf{B}' = \mathbf{D} \times \mathbf{B} = -\tau \, \mathbf{N}\,.
\end{equation}
where $\kappa$ and $\tau$ are the curvature and torsion, whereas $\mathbf{D} = \tau \, \mathbf{T} + \kappa \, \mathbf{B}$ is the Darboux vector.
\\
The material frame $\{\bfe_1, \bfe_2, \, \bfe_3 \}$, is related to the FS frame by a clockwise rotation about the tangent $\mathbf{T}$,  which is chosen as $\bfe_3$. Thus the normal basis on the orthogonal plane $\{\bfe_1, \bfe_2\}$, is related to the FS normal basis $\{\mathbf{N}, \mathbf{B}\}$ by
\begin{equation} \label{def:eibasis}
\bfe_1 = \cos \theta \, \mathbf{N} + \sin \theta \, \mathbf{B}  \,, \quad \bfe_2 = -\sin \theta \, \mathbf{N} + \cos \theta \, \mathbf{B} \,, \quad \bfe_3 = \mathbf{T} \,.
\end{equation}
Since there is a freedom in the description of the orthogonal space to the curve, it is important to bear in mind that $\theta$ is not a physical degree of freedom, but rather a gauge function in the selection of the basis in plane orthogonal to the curve. This gauge could be fixed by choosing a curve $\gamma$ on the surface of the rod, in which case $\bfe_1$ can be chosen as the radial vector field centered on $\Gamma$ and pointing towards $\gamma$.
\\
By comparing the structure equations of the material frame (Eqs. (\ref{def:Streqs})) with the FS formulas (Eqs. (\ref{def:FSeqs})), it is straightforward to check that the material curvatures are related to the curvature and torsion by the following relations
\begin{equation} \label{eq:MatFS}
\kappa_1 =\kappa \sin \theta \,, \quad \kappa_2 = \kappa \cos \theta\,, \quad \kappa_3 =  \tau + \theta'\,.
\end{equation}
$\kappa_1$ and $\kappa_2$ provide a decomposition of the curvature, whereas the local twist $\kappa_3$ is the sum of the torsion of the curve and the rate of rotation of the material frame with respect to the FS frame. Conversely we have that $\kappa$, $\tau$ and $\theta$ can be expressed in terms of the material curvatures $\kappa_i$ as
\begin{equation} \label{eq:FSMat}
\kappa^2 = \kappa_1^2 + \kappa_2^2\,, \quad \tau = \frac{\kappa_1 \, \kappa'_2 - \kappa_2 \, \kappa'_1 }{\kappa_1^2 + \kappa_2^2} + \kappa_3 \,, \quad  \theta = \arctan \frac{\kappa_1}{\kappa_2}\,.  
\end{equation}
These equations allow us to change from one set of variables to the other.
\\
The energy can be regarded as a functional depending on the curvature, torsion and the angle between the both frames, i. e. $E = \int \calF(\kappa, \tau, \theta) \rmd s$. The EL derivatives of the energy with respect to $\kappa$, $\tau$ and $\theta$ are
\begin{equation} \label{eq:FSELders}
 \calF_\kappa := \frac{\delta E}{\delta \kappa} = \frac{ \partial \calF }{\partial \kappa} - \left(\frac{ \partial \calF }{\partial \kappa'}\right)' \,, 
 \quad
 \calF_\tau := \frac{\delta \calF}{\delta \tau} = \frac{ \partial \calF }{\partial \tau} - \left(\frac{ \partial \calF }{\partial \tau'}\right)'\,, 
 \quad 
 \calF_\theta := \frac{\delta \calF}{\delta \theta} = \frac{ \partial \calF }{\partial \theta} - \left(\frac{\partial \calF }{\partial \theta'} \right)' + \left(\frac{\partial \calF }{\partial \theta''}\right)'' \,.
\end{equation}
Using relations (\ref{eq:MatFS}), we can express these FS EL derivatives in terms of the material EL derivatives defined in Eq. (\ref{def:Si}), 
\begin{equation} \label{eq:FSMatELders}
 \kappa \calF_\kappa = \kappa_1 \rmS_1 + \kappa_2 \rmS_2  \,, 
 \quad
 \calF_\tau = \rmS_3 \,, 
 \quad 
 \calF_\theta = \kappa_2 \rmS_1 - \kappa_1 \rmS_2 - \rmS'_3\,.
\end{equation}
As for the $\theta$ dependence, since it is a scalar field along the curve independent of the other geometrical degrees of freedom, the variation of the energy with respect to $\theta$ ought to hold independently, such that the EL equation $\calF_\theta =0$ is satisfied \cite{Tu2008, Starostin2009}. This reproduces Eq. (\ref{eq:sym12}), which stems from the fact that the arc length derivative of the intrinsic torque is orthogonal to the tangent vector, as shown in Sec. (\ref{sec:FMEucInv}). 
\\
Inverting Eqs. (\ref{eq:FSELders}) (or using Eqs. (\ref{eq:FSMat})) we describe the normal material EL derivatives of the energy in terms of the FS EL derivatives, 
\begin{equation} \label{eq:ELdersmathtoFS}
 \rmS_1 =  \sin \theta \, \calF_\kappa + \cos \theta \, \frac{\calF_\tau'}{\kappa} \,, \quad \rmS_2  =  \cos \theta \, \calF_\kappa - \sin \theta \, \frac{\calF_\tau'}{\kappa} \,.
\end{equation}
The force and intrinsic torque vectors in the FS frame are \cite{CapoChryssGuv2002, Starostin2009}
\begin{subequations} \label{FSvectFSfr}
\begin{eqnarray}
\bfF_{FS} &=& -\left( \kappa \calF_\kappa + \tau \calF_\tau + \calH - \mu \right) \, \mathbf{T} 
- \left( \calF_\kappa^{'} + \frac{\tau}{\kappa} \calF_\tau^{'} \right) \mathbf{N} 
+\left( \left( \frac{{\calF}_\tau^{'}}{\kappa} \right)^{'} + \kappa {\calF}_\tau - \tau {\calF}_\kappa \right) \mathbf{B} \,, \label{FvectFSfr}\\
\bfS_{FS} &=& \calF_\tau \bfT + \frac{\calF_\tau'}{\kappa}\bfN +\calF_\kappa \bfB\,. \label{SvectFSfr}
\end{eqnarray}
\end{subequations}
where in this case the Hamiltonian is defined by
\begin{equation}
\calH = \kappa'  p_\kappa + \tau'  p_\tau + \theta' p_\theta + \theta'' P_{\theta'} -\calF \,, \quad p_\kappa = \frac{\partial \calF}{\partial \kappa'}\,, \quad p_\tau = \frac{\partial \calF}{\partial \tau' } \,, \quad  p_\theta = \frac{\partial \calF}{\partial \theta'} - \left(  \frac{\partial \calF}{\partial \theta''} \right)'\,, \quad P_{\theta'} =  \frac{\partial \calF}{\partial \theta''}\, .
\end{equation}
Although the EL derivative with respect to $\theta$ does not enter the tangential component, since $\calF_\theta=0$, the dependence on $\theta$ enters through the Hamiltonian. In general, these terms account for the twisting degree of freedom of the curve by quantifying the rate of rotation of the FS frame. 
\\
Since $\theta'$ and $\tau$ are assumed to enter the energy only through their sum, the FS momenta are related to the material momenta by the following relations
 \begin{equation}
\kappa p_\kappa = \kappa_1 p_1 + \kappa_2 p_2 \,, \quad p_\tau = P_{\theta'} = p_3\,, \quad p_\theta =  \kappa_2 p_1 - \kappa_1 p_2 +\rmS_3\,.  
 \end{equation}
Substituting these relations between the two sets of momenta along with Eqs. (\ref{eq:FSMatELders}) in the FS force and intrinsic torque vectors, Eqs. (\ref{FSvectFSfr}), we recover their expressions in the material frame given by Eqs. (\ref{eq:vecF}) and (\ref{def:MvecMF}).
\\
Expressing the derivative of the force vector in the form $\bfF' = \mathcal{E}_\mathbf{N} \, \mathbf{N} + \mathcal{E}_\mathbf{B} \, \mathbf{B}$, and taking into account that it is conserved, we obtain the EL equations along the principal normal and binormal \cite{CapoChryssGuv2002, Tu2008, Thamwattana2008A, Thamwattana2008B, Starostin2009}:
\begin{subequations} \label{eq:ELFSframe}
\begin{eqnarray} 
\mathcal{E}_\mathbf{N} &=& - \calF_\kappa'' - 2 \tau \left(\frac{\calF_\tau'}{\kappa}\right)' - \tau' \, \frac{\calF_\tau'}{\kappa} - \kappa \left(\kappa \calF_\kappa + 2 \, \tau \, \calF_\tau + \calH - \mu\right) + \tau^2 \, \calF_\kappa=0\,,\\
\mathcal{E}_\mathbf{B} &=&  \left(\frac{\calF_\tau'}{\kappa}\right)'' + \left(\kappa \, \calF_\tau\right)' - \,\frac{ \tau^2}{\kappa} \calF_\tau' - 2 \tau \calF_\kappa' - \tau ' \, \calF_\kappa =0\,.
\end{eqnarray}
\end{subequations}
Note that these FS EL equations are one order higher in arc length derivatives and are not as symmetric as the material EL eqs. (\ref{eq:EL12})
\\
Substituting Eqs. (\ref{eq:ELdersmathtoFS}) in Eqs. (\ref{eq:EL12}), and using Eqs. (\ref{eq:ELFSframe}), we can express the material EL derivatives as linear combinations of the FS EL derivatives,
\begin{subequations}
\begin{eqnarray} \label{eq:eps12epsNB}
\mathcal{E}_1 &=& \cos \theta \, \mathcal{E}_\mathbf{N} + \sin \theta \, \mathcal{E}_\mathbf{B}\,, \\
\mathcal{E}_2 &=& -\sin \theta \, \mathcal{E}_\mathbf{N}+ \cos \theta \, \mathcal{E}_\mathbf{B}\,.
\end{eqnarray}
\end{subequations}
Therefore, the vanishing of the FS EL derivatives implies the vanishing of the material EL derivatives.
\\
In these expressions it is manifest that any material frame is equivalent to the FS only if it is obtained by a constant rotation, i.e. when $\theta' = 0$.
\\
For the energy (\ref{eq:BendEnDens}) of isotropic rods, the EL equations (\ref{eq:ELFSframe}) read
\begin{subequations}
 \begin{eqnarray}
-\kappa ''- \kappa  \left(\frac{1}{2} \left(\kappa ^2-c_0^2\right)-\zeta \right)+\tau \left( \tau (\kappa-c_0) -\rms_3 \kappa \right)&=&0\,,\\
(\rms_3 -2 \tau ) \kappa' - (\kappa - c_0) \tau'  &=&0\,.
\end{eqnarray}
\end{subequations}
Integration of the second equation reproduces the first integral (\ref{eq:tauIsoKR}), which substituted in the first one yields
\begin{equation}
 -\kappa ''- \kappa  \left(\frac{1}{2} \left(\kappa ^2-c_0^2\right)-\zeta \right)+\frac{\upsilon ^2}{\left(\kappa -c_0\right){}^3} + \frac{c_0 s_3 \upsilon }{\left(\kappa -c_0\right){}^2}  -\frac{s_3^2}{4} (\kappa +c_0)=0\,.
\end{equation}
Integrating this equation reproduces the first integral (\ref{eq:Fstintkappaisorod}).

\section{Alternative derivation of EL equations} \label{App:altderELeqs}

Consider a local deformation of the rod spanned in the material frame $\delta \bfY = \delta \psi_i \bfe_i$. The change in arc-length element is given by $\delta \rmd s = \bfe_3 \cdot \delta \bfY' \rmd s$. Thus, the fixed length constraint, implying that $\delta \bfY' \cdot \bfe_3 =0$, establishes the following relation between the tangential and normal components of the deformation
\begin{equation} \label{eq:Fixlengthcond}
\delta \psi_3' + \kappa_1 \, \delta \psi_2 - \kappa_2 \, \delta \psi_1 = 0\,.
\end{equation}
We can use this relation to eliminate the derivative of the tangential deformation in favor of the normal deformations.
Moreover, under this deformation the material frame changes as 
\begin{equation} \label{def:deltaei}
\delta \bfe_i = \delta \bm{\Theta} \times \bfe_i = -\epsilon_{ijk} \delta {\Theta}_j \bfe_k \,, 
\end{equation}
where  $\delta \bm{\Theta} = \delta {\Theta}_i \bfe_i $ is the vector quantifying the rate of rotation of the adapted basis due to the variation. Its components are given by
\begin{equation}
 \delta \Theta_i = -\frac{1}{2} \epsilon_{ijk} \bfe_j \cdot \delta \bfe_k\,.
\end{equation}
From Eqs. (\ref{def:Streqs}) and (\ref{def:deltaei}), and taking into account the commutativity of arc-length derivation and variation, $\partial_s \delta = \delta \partial_s$, we obtain the variation of the Darboux vector $\delta \bfD$ in terms of $\delta \bm{\Theta}$
\begin{equation} \label{deltaD}
 \delta \bfD = \delta \bm{\Theta}' + \delta \bm \Theta \times \bfD\,.
\end{equation}
From the identity $\delta \bfY' = \delta \bfe_3= (\delta \psi_k' + \varepsilon_{ijk} \kappa_i \delta \psi_j) \bfe_k$, implying the isometry condition (\ref{eq:Fixlengthcond}), we can express the normal components of $\delta \bm{\Theta}$ in terms of the components of $\delta \bfY$
\begin{subequations} \label{eq:deltaTheta12}
\begin{eqnarray}
\delta \Theta_1 = \bfe_3 \cdot \delta \bfe_2 &=& - \nabla_s \delta \psi_2 + \kappa_1 \, \delta \psi_3 \,,\\ 
\delta \Theta_2 =\bfe_1 \cdot \delta \bfe_3 &=& \nabla_s \delta \psi_1 +  \kappa_2 \, \delta \psi_3 \,,
\end{eqnarray}
\end{subequations}
From the projections of Eq. (\ref{deltaD}) onto the material frame we obtain that the variations of the material curvatures are given by 
\begin{equation} \label{deltakappai}
\delta \kappa_i = \delta \Theta_i ' + \mathbf{D} \cdot \delta \bfe_i= \delta \Theta_i' + \epsilon_{ijk} \kappa_j \delta \Theta_k \,.
\end{equation}
The normal projections provide the two equations
\begin{subequations}
\begin{eqnarray}
\delta \kappa_1 - \kappa_2 \delta \Theta_3 &=& \nabla_s \delta \Theta_1 \,,\\
\delta \kappa_2 + \kappa_1 \delta \Theta_3 &=& \nabla_s \delta \Theta_2 \,.
\end{eqnarray}
\end{subequations}
Taking a linear combination of these equations and substituting Eqs. (\ref{eq:deltaTheta12}), we determine the tangential component of $\delta \bm{\Theta}$
\begin{equation}
\delta \Theta_3 = \bfe_2 \cdot \delta \bfe_1 =  
\frac{1}{\kappa^2} \left[\kappa_1 \, \nabla^2_s \delta \psi_1 + \kappa_2 \, \nabla^2_s \delta \psi_2 + \left( \kappa_1 \nabla_s \kappa_2 - \kappa_2 \nabla_s \kappa_1\right) \delta \psi_3 \right] + \delta \theta   \,, 
\end{equation}
where 
\begin{equation}
\delta \theta = \frac{1}{\kappa^2} \left(\kappa_2 \, \delta \kappa_1 - \kappa_1 \, \delta \kappa_2  \right)\,.
\end{equation}
 is the change of the angle measuring the rotation of the rod, see Appendix \ref{App:FStoMF}. These equations determine the variation of the material frame, while all the other components are obtained by using the orthonormality of the basis.
\\
Now, we can express the variations of the material curvatures, given in Eq. (\ref{deltakappai}), as
\begin{subequations}
\begin{eqnarray} \label{eq:deltakappai}
\delta \kappa_1 &=& \nabla_s \delta \Theta_1 + \kappa_2 \delta \Theta_3 = \frac{\kappa_1}{\kappa} \, \delta \Psi + \kappa_2 \, \delta \theta  \,, \\
\delta \kappa_2 &=& \nabla_s \delta \Theta_2 - \kappa_1 \delta \Theta_3 =\frac{\kappa_2}{\kappa} \, \delta \Psi - \kappa_1 \, \delta \theta \,, \\
\delta \kappa_3 &=& \delta \Theta_3' + \kappa_1 \, \delta \Theta_2 - \kappa_2 \, \delta \Theta_1 = \delta \Theta_3'  + \kappa_1 \, \nabla_s \delta \psi_1 + \kappa_2 \, \nabla_s \delta \psi_2\,,
\end{eqnarray}
\end{subequations}
where $\delta \Psi$ is defined by
\begin{equation} \label{def:eta}
\delta \Psi = \frac{1}{\kappa} \left( \kappa_2 \nabla_s^2 \delta \psi_1 - \kappa_1 \nabla_s^2 \delta \psi_2\right) + (\kappa \delta \psi_3)' \,.
\end{equation}
As shown in Ref. \cite{Powers2010} for the case of total curvature and twisting, by using these expressions in the variational principle and isolating $\delta \psi_1$ and $\delta \psi_2$, as well as $\delta \theta$, through integrations by parts, their coefficients reproduce the normal EL equations, (\ref{eq:EL12}) and  Eq. (\ref{eq:sym12}) respectively. The variation $\delta \psi_3$ appears within a total derivative, so it only contributes to the boundary terms.

\end{appendix}

\bibliography{BibKirchhoff}

\end{document}